%% file: etaprime.tex
\documentclass[twocolumn,epjc3]{svjour3mod}  
\RequirePackage[T1]{fontenc}
\RequirePackage[utf8]{inputenc}
\smartqed
\RequirePackage{graphicx}
\RequirePackage{mathptmx}
\RequirePackage{flushend}
\RequirePackage{cite}
\RequirePackage{latexsym}
\RequirePackage[colorlinks,citecolor=blue,urlcolor=blue,linkcolor=blue]{hyperref}
\RequirePackage{amsmath,amssymb}
\RequirePackage{psfrag,booktabs}
\RequirePackage{textcomp,gensymb}

\usepackage[figure,table]{hypcap}

\makeatletter
  \def\my@tag@font{\normalsize}
  \def\maketag@@@#1{\hbox{\m@th\normalfont\my@tag@font#1}}
  \let\amsmath@eqref\eqref
  \renewcommand\eqref[1]{{\let\my@tag@font\relax\amsmath@eqref{#1}}}
\makeatother

\DeclareUnicodeCharacter{00B0}{\degree}

\newcommand{\mpc}{M_\pi}
\newcommand{\mpn}{M_{\pi^0}}
\newcommand{\meta}{M_{\eta}}
\newcommand{\metap}{M_{\eta'}}
\newcommand{\setap}{s_{\eta'}}
\newcommand{\Qetap}{Q_{\eta'}}
\newcommand{\M}{\mathcal{M}}
\newcommand{\N}{\mathcal{N}}
\newcommand{\T}{\mathcal{T}}
\newcommand{\BR}{\mathcal{B}}
\newcommand{\disc}{{\rm disc}\,}
\renewcommand{\Re}{{\rm Re}\,}
\renewcommand{\Im}{{\rm Im}\,}
\providecommand{\order}{\mathcal{O}}
\newcommand{\GeV}{\,{\rm GeV}}

\newcommand{\nn}{\nonumber\\}
\newcommand{\chpt}{$\chi$PT}
\newcommand{\rcht}{R$\chi$T}
\newcommand{\bsp}{\begin{sloppypar}}
\newcommand{\esp}{\end{sloppypar}}
\newcommand{\diff}{\text{d}}

\begin{document}

\title{Dispersion relations for \boldmath{$\eta'\to \eta\pi\pi$}
}

\author{Tobias~Isken\thanksref{e1,hiskp}
	\and
	Bastian~Kubis\thanksref{e2,hiskp}
	\and
	Sebastian~P.~Schneider\thanksref{e3,hiskp}
	\and
	Peter~Stoffer\thanksref{e4,hiskp,ucsd}
}

\thankstext{e1}{e-mail: isken@hiskp.uni-bonn.de}
\thankstext{e2}{e-mail: kubis@hiskp.uni-bonn.de}
\thankstext{e3}{e-mail: schneider@hiskp.uni-bonn.de}
\thankstext{e4}{e-mail: pstoffer@ucsd.edu}

\institute{Helmholtz-Institut f\"ur Strahlen- und Kernphysik and Bethe Center for Theoretical Physics,
           Universit\"at Bonn, D-53115  Bonn, Germany \label{hiskp}
           \and
           Department of Physics, University of California at San Diego, La Jolla, CA 92093, USA \label{ucsd}
}

\date{}

\maketitle


\begin{abstract}
\bsp
We present a dispersive analysis of the decay amplitude for $\eta'\to\eta\pi\pi$ that is based on the
fundamental principles of analyticity and unitarity. In this framework,
final-state interactions are fully taken into account. Our dispersive representation relies only on
input for the $\pi\pi$ and $\pi\eta$ scattering phase shifts. Isospin symmetry allows us to
describe both the charged and neutral decay channel in terms of the same function.
The dispersion relation contains subtraction constants that cannot be fixed by unitarity.
We determine these parameters by a fit to Dalitz-plot data from the VES and BES-III
experiments. We study the prediction of a low-energy theorem and compare the dispersive fit
to variants of chiral perturbation theory.
\esp
\end{abstract}


\section{Introduction}

\begin{sloppypar}
The treatment of hadronic three-body decays using dispersion relations is a classic subject. Already in the 1960s, Khuri and Treiman developed a framework in the context of 
$K\to 3\pi$ decays~\cite{Khuri:1960zz}. One of its main virtues is the fact that the most important final-state interactions among the three pions are fully taken into account, in contrast to perturbative, 
field-theory-based approaches: analyticity and unitarity are respected exactly. This becomes the more important, the higher the mass of the decaying 
particle, hence the higher the possible energies of the two-pion subsystems within the Dalitz plot. But even in decays of relatively light pseudo\-scalar mesons like $\eta\to3\pi$, final-state interactions 
strongly perturb the spectrum. In such a case, a dispersive approach that resums final-state rescattering effects is essential to reach high precision; see Refs.~\cite{Kambor:1995yc,Anisovich:1996tx,Kampf:2011wr,Lanz:2013ku,Guo:2015zqa,Guo:2016wsi,Colangelo:2016jmc,Albaladejo:2017hhj}. In this article, we present the application of these techniques to the decay $\eta'\to\eta\pi\pi$.

The decay $\eta'\to\eta\pi\pi$ has received considerable interest in past years for several reasons. Due to the U(1)$_A$ anomaly the $\eta'$ is not a Goldstone boson and therefore 
``standard'' chiral perturbation theory (\chpt) based on the spontaneous breaking of SU(3)$\times$SU(3) chiral symmetry fails to adequately describe processes involving the $\eta'$. In the limit of the number of colors $N_c$ becoming large (``large-$N_c$ limit'') the
axial anomaly vanishes, which leads to a U(3)$_L\times$U(3)$_R$ symmetry, so that a simultaneous expansion in small momenta, small quark masses, and large $N_c$ gives rise to a power counting scheme
that in principle allows one to describe interactions of the pseudoscalar nonet ($\pi,K,\eta,\eta'$). However, the question whether this framework dubbed large-$N_c$ \chpt{}~\cite{Kaiser:2000gs,Escribano:2010wt} 
is actually well-established remains under discussion, mainly due to the large $\eta'$ mass. This is an issue that can in principle be addressed by a study of $\eta'\to\eta\pi\pi$. So far there 
are indications that a large-$N_c$ \chpt{} treatment alone is not sufficient to describe the decay, as final-state interactions play a rather important role, see Refs.~\cite{Escribano:2010wt,Riazuddin:1971ie}.

Furthermore, the $\eta'\to\eta\pi\pi$ decay channel could be used to constrain $\pi\eta$ scattering: the $\eta'$ mass is sufficiently small 
so that the channel is not polluted by nonvirtual intermediate states other than the rather well-constrained $\pi\pi$ scattering. In the past claims were made that the mechanism via the intermediate scalar resonance $a_0(980)\to\pi\eta$ even dominates 
the decay~\cite{Deshpande:1978iv,Singh:1975aq,Fariborz:1999gr}. These claims are based on effective Lagrangian models with the explicit inclusion of a scalar nonet incorporating the $a_0(980)$, $f_0(980)$, and 
$\sigma$ [$f_0(500)$] resonances. They were further supported by Refs.~\cite{Beisert:2002ad,Beisert:2003zs}: a chiral unitary approach shows large corrections in the $\pi\eta$ channel and there is a dominant low-energy
constant in the U(3) \chpt{} calculation that is saturated mostly by the $a_0(980)$. The $\pi\eta$ $P$-wave, however, was found to be strongly suppressed~\cite{Borasoy:2005du,RobinDiss,Kubis:2009sb}.
\end{sloppypar}

The $\eta'\to\eta\pi^0\pi^0$ decay channel is expected to show a cusp effect at the charged-pion threshold~\cite{Kubis:2009sb} that in principle can be used to obtain information on $\pi\pi$ scattering lengths.
So far this phenomenon has not been observed: the most recent measurement with the GAMS-$4\pi$ 
spectrometer did not have sufficient statistics to resolve this subtle effect~\cite{Blik:2009zz}.

The extraction of $\pi\eta$ scattering parameters such as the scattering length and the effective-range parameter is a more complicated subject compared to $\pi\pi$ scattering. There is no one-loop 
cusp effect as in the $\pi\pi$ channel, since the $\pi\eta$ threshold sits on the border of the physical region and not inside. The hope of extracting scattering parameters from a two-loop cusp is 
shattered likewise: there is a rather subtle cancellation of this effect at threshold (see Refs.~\cite{Schneider:2009rz,Diplom} for an elaborate discussion).

\begin{sloppypar}
Measurements of the Dalitz plot of the charged channel have been performed by the VES~\cite{Dorofeev:2006fb} and BES-III~\cite{Ablikim:2010kp} collaborations, while earlier measurements at rather low 
statistics have been reported in Refs.~\cite{Kalbfleisch:1974ku,Briere:1999bp}. The more recent measurements seem to disagree considerably with regard to the values of the Dalitz-plot parameters, and also in 
comparison with the GAMS-$4\pi$ measurement~\cite{Blik:2009zz} the picture remains inconsistent.
\end{sloppypar}

This article is structured as follows. We will start by discussing the necessary kinematics as well as the resulting analytic structure of $\eta'\to\eta\pi\pi$ in Sect.~\ref{sec:etapkinematics}, before deriving and analyzing dispersion relations 
for the decay in Sect.~\ref{sec:etapdisprel}. In Sect.~\ref{sec:numsol}, we will discuss the numerical solution of the dispersion relation. The results of the fits to data will be discussed in Sect.~\ref{sec:subconsts}.  Predictions for higher Dalitz-plot parameters, the occurrence of Adler zeros close to the soft-pion points, and predictions for the decay into the neutral final state are discussed in Sect.~\ref{sec:predictions}.
Finally, we perform a matching of the free parameters to extensions of \chpt{} in Sect.~\ref{sec:matching}.
Some technical details are relegated to the appendices.

\section{Kinematics}\label{sec:etapkinematics}
We define transition amplitude and kinematic variables of the $\eta'\to\eta\pi\pi$ decay in the usual fashion, 
\begin{align}
&\langle\pi^i(p_1)\pi^j(p_2)\eta(p_3)|T|\eta'(P_{\eta'})\rangle \nn &\quad = (2\pi)^4\delta^{(4)}(P_{\eta'}-p_1-p_2-p_3) \delta^{ij}\M(s,t,u)\,,
\end{align}
where $i,\,j$ refer to the pion isospin indices.\footnote{In the following, we will consider both the charged decay channel $\eta'\to\eta\pi^+\pi^-$ and the neutral channel $\eta'\to\eta\pi^0\pi^0$. They differ only by isospin-breaking effects.}  
We define the Mandelstam variables for the three-particle decay processes according to
\begin{equation}
s = (P_{\eta'} - p_3)^2\,,\quad t = (P_{\eta'} - p_1)^2\,,\quad u = (P_{\eta'} - p_2)^2\,,
\end{equation}
which fulfill  the relation
\begin{equation}\label{eq:etapstu}
s+t+u = \metap^2 + \meta^2 + 2\mpc^2 =: 3\setap\,.
\end{equation}
The process is invariant under exchange of the pions, that is, under $t\leftrightarrow u$. In the center-of-mass system of the two pions, one has
\begin{equation}
t(s,z_s) ,\, u(s,z_s) = \frac{1}{2}\big(3\setap - s \pm \kappa_{\pi\pi}(s)z_s\big)\,,
\end{equation}
where $z_s=\cos\theta_s$ refers to the scattering angle, 
\begin{align}
z_s = \cos\theta_s = \frac{t - u}{\kappa_{\pi\pi}(s)} \,, \quad
\kappa_{\pi\pi}(s) = \sigma(s)\lambda^{1/2}(\metap^2,\meta^2,s) \,,
\end{align}
with the K\"all\'en function $\lambda(x,y,z) = x^2 + y^2 + z^2 - 2(xy + xz + yz)$ 
and $\sigma(s) = \sqrt{1-{4\mpc^2}/{s}}$.
Similarly, in the center-of-mass system of the $t$-channel, one finds
\begin{equation}
s(t,z_t),\, u(t,z_t) = \frac{1}{2}\Bigl(3\setap - t \mp \frac{\Delta}{t} \mp \kappa_{\pi\eta}(t)z_t\Bigr)\,,
\end{equation}
with $\Delta:=(\metap^2-\mpc^2)(\meta^2-\mpc^2)$ and
\begin{align}
 z_t &= \cos\theta_t = \frac{t(u - s) - \Delta}{t\,\kappa_{\pi\eta}(t)}\,,\nn
\kappa_{\pi\eta}(t) &= \frac{\lambda^{1/2}(\meta^2,\mpc^2,t)\lambda^{1/2}(\metap^2,\mpc^2,t)}{t} \,.
\end{align}
Due to crossing symmetry, the $u$-channel relations follow from $t \leftrightarrow u$, $z_t \leftrightarrow -z_u$.

The physical thresholds in the three channels are given by
\begin{align}
	s_0 = 4\mpc^2 \, , \quad t_0 = u_0 = (\meta+\mpc)^2 .
\end{align}

\section{Dispersion relations for \boldmath{$\eta'\to\eta\pi\pi$}}
\label{sec:etapdisprel}

In this section, we set up dispersion relations for the decay process $\eta'\to\eta\pi\pi$, in analogy to previous work
on different decays into three pions~\cite{Khuri:1960zz,Anisovich:1996tx,Kambor:1995yc,Niecknig:2012sj}.
The idea is to derive a set of integral equations for the scattering processes $\eta'\eta\to\pi\pi$ and $\pi\eta'\to\pi\eta$ with hypothetical mass assignments that make these (quasi-)elastic: 
in such a  kinematic regime the derivation is straightforward.
The dispersion relation for the decay channel is then obtained by analytic continuation of the scattering processes to the decay region. 

\begin{sloppypar}
We will begin our discussion by decomposing the amplitude in terms of functions of one Mandelstam variable only. This form will prove 
very convenient in the derivation of the integral equations and their numerical solution at a later stage.
Such a decomposition goes under the name of ``reconstruction theorem'' and was proven to hold in the context of chiral perturbation theory up to two-loop order for pion--pion scattering~\cite{Stern:1993rg}. It was subsequently generalized to the case of unequal masses~\cite{Ananthanarayan:2000cp} and to general scattering of pseudoscalar octet mesons~\cite{Zdrahal:2008bd}.  We derive it in \ref{app:decomposition}, finding the form
\begin{align}
	\label{eq:decomposition}
	\M(s,t,u) = \M_0^0(s) &+ \Big[ \M_0^1(t)+ \{(s - u)t + \Delta\}\M_1^1(t) \nn
		&\quad + (t \leftrightarrow u) \Big] \,,
\end{align}
where $\M_\ell^I(s)$ are functions of one variable that only possess a right-hand cut. 
Here, $\ell$ refers to angular momentum and $I$ to isospin: 
isospin conservation of the decay constrains the total isospin of the final-state pion pair to $I=0$, 
while the $\pi\eta$ system always has $I=1$.
Equation~\eqref{eq:decomposition} follows from a partial-wave expansion of the discontinuities in fixed-$s$, -$t$, and -$u$ dispersion relations, symmetrized with respect to the three channels.
Given the smallness of the available phase space, the partial-wave expansion is truncated after $S$- and $P$-waves. 
A $\pi\pi$ $P$-wave contribution is forbidden by charge conjugation symmetry.
We stress that the truncation only neglects the discontinuities or rescattering phases in partial waves
of angular momentum $\ell\geq 2$: projecting $\M(s,t,u)$ of Eq.~\eqref{eq:decomposition} on the $\pi\pi$ $D$-wave (in the $s$-channel) 
yields a nonvanishing result, however, this $D$-wave is bound to be real apart from three-particle-cut contributions.

We will briefly discuss the final-state scattering amplitudes that are involved in $\eta'\to\eta\pi\pi$, namely $\pi\pi\to\pi\pi$ and $\pi\eta\to\pi\eta$.
Given again the maximum energies accessible in the decay, both rescattering channels are treated in the elastic approximation,
such that the corresponding partial waves can be parametrized in terms of a phase shift only, without any inelasticity effects.
The $\pi\pi$ scattering amplitude (confined to $I=0$) is approximated by
\begin{equation}\label{eq:pipiPWamp}
 \T^0(s,z_s)=\frac{32\pi}{\sigma(s)}\sin\delta_0^0(s)e^{i\delta_0^0(s)}\,,
\end{equation}
with $\delta_0^0$ denoting the $S$-wave phase shift. 
Analogously, the $\pi\eta$ scattering amplitude can be represented, neglecting
$D$- and higher waves, according to
\begin{align}\label{eq:pietaPWamp}
\T^1(t,z_t) = \frac{16\pi\,t}{\lambda(t,\meta^2,\mpc^2)^{1/2}}&\Bigl(\sin\delta_0^1(t)e^{i\delta_0^1(t)} \nn
& +3z_t\sin\delta_1^1(t)e^{i\delta_1^1(t)}\Bigr)\,,
\end{align}
where $\delta_\ell^1$ is the $\pi\eta$ phase shift of angular momentum $\ell$. 

The unitarity condition for the decay of the $\eta'$ to a generic three-body final state $n$ can be written as
\begin{equation}\label{eq:unitarity-relation}
\disc \M_n = i \sum_{n'} (2\pi)^{4}\delta^{(4)}(p_n-p_{n'}) \T_{n'n}^*\M_{n'}\,,
\end{equation}
where $\M_{n'}$ denotes the $\eta'\to n'$ decay amplitude and $\T_{n'n}$ describes the $n'\to n$ transition, while the sum runs over all possible intermediate states $n'$.\footnote{Here and in the following, relations that involve the discontinuity are always thought to contain 
an implicit $\theta$-function that denotes the opening of the respective threshold, i.e.\ $\theta(s-s_0)$ for the $\pi\pi$ channel and $\theta(t-t_0)$ for the $\pi\eta$ channel.} The
integration over the intermediate-state momenta is implied in this short-hand notation. 
Limiting the sum to $\pi\pi$ and $\pi\eta$ rescattering, carrying out the phase-space integration, 
and inserting the partial-wave expansion for the $\pi\pi$ and $\pi\eta$ amplitudes 
Eqs.~\eqref{eq:pipiPWamp} and \eqref{eq:pietaPWamp}, we find
\begin{align}\label{eq:unitarity-relation-2}
&\disc \M(s,t,u) = \frac{i}{2\pi}\biggl\{\int \diff\Omega_s' \sin\delta_0^0(s)e^{-i\delta_0^0(s)}\M(s,t',u')\nn
&\quad + \int \diff\Omega_t'\Bigl( \sin\delta_0^1(t)e^{-i\delta_0^1(t)} \nn
& \qquad\qquad\quad + 3z_t''\sin\delta_1^1(t)e^{-i\delta_1^1(t)}\Bigr)\M(s',t,u')\nn
&\quad + \int \diff\Omega_u'\Bigl( \sin\delta_0^1(u)e^{-i\delta_0^1(u)} \nn
& \qquad\qquad\quad + 3z_u''\sin\delta_1^1(u)e^{-i\delta_1^1(u)}\Bigr)\M(s',t',u)\biggr\}\,,
\end{align}
where $\diff\Omega_{s,t,u}'$ denotes the angular integration between the initial and intermediate state of the respective $s$-, $t$-, $u$-channel subsystem, while $z_{s,t,u}''$ refers to the center-of-mass scattering angles between the intermediate and final state. 
Finally, we can insert the decomposition of the decay amplitude~\eqref{eq:decomposition} on the left- and right-hand side of Eq.~\eqref{eq:unitarity-relation-2} and find the unitarity relations 
for the single-variable functions~$\M_\ell^I$:
\begin{align}\label{eq:unireletap}
\disc\M_0^0(s)&=2i\bigl\{\M_0^0(s)+\hat\M_0^0(s)\bigr\}\sin\delta_0^0(s)e^{-i\delta_0^0(s)}\,,\nn
\disc\M_\ell^1(t)&=2i\bigl\{\M_\ell^1(t)+\hat\M_\ell^1(t)\bigr\}\sin\delta_\ell^1(t)e^{-i\delta_\ell^1(t)}\,,
\end{align}
where $\ell=0,\,1$. The \emph{inhomogeneities} $\hat\M_\ell^I$ are given by 
\begin{align}\label{eq:inhomogeneitiesz}
\hat \M_0^0(s) &= 2\langle\M_0^1\rangle + \Big[\frac{3}{2}(s-\setap)(3\setap-s)+2\Delta\Big]\langle\M_1^1\rangle \nn & +s\kappa_{\pi\pi}\langle z_s\M_1^1\rangle+\frac{\kappa_{\pi\pi}^2}{2}\langle z_s^2\M_1^1\rangle\,,\nn
\hat \M_0^1(t) &= \langle\M_0^0\rangle^- + \langle\M_0^1\rangle^+ \nn
& + \frac{1}{4}\Bigl[3(\setap-t)(3\setap-t) +\Delta\Bigl(2-\frac{\Delta}{t^2}\Bigr)\Bigl] \langle\M_1^1\rangle^+\nn
& -\frac{\kappa_{\pi\eta}}{2}\Bigl[t+\frac{\Delta}{t}\Bigr]\langle z_t\M_1^1\rangle^+-\frac{\kappa_{\pi\eta}^2}{4}\langle z_t^2\M_1^1\rangle^+\,,\nn
\hat \M_1^1(t) &= \frac{3}{t\kappa_{\pi\eta}}\biggl\{\langle z_t\M_0^0\rangle^- - \langle z_t\M_0^1\rangle^+  \nn
& - \frac{1}{4}\Bigl[3(\setap-t)(3\setap-t)+\Delta\Bigl(2-\frac{\Delta}{t^2}\Bigr)\Bigl] \langle z_t\M_1^1\rangle^+\nn
& + \frac{\kappa_{\pi\eta}}{2}\Bigl[t+\frac{\Delta}{t}\Bigr]\langle z_t^2\M_1^1\rangle^+-\frac{\kappa_{\pi\eta}^2}{4}\langle z_t^3\M_1^1\rangle^+\biggr\}\,,
\end{align}
where we have defined the short-hand notation
\begin{align}\label{eq:angularav}
 \langle z^n f\rangle &:=\frac{1}{2}\int_{-1}^{1}\diff z\,z^n f\Bigl(\frac{3\setap-s+z\kappa_{\pi\pi}(s)}{2}\Bigr)\,, \nn
 \langle z^n f\rangle^{\pm} &:=\frac{1}{2}\int_{-1}^{1}\diff z\,z^n f\Bigl(\frac{3\setap-t+z\kappa_{\pi\eta}(t)\pm\Delta/t}{2}\Bigr)\,.
\end{align}
Note that the analytic continuation of Eqs.~\eqref{eq:inhomogeneitiesz} and \eqref{eq:angularav} both in the Mandelstam variables and the decay mass $\metap$ involves several subtleties. 
This is discussed in detail in Refs.~\cite{Kambor:1995yc,Anisovich:1996tx,Walker,StefanDiss} for $\eta\to3\pi$, 
as well as in Ref.~\cite{Niecknig:2012sj} for $\omega/\phi\to3\pi$ and in Ref.~\cite{Schneider:2012nng} specifically for $\eta'\to\eta\pi\pi$.
One important consequence is the generation of three-particle-cut contributions in the decay kinematics considered here.

The solutions of the unitarity relations Eq.~\eqref{eq:unireletap} can be written as
\begin{align}\label{eq:dispreletap1}
 \M_0^0(s) &= \Omega_0^0(s)\biggl\{P_0^0(s) + \frac{s^n}{\pi}\int\limits_{s_0}^{\infty} \frac{\diff s'}{{s'}^n} \frac{\hat \M_0^0(s')\sin\delta_0^0(s')}{|\Omega_0^0(s')|(s'-s)}\biggr\}\,,\nn
 \M_\ell^1(t) &= \Omega_\ell^1(t)\biggl\{P_\ell^1(t) + \frac{t^n}{\pi}\int\limits_{t_0}^{\infty}\frac{\diff t'}{{t'}^n} \frac{\hat \M_\ell^1(t')\sin\delta_\ell^1(t')}{|\Omega_\ell^1(t')|(t'-t)}\biggr\}\,,
\end{align}
where the Omn\`es functions $\Omega_\ell^I$ are given as
\begin{equation}
\Omega_\ell^I(s)=\exp\biggl\{\frac{s}{\pi}\int_{\rm thr}^\infty \diff s'\frac{\delta_\ell^I(s')}{s'(s'-s)}\biggr\} 
\end{equation}
(with the appropriate thresholds ${\rm thr}=s_0$ or $t_0$).
The order $n$ of the subtraction polynomials in the dispersion relations is determined such that the dispersion integrals are convergent. However, we can always ``oversubtract'' a dispersion integral
at the expense of having to fix the additional subtraction constants and possible ramifications for the high-energy behavior of our amplitude.\footnote{Each additional subtraction constant contributes
an additional power of $s$ to the asymptotic behavior of the amplitude if the corresponding sum rule for the subtraction constant is not fulfilled exactly. This can lead to a violation of the 
Froissart--Martin bound.} To study the convergence behavior of the integrand we have to make assumptions as regards the asymptotic behavior of the phase shifts. We assume
\begin{equation}\label{eq:asymptoticphase}
\delta_0^0(s)\to\pi\,,\quad \delta_0^1(t)\to\pi\,,\quad \delta_1^1(t)\to 0\,,
\end{equation}
as $s,\,t\to\infty$.  Note that an asymptotic behavior of $\delta(s)\to k\pi$ implies that the corresponding Omn\`es function behaves like $s^{-k}$ in the same limit.
\end{sloppypar}

Finally, we assume an asymptotic behavior of the amplitude inspired by the Froissart--Martin bound~\cite{Froissart:1961ux,Martin:1962rt},
\begin{align}\label{eq:asymptoticsSingleVarFuncs}
\M_0^0(s)=\order(s)\,, \;\;\; \M_0^1(t)=\order(t)\,, \;\;\; \M_1^1(t)=\order(t^{-1})\,,
\end{align}
which allows the following choice for the subtraction polynomials:
\begin{align}
 P_0^0(s)&=\alpha_0+\beta_0 \frac{s}{\metap^2} + \gamma_0 \frac{s^2}{\metap^4} \,,\nn
 P_0^1(t)&=\alpha_1+\beta_1 \frac{t}{\metap^2} + \gamma_1 \frac{t^2}{\metap^4} \,,\qquad
 P_1^1(t)=0\,.
\end{align}
The subtraction constants thus defined are correlated since the decomposition Eq.~\eqref{eq:decomposition} is not unique. By virtue of Eq.~\eqref{eq:etapstu}, there exists a four-parameter polynomial transformation of the single-variable functions $\M_\ell^I$ that leaves $\M(s,t,u)$ invariant. Restricting the asymptotic behavior to Eq.~\eqref{eq:asymptoticsSingleVarFuncs} reduces it to the following two-parameter transformation:
\begin{align}\label{eq:transformation}
 \M_0^0(s) &\to \M_0^0(s) + c_1 + c_2 \frac{s-\setap}{\metap^2} \,,\nn
 \M_0^1(t) &\to \M_0^1(t) - \frac{1}{2}c_1 + c_2 \frac{t-\setap}{\metap^2} \,,
\end{align}
which can be used to set the first two coefficients in the Taylor expansion of $\M_0^1(t)$ around $t=0$ to zero. Since the transformation polynomial is a trivial solution of the dispersion relation (with vanishing discontinuity),
the transformed representation still can be cast in the form of Eq.~\eqref{eq:dispreletap1}. Relabeling the transformed subtraction constants and inhomogeneities to the original names, we obtain the following form of the integral equations:
\begin{align}\label{eq:Minteq}
 \M_0^0(s) &= \Omega_0^0(s)\biggl\{\alpha_0+\beta_0 \frac{s}{\metap^2} + \gamma_0 \frac{s^2}{\metap^4} \nn
& \qquad\qquad + \frac{s^3}{\pi}\int\limits_{s_0}^{\infty}\frac{\diff s'}{{s'}^3} \frac{\hat \M_0^0(s')\sin\delta_0^0(s')}{|\Omega_0^0(s')|(s'-s)}\biggr\}\,,\nn
 \M_0^1(t) &= \Omega_0^1(t)\biggl\{\gamma_1 \frac{t^2}{\metap^4} +\frac{t^3}{\pi}\int\limits_{t_0}^{\infty}\frac{\diff t'}{{t'}^3} \frac{\hat \M_0^1(t')\sin\delta_0^1(t')}{|\Omega_0^1(t')|(t'-t)}\biggr\}\,,\nn
 \M_1^1(t) &= \frac{\Omega_1^1(t)}{\pi}\int\limits_{t_0}^{\infty}\diff t' \frac{\hat \M_1^1(t')\sin\delta_1^1(t')}{|\Omega_1^1(t')|(t'-t)}\,.
\end{align}

\begin{sloppypar}
In the following, we will neglect the $\pi\eta$ $P$-wave as it is strongly suppressed with respect to the $S$-wave of $\pi\pi$ and $\pi\eta$ scattering; see for example Refs.~\cite{Kubis:2009sb,Bernard:1991xb}. 
In fact, the $\pi\eta$ $P$-wave has exotic quantum numbers, such that the phase shift is expected to be very small at low energies.  In chiral perturbation theory, this phase (or the corresponding discontinuity)
only starts at $\order(p^8)$ (three loops) and is therefore, in this respect, as suppressed as all higher partial waves.
\end{sloppypar}

The decomposition of the amplitude in this case simply reads
\begin{equation}
	\label{eq:decompositionSWaves}
	\M(s,t,u) = \M_0^0(s) + \M_0^1(t) + \M_0^1(u) \, .
\end{equation}
We will call the dispersive representation as outlined above ``DR$_4$'', as it depends on four subtraction constants. In our numerical analysis, we compare it to a representation where we further reduce the number of free parameters by assuming a more restrictive asymptotic behavior of the amplitude: $\M(s,t,u) = \order(s^0,t^0,u^0)$ for large values of $s$, $t$, $u$, respectively. In this case, the symmetrization procedure in the reconstruction theorem is possible for $S$-waves only and the single-variable functions fulfill the integral equations
\begin{align}\label{eq:MinteqSWaves}
 \M_0^0(s) &= \Omega_0^0(s)\biggl\{\alpha+\beta \frac{s}{\metap^2} \nn
 	&\qquad\qquad + \frac{s^2}{\pi}\int\limits_{s_0}^{\infty}\frac{\diff s'}{{s'}^2} \frac{\hat \M_0^0(s')\sin\delta_0^0(s')}{|\Omega_0^0(s')|(s'-s)}\biggr\}\,,\nn
 \M_0^1(t) &= \Omega_0^1(t)\biggl\{\gamma \frac{t}{\metap^2} + \frac{t^2}{\pi}\int\limits_{t_0}^{\infty}\frac{\diff t'}{{t'}^2} \frac{\hat \M_0^1(t')\sin\delta_0^1(t')}{|\Omega_0^1(t')|(t'-t)}\biggr\} \,. 
\end{align}
Note that the transformation~\eqref{eq:transformation} still allows us to set the first subtraction constant in $\M_0^1$ to zero. The second subtraction constant cannot be removed without changing the asymptotic behavior. As there are three subtraction constants $\alpha$, $\beta$, and $\gamma$, we refer to this setup as ``DR$_3$''.

\bsp
The representation DR$_4$~\eqref{eq:Minteq} (with $\delta_1^1(t)\equiv0$) is equivalent to DR$_3$~\eqref{eq:MinteqSWaves}, provided that the subtraction constants $\gamma_0$ and $\gamma_1$ fulfill a certain sum rule in order to guarantee the constraint of the asymptotic behavior. Explicitly, the relation between the DR$_4$ and DR$_3$ subtraction constants is given by
\begin{align}
	\label{eq:sumruleDR4}
	\alpha_0 &= \alpha + \gamma \frac{3\setap}{\metap^2} \, , \quad \beta_0 = \beta - \gamma \big( 1 + 3\setap \omega_0^0 \big) \, , \nn
	\gamma_0 &= I_0^0 + \gamma \metap^2 \left( \omega_0^0 - \frac{3\setap}{2} \tilde\omega_0^0 \right) \, , \quad \gamma_1 = I_0^1 + \gamma \metap^2 \omega_0^1 \,,
\end{align}
where $\omega_\ell^I := {\Omega_\ell^I}'(0)$, $\tilde\omega_\ell^I := {\Omega_\ell^I}''(0) - 2 ({\Omega_\ell^I}'(0))^2$, and the sum rule is encoded by the integrals
\begin{equation}
	\label{eq:sumruleDR4integral}
	I_\ell^I := \frac{\metap^4}{\pi} \int^\infty_{\mathrm{thr}} \frac{\diff s'}{{s'}^3} \frac{\hat \M_\ell^I(s') \sin\delta_\ell^I(s')}{|\Omega_\ell^I(s')|} \, .
\end{equation}
\esp

Since the Froissart--Martin bound is strictly valid only for elastic scattering, a pragmatic approach concerning the number of subtractions is advisable. On the one hand, we try to work with the minimal number of subtractions allowing for a good fit of the data. On the other hand, additional subtractions help to reduce the dependence on the high-energy behavior of the phase shifts. Therefore, in our analysis we compare both representations DR$_3$ and DR$_4$.

As we will show, the representation DR$_3$ of Eqs.~\eqref{eq:decompositionSWaves} and \eqref{eq:MinteqSWaves} indeed allows for a perfect fit of the data from the VES~\cite{Dorofeev:2006fb} and BES-III~\cite{Ablikim:2010kp} experiments. With data of even higher statistics, it might become possible to include the $P$-wave in the fit. In this case, one needs to determine the four subtraction constants of the DR$_4$ representation in Eq.~\eqref{eq:Minteq}. A fifth subtraction constant would be introduced if we assumed a different high-energy behavior of $\delta_1^1$:
if a resonance with exotic quantum numbers $J^{PC}=1^{-+}$ coupling to $\pi\eta$ exists (the search for which seems
inconclusive so far~\cite{Adolph:2014rpp,Schott:2012wqa}) and we assume the $P$-wave phase to approach $\pi$ instead of 
$0$ asymptotically, the $P$-wave would allow for a nonvanishing (constant) subtraction polynomial in Eq.~\eqref{eq:dispreletap1}, which cannot be removed by the transformation~\eqref{eq:transformation}.

The inclusion of a $\pi\pi$ $D$-wave contribution, which has been suggested in Ref.~\cite{Escribano:2010wt}
(in the form of the $f_2(1270)$ resonance) would require even higher-order subtraction polynomials.

\section{Numerical solution of the dispersion relation}\label{sec:numsol}

\subsection{Iteration procedure}

\begin{sloppypar}
The numerical treatment of the integral equations~\eqref{eq:Minteq} or~\eqref{eq:MinteqSWaves} is a rather nontrivial matter, and specific care has to 
be taken in calculating the Omn\`es functions, the inhomogeneities with their complicated structure, as well as the dispersion
integrals over singular integrands.  All the details can be found in Ref.~\cite{Schneider:2012nng}.
\end{sloppypar}

The solution of the integral equations is obtained by an iteration procedure: 
we start with arbitrary functions $\M_0^0$ and $\M_0^1$, which we choose to be the respective Omn\`es functions; the final result is of course independent of the particular choice of these starting points. Then we calculate the inhomogeneities $\hat\M_0^0$ and $\hat\M_0^1$, and insert them into the dispersion integrals~\eqref{eq:Minteq} or~\eqref{eq:MinteqSWaves}. This procedure is repeated until sufficient convergence with respect to the input functions is reached.  The iteration is observed to converge rather quickly after only a few steps.

The integral equations have a remarkable property that greatly reduces the numerical cost of the calculations: they are linear in the subtraction constants. 
Thus we can write (for the DR$_3$ representation)
\begin{align}\label{eq:Mbasisfunctions}
 \M(s,t,u) =\alpha\M_\alpha(s,t,u)+\beta\M_\beta(s,t,u) +\gamma\M_\gamma(s,t,u)\,,
\end{align}
where we have defined
\begin{equation}
\M_\alpha(s,t,u):=\M(s,t,u)\bigr|_{\alpha=1,\,\beta=\gamma=0}\,,
\end{equation}
and analogously for the remaining \emph{basis functions}.
Each of the basis functions fulfills the decomposition Eq.~\eqref{eq:decompositionSWaves}, and we denote the corresponding single-variable functions
by $\M_\alpha^0(s)$, $\M_\alpha^1(t)$, etc., i.e.\ 
\begin{equation}
 \M_\alpha(s,t,u)=\M_\alpha^0(s)+\M_\alpha^1(t)+\M_\alpha^1(u) \,.
\end{equation}
We can perform the iteration procedure separately for each of these while fixing the subtraction constants after the iteration.

\subsection{Phase input}\label{sec:phaseshifts}

\begin{figure*}[t!]
	\input{plots/delta00}
	\input{plots/delta00physreg}
  \caption{The isospin-zero $\pi\pi$ $S$-wave effective phase shift $\delta_0^0(s)$, constructed with input from~\cite{Caprini:2011ky}. In the left panel, the physical region of the decay $\eta'\to\eta\pi\pi$ is indicated by dashed vertical lines at the threshold $s=4\mpc^2$ and at $s=(\metap-\meta)^2$. The right panel shows a magnification of the physical region.}
\label{fig:delta00}
\end{figure*}
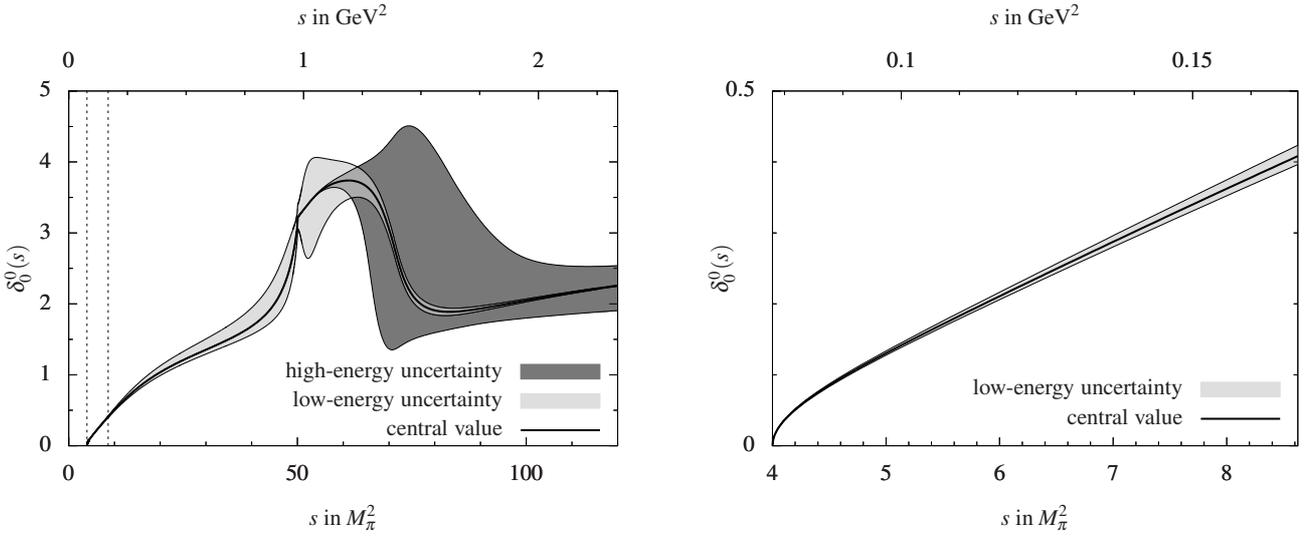

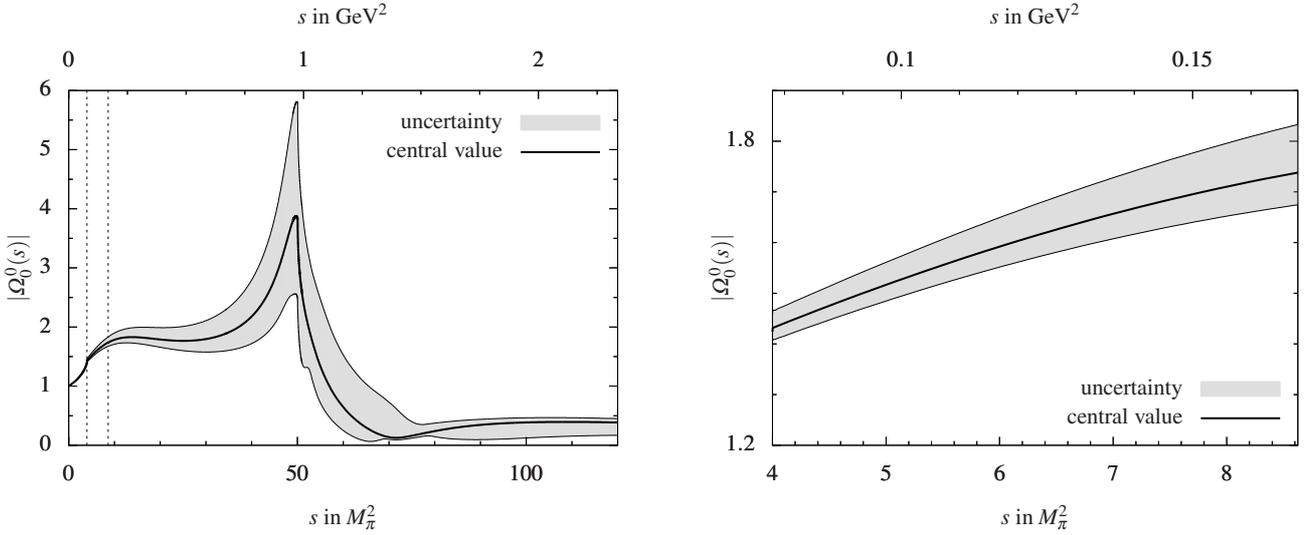
\begin{figure*}[t!]
	\input{plots/omega00}
	\input{plots/omega00physreg}
  \caption{The absolute value of the Omnès function $|\Omega_0^0(s)|$, calculated from the $\pi\pi$ $S$-wave effective phase shift $\delta_0^0(s)$. The uncertainty band includes both uncertainties in the phase, combined in quadrature.}
\label{fig:omega00}
\end{figure*}

\begin{figure*}[t!]
	\input{plots/delta01}
	\input{plots/delta01physreg}
  \caption{The $\pi\eta$ $S$-wave effective phase shift $\delta_0^1(t)$~\cite{Albaladejo:2015aca}. In the left panel, the physical region of the decay $\eta'\to\eta\pi\pi$ is indicated by dashed vertical lines at the threshold $t=(\meta+\mpc)^2$ and at $t=(\metap-\mpc)^2$. The right panel magnifies the physical region. The restricted uncertainty band is generated by varying the parameter in the range $105° \le \delta_{12} \le 125°$, while the full uncertainty band is obtained from the parameter range $90° \le \delta_{12} \le 125°$.}
\label{fig:delta01}
\end{figure*}
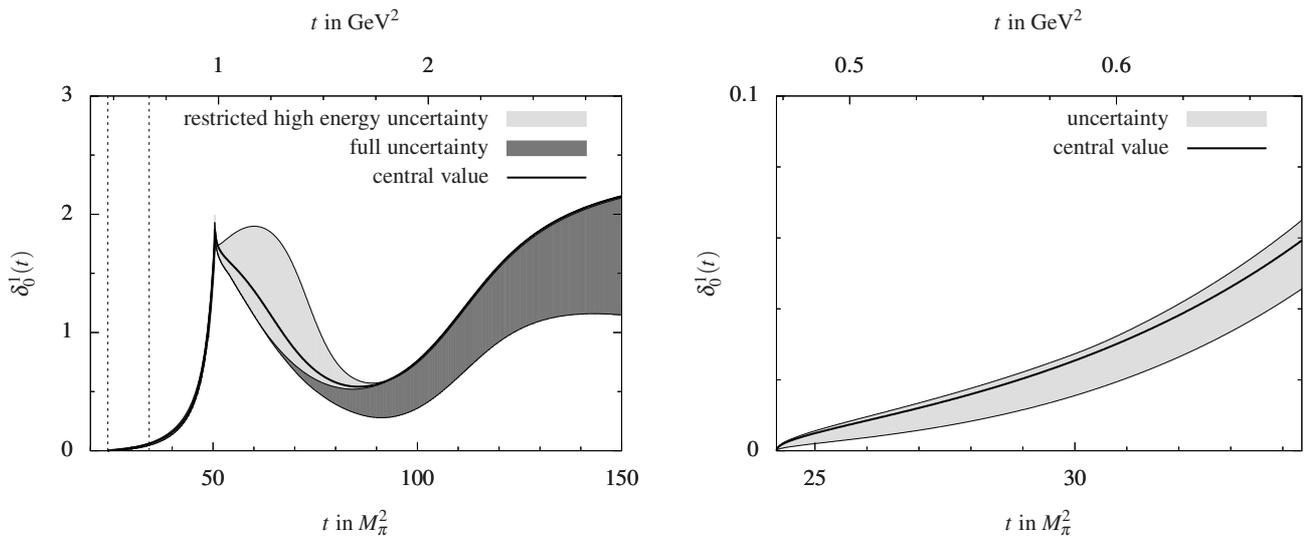

\begin{figure*}[t!]
	\input{plots/omega01}
	\input{plots/omega01physreg}
  \caption{The absolute value of the Omnès function $|\Omega_0^1(t)|$, calculated from the $\pi\eta$ $S$-wave effective phase shift $\delta_0^1(t)$. The uncertainty band corresponds to the full parameter variation, $90° \le \delta_{12} \le 125°$.}
\label{fig:omega11}
\end{figure*}
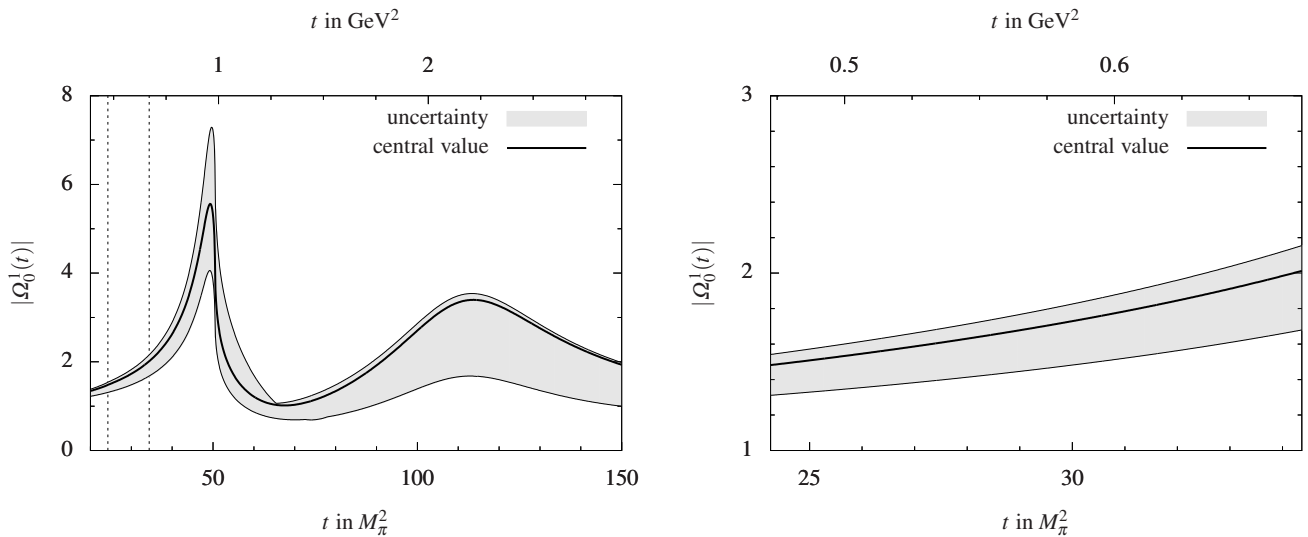

The crucial input in the dispersion relation consists of the $\pi\pi$ and $\pi\eta$ $S$-wave phase shifts $\delta_0^0$ and $\delta_0^1$. Below the threshold of inelastic channels, the dispersion relation correctly describes rescattering effects according to Watson's final-state theorem with phases $\delta_0^0$ and $\delta_0^1$ that are equal to the phase shifts of elastic scattering. In principle, a coupled-channel analysis could be used to describe the process above the opening of inelastic channels, e.g.\ the explicit inclusion of $K\bar K$ intermediate states would provide a fully consistent description in the region of the $f_0(980)$ and $a_0(980)$ resonances; such a coupled-channel generalization of the Khuri--Treiman equations has recently been investigated for $\eta\to3\pi$~\cite{Albaladejo:2017hhj}. Alternatively, the single-channel equations~\eqref{eq:dispreletap1} remain valid if we promote $\delta_0^0$ and $\delta_0^1$ to \textit{effective} phase shifts for this decay. As a full coupled-channel analysis is beyond the goal of this work, we construct effective phase shifts and quantify the uncertainties above the inelastic threshold. In such an effective one-channel problem, there are two extreme scenarios of the phase motion of $\delta_0^0$ at the $f_0(980)$ resonance, depending on how strongly the system couples to strangeness~\cite{Donoghue:1990xh,Ananthanarayan:2004xy}: large strangeness production manifests itself as a peak at the position of the $f_0(980)$ in the corresponding Omn\`es function, and thus the phase shift is increased by about $\pi$ while running through the resonance (this scenario is also realized in the elastic $\pi\pi$ scattering phase shift). If the coupling to the channel with strangeness is weak, the corresponding Omn\`es function has a dip at the resonance position and the corresponding phase shift decreases (this is realized in the phase of the nonstrange scalar form factor of the pion). Scenarios in between these two extremes are conceivable.

For the input on the elastic $\pi\pi$ phase shift, we use the results of very sophisticated analyses of the Roy (and similar) equations~\cite{Caprini:2011ky,GarciaMartin:2011cn}. As both analyses agree rather well, we only take one of these parametrizations~\cite{Caprini:2011ky} into account. In Fig.~\ref{fig:delta00}, we show our phase $\delta_0^0(s)$, which agrees with the Roy solution~\cite{Caprini:2011ky} below the inelastic threshold. The uncertainty due to the variation of the parameters in the Roy solution is shown as a light gray band labeled ``low-energy uncertainty.''

Now, the continuation into the inelastic region is modeled as follows. We calculate the $S$-waves for $\eta'\eta\to\pi\pi$ and $\eta'\eta\to K\bar K$ in large-$N_c$ \chpt{} at next-to-leading order (tree level) and unitarize this coupled-channel system with an Omnès matrix taken from Ref.~\cite{Daub:2015xja}. The large-$N_c$ \chpt{} representation depends on the low-energy constants (LECs) $L_2$ and $L_3$. We take their values from the most recent dispersive analysis of $K_{\ell4}$ decays~\cite{Colangelo:2015kha},
\begin{equation}
	\label{eq:kl4lecs}
	L_2^r = 0.63(13) \times 10^{-3} \,, \quad L_3^r = -2.63(46) \times 10^{-3} \,,
\end{equation}
and vary each of them within its uncertainty. Adding the variations of the phase in quadrature generates the broad dark gray band labeled ``high-energy uncertainty'' in Fig.~\ref{fig:delta00}. This treatment correctly generates a smooth phase drop by $\pi$ with respect to the elastic $\pi\pi$ scattering phase, and the uncertainty band covers a broad energy range for the position of this decrease. Still, the phase drops at sufficiently high energies such that the corresponding Omnès function, shown in Fig.~\ref{fig:omega00}, exhibits a peak at the position of the $f_0(980)$ resonance. Asymptotically, we smoothly drive $\delta_0^0$ to a value of $\pi$. We wish to emphasize that the role of the large-$N_c$ input is not essential and only that of an auxiliary tool, which allows for a smooth construction that obeys the two desired features: the occurrence of an $f_0(980)$ peak in accordance with expectations from scalar-resonance models (see, e.g., Sect.~\ref{sec:chiralamps}), and an asymptotic phase of $\pi$ (as opposed to $2\pi$, say).  The large high-energy uncertainty in Fig.~\ref{fig:delta00} should safely cover a large variety of phases obeying these constraints.
Note, finally, that in the physical region of the decay $\eta'\to\eta\pi\pi$, the uncertainties of the phase and the Omnès function are small.

In the same spirit of an effective one-channel treatment, we consider isospin-breaking effects. In the isospin limit, our formalism applies identically to both the charged and the neutral processes $\eta'\to\eta\pi^+\pi^-$ and $\eta'\to\eta\pi^0\pi^0$. In order to account for the most important isospin-breaking effects, we construct effective phase shifts for the neutral decay mode that have the correct thresholds and reproduce the expected nonanalytic cusp behavior, as we explain in detail in \ref{app:isospinbreaking}.

For the $\pi\eta$ phase shift $\delta_0^1$, we take the phase of the scalar form factor $F_S^{\eta\pi}$ of Ref.~\cite{Albaladejo:2015aca} as input, shown in Fig.~\ref{fig:delta01}. In that reference, a $\pi\eta$--$K\bar K$ $S$-wave coupled-channel $T$-matrix has been constructed, to which chiral constraints~\cite{Gasser:1984gg} have been imposed as well as experimental information on the $a_0(980)$ and $a_0(1450)$ resonances. The remaining model uncertainty has been subsumed in the dependence on one single phase $\delta_{12}$.  
The ``central'' solution corresponds to a parameter value of $\delta_{12} = 107.5°$, while the uncertainty band is generated as an envelope of the solutions obtained by varying the parameter in the restricted range $90° \le \delta_{12} \le 125°$, which is compatible with chiral predictions for the scalar radius. The largest part of the high-energy uncertainty is generated by values of the parameter in the range $90° \le \delta_{12} \le 105°$, as shown in Fig.~\ref{fig:delta01}: while for all $\delta_{12} \geq 105°$, the phase approaches its asymptotic value of $\pi$ very quickly above the $a_0(1450)$ resonance, this convergence becomes very slow and is extended over a vast energy range for $\delta_{12}<105°$. As this high-energy behavior turns out to affect the uncertainties in some (but not all) quantities extracted from data fits rather strongly, we will occasionally also refer to the reduced uncertainty, induced by the more restricted range $105° \le \delta_{12} \le 125°$. The Omnès function with an uncertainty band generated by the full variation $90° \le \delta_{12} \le 125°$ is shown in Fig.~\ref{fig:omega11}. In particular, we observe a pronounced peak at the position of the $a_0(980)$ resonance for all parameter values.

\section{Determination of the subtraction constants}\label{sec:subconsts}

\begin{sloppypar}
After having solved the integral equations numerically, we have to determine the free parameters in the dispersion relation, i.e.\ the subtraction constants $\alpha$, $\beta$, and $\gamma$ in the case of the DR$_3$ representation, or $\alpha_0$, $\beta_0$, $\gamma_0$, and $\gamma_1$ in the case of DR$_4$. We summarize the experimental situation on $\eta'\to\eta\pi\pi$ Dalitz plots in Sect.~\ref{sec:experiments}. In Sect.~\ref{sec:Fits}, we discuss the results of fitting the subtraction constants to the most recent data sets. 
\end{sloppypar}

\subsection{Sampling of experimental Dalitz plots}\label{sec:experiments}

\begin{sloppypar}
In experimental analyses of the $\eta'\to\eta\pi\pi$ Dalitz plot, one 
defines symmetrized coordinates $x$ and $y$ according to
\begin{align}\label{eq:xyetapdef}
 x=\frac{\sqrt{3}(t-u)}{2\metap \Qetap}\,,\quad
 y=(\meta+2\mpc)\frac{(\metap-\meta)^2-s}{2\mpc \metap \Qetap}-1\,,
\end{align}
where $Q_{\eta'} := M_{\eta'}-M_{\eta}-2M_{\pi}$. The squared amplitude of the decay is then expanded in terms of these variables,
\begin{align}\label{eq:dalitzparametrization}
 |\M_\mathrm{exp}(x,y)|^2 =|\N_\mathrm{exp}|^2 \big\{1+ay+by^2+cx+dx^2 + \ldots \big\}\,,
\end{align}
and the parameters $a$, $b$, $c$, $d$ are fitted to experimental data. Note that a nonzero value for the parameter $c$ (in $\eta'\to\eta\pi^+\pi^-$) would indicate violation of charge conjugation symmetry; there is no indication of a nonzero $c$ up to this point. In the following we consider two recent measurements 
of the charged final state $\eta'\to\eta\pi^+\pi^-$. These determinations of the Dalitz-plot parameters by the BES-III~\cite{Ablikim:2010kp} and the VES~\cite{Dorofeev:2006fb} collaborations currently feature the highest 
statistics. In Table~\ref{tab:etapexp} we have summarized some details and results of the two experiments.

\begin{table}
\centering
\renewcommand{\arraystretch}{1.4}
\begin{tabular}{c c c}
\toprule
		&BES-III~\cite{Ablikim:2010kp}   		& VES~\cite{Dorofeev:2006fb}		\\
\midrule
$a$		&$-47\pm11\pm3$		& $-127\pm16\pm8$	\\
$b$		&$-69\pm19\pm9$ 		& $-106\pm28\pm14$	\\
$c$		&$+19\pm11\pm3$ 		& $+15\pm11\pm14$  	\\
$d$		&$-73\pm12\pm3$ 		& $-82\pm17\pm8$	\\
\midrule
\# events	&$43\,826\pm211$			& $\simeq 8623$	 		\\
\# $x$ bins	&$26$					& $8$				\\
\# $y$ bins	&$22$					& $8$				\\
\bottomrule
\end{tabular}
\renewcommand{\arraystretch}{1.0}
\caption{Dalitz-plot parameter measurements by the BES-III and VES collaborations in units of $10^{-3}$. The first error on the Dalitz-plot parameters is always statistical, the second systematic. We have estimated the number of 
$\eta'\to\eta\pi^+\pi^-$ events for the VES collaboration from the total number of $\eta'$ events and the branching ratio $\BR(\eta'\to\eta\pi^+\pi^-)=(42.9\pm0.7)\%$~\cite{PDG}.}
\label{tab:etapexp}
\end{table}

\begin{figure}[t]
 \centering
  \vspace{-0.5cm}\includegraphics*[width= 0.9\linewidth]{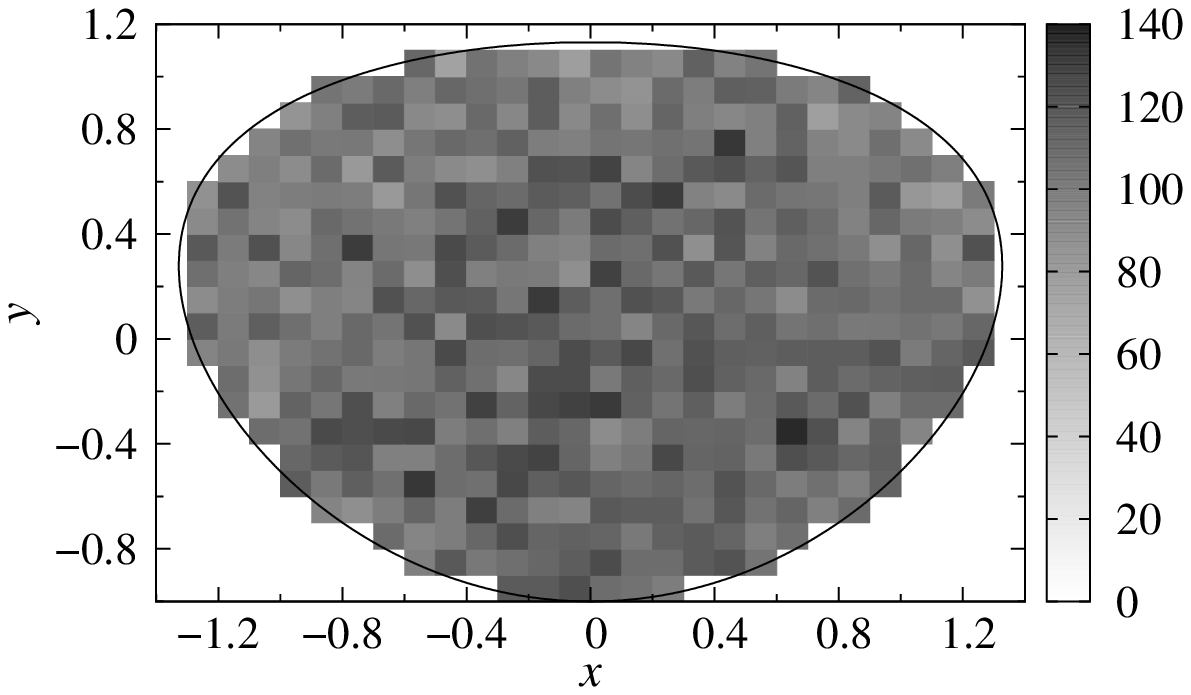} \\[-0.5cm]
  \includegraphics*[width= 0.9\linewidth]{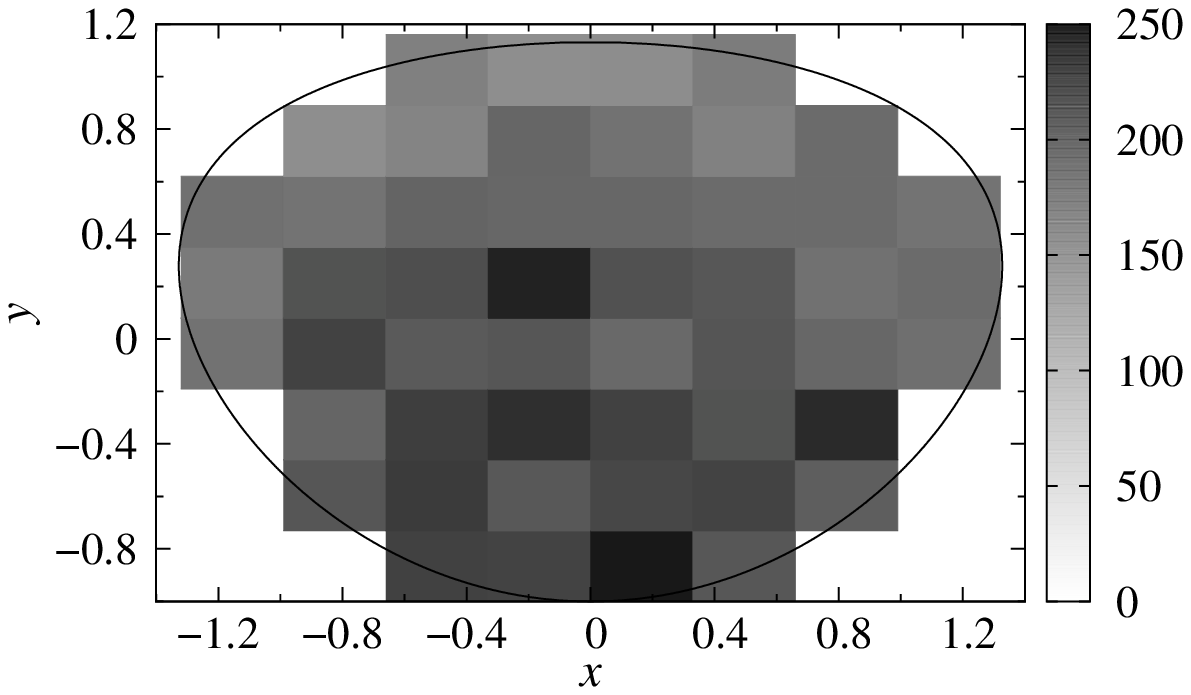}
  \caption{Dalitz-plot samples for $\eta'\to\eta\pi^+\pi^-$ from the experimental Dalitz-plot distribution in Table~\ref{tab:etapexp} for the BES-III (top panel) and the VES (bottom panel) experiment.}
\label{fig:DalitzMC}
\end{figure}

For our analysis, we have generated pseudodata samples from the Dalitz-plot distributions as measured by the two groups~\cite{Kupscpriv};
the resulting Dalitz-plot distributions are shown
in Fig.~\ref{fig:DalitzMC}. 
To check our results we have refitted the parametrization~\eqref{eq:dalitzparametrization} to the synthesized data sets, and find agreement with the fit parameters of Table~\ref{tab:etapexp} within statistical uncertainties, as well as with the correlation matrix quoted in Ref.~\cite{Ablikim:2010kp}.
We note that the two data sets disagree on the parameter $a$ at the $2\sigma$ level; of course, it would be desirable that this experimental disagreement be resolved by future measurements. 

\end{sloppypar}

\subsection{Fitting experimental data}\label{sec:Fits}

\begin{sloppypar}
We proceed by fitting the subtraction constants, which are the free parameters in our dispersive representation of the amplitude, to the following data.
\begin{itemize}
	\item Dalitz-plot distribution for the charged channel from \mbox{BES-III}~\cite{Ablikim:2010kp} and VES~\cite{Dorofeev:2006fb} experiments.
	\item The partial decay width~\cite{PDG}.
\end{itemize}
Note that we use \emph{real} fit parameters: in principle the subtraction constants can have imaginary parts due to the complex discontinuity~\eqref{eq:unireletap}.
However, since the imaginary parts of the subtraction constants are proportional to 
three-particle-cut contributions, and the available decay phase space of $\eta'\to\eta\pi\pi$ is small, the imaginary parts are so tiny that---given the precision of the data sets---their effect is entirely negligible (this is not the case for processes involving the decay of heavier mesons, compare e.g.\ Refs.~\cite{Niecknig:2012sj,Niecknig:2015ija}).
\end{sloppypar}

In the following, we set up a scheme that allows us to fit both the experimental Dalitz-plot distribution and the partial decay width simultaneously and avoids some strong correlations between the fitting parameters.\footnote{We write down formulae for the DR$_3$ representation with subtraction constants $\alpha$, $\beta$, and $\gamma$. The fitting procedure for the DR$_4$ representation is completely analogous.}

\begin{sloppypar}
First, we perform the following transformation of the fit parameters:
\begin{equation}
	\alpha = \bar\N \bar\alpha \,, \quad \beta = \bar\N \bar\beta \,, \quad \gamma = \bar\N \bar\gamma \,.
\end{equation}
Hence, we write the squared amplitude as
\begin{equation}
	| \M(s,t,u) |^2 = |\bar\N|^2 |\bar\M(s,t,u)|^2 \,,
\end{equation}
where
\begin{align}
	\bar\M(s,t,u) = \bar\alpha \M_\alpha(s,t,u) +  \bar\beta \M_\beta(s,t,u) +  \bar\gamma \M_\gamma(s,t,u) \,,
\end{align}
and fix the arbitrary normalization of $\bar\M$ to
\begin{equation}
	\int \diff x\,\diff y\,|\bar\M(x,y)|^2 = 1 \,.
	\label{eq:barMnorm}
\end{equation}
This condition results in a quadratic equation for the re\-scaled subtraction constants $\bar\alpha$, $\bar\beta$, and $\bar\gamma$. We choose to express $\bar\gamma$ in terms of $\bar\alpha$ and $\bar\beta$. The experimental partial decay width now directly determines the normalization $\bar\N$ and has no influence on the parameters $\bar\alpha$ and $\bar\beta$.
\end{sloppypar}

On the other hand, the experimental Dalitz-plot distribution~\eqref{eq:dalitzparametrization} has again an arbitrary normalization. Hence, we have to fit the Dalitz-plot data according to
\begin{equation}
	|\M_\mathrm{exp}|^2 = |\N_\mathrm{exp}|^2 \frac{|\bar\M|^2}{|\bar\M(x=y=0)|^2} =: |\N|^2 |\bar\M|^2 \,.
\end{equation}
The Dalitz-plot distribution therefore determines the fitting parameters $\bar\alpha$, $\bar\beta$, and the irrelevant normalization $\N_\mathrm{exp}$ or $\N$.

\bsp
Note that the experimental Dalitz-plot distribution is effectively described by three Dalitz-plot parameters $a$, $b$, and $d$. In the dispersive representation DR$_3$, the shape of the Dalitz-plot distribution depends only on the two fitting parameters $\bar\alpha$ and $\bar\beta$. Therefore, the parametrization DR$_3$ has predictive power. The representation DR$_4$ describes the shape of the Dalitz-plot distribution again in terms of three parameters.
\esp

The result of the fit to data provides us with a representation of the amplitude that fulfills the
strong constraints of analyticity and unitarity. This will be an essential input for a forthcoming dispersive analysis of $\eta'\to3\pi$~\cite{etaprimeto3pi}.\footnote{Notice that the decay $\eta'\to3\pi$
can proceed via $\eta'\to\eta\pi\pi$ and isospin-breaking rescattering $\eta\pi\to\pi\pi$ (which can be extracted from analytic continuation of the dispersive amplitude $\eta\to3\pi$~\cite{Descotes-Genon:2014tla}) and direct isospin breaking
$\eta'\to3\pi$.}

The experimental partial decay width~\cite{PDG}
\begin{align}
	\Gamma&(\eta'\to\eta\pi^+\pi^-) = \frac{1}{32(2\pi)^3\,\metap^3} \int \diff s \, \diff t | \M(s,t,u) |^2 \nn
		&= \frac{\mpc Q_{\eta'}^2}{ 128 \pi^3 \sqrt{3} \metap(\meta+2\mpc)} \int \diff x \, \diff y | \M(x,y) |^2 \nn
		&= (84.5\pm4.1) \times 10^{-6} \GeV
\end{align}
fixes the normalization to
\begin{equation}
	\bar{\N} = 13.88\pm0.34\,.
\end{equation}
For the rescaled subtraction constants in the DR$_3$ representation, the fit of the dispersive representation with central values of the phase input to the BES-III data sample~\cite{Ablikim:2010kp} leads to
\begin{equation}\label{eq:DispfitcorBES}
\begin{aligned}
&\bordermatrix{
	& \bar\alpha & \bar\beta \cr
	\bar\alpha = -2.34 \pm 0.26 & 1.00 & -1.00 \cr
	\bar\beta = \phantom{+}6.70 \pm 0.83 & & \phantom{+}1.00 \cr
} \,, \\[.2cm]
&\bar{\gamma}\big(\bar{\alpha},\bar{\beta}\big) = 1.12\pm0.14\,,
\end{aligned}
\end{equation}
while the fit to the VES data~\cite{Dorofeev:2006fb} gives
\begin{equation}\label{eq:DispfitcorVES}
\begin{aligned}
&\bordermatrix{
	& \bar\alpha & \bar\beta \cr
	\bar\alpha = -2.63 \pm 0.54 & 1.00 & -1.00 \cr
	\bar\beta = \phantom{+}7.41 \pm 1.73 & & \phantom{+}1.00 \cr
} \,, \\[.2cm]
&\bar{\gamma}\big(\bar{\alpha},\bar{\beta}\big) = 1.29\pm0.29\,.
\end{aligned}
\end{equation}
The uncertainties and correlations are the statistical ones due to the fitted data. We observe a strong anti-correlation between $\bar{\alpha}$ and $\bar{\beta}$. Note that choosing one or the other of the two solutions of the quadratic constraint on $\bar\gamma$ just results in an irrelevant overall sign change of the amplitude.

Similarly, the fit of the dispersive representation DR$_4$ to BES-III results in
\begin{equation}\label{eq:Dispfit4corBES}
\begin{aligned}
&\bordermatrix{
	& \bar\alpha_0 & \bar\beta_0 & \bar\gamma_0 \cr
	\bar\alpha_0 = -0.84 \pm 0.08 & 1.00 & -0.86 & \phantom{+}0.28 \cr
	\bar\beta_0 = \phantom{+}2.01 \pm 0.46 & & \phantom{+}1.00 & -0.73 \cr
	\,\bar\gamma_0 = \phantom{+}1.79 \pm 1.25 & & & \phantom{+}1.00 \cr
} \,, \\[.2cm]
&\bar{\gamma}_1\big(\bar{\alpha}_0,\bar{\beta}_0,\bar\gamma_0\big) = 0.38\pm0.05\,,
\end{aligned}
\end{equation}
while the DR$_4$ fit to the VES data gives
\begin{equation}\label{eq:Dispfit4corVES}
\begin{aligned}
&\bordermatrix{
	& \bar\alpha_0 & \bar\beta_0 & \bar\gamma_0 \cr
	\bar\alpha_0 = -0.79 \pm 0.16 & 1.00 & -0.85 & \phantom{+}0.19 \cr
	\bar\beta_0 = \phantom{+}1.03 \pm 0.86 & & \phantom{+}1.00 & -0.67 \cr
	\,\bar\gamma_0 = \phantom{+}5.02 \pm 2.34 & & & \phantom{+}1.00 \cr
} \,, \\[.2cm]
&\bar{\gamma}_1\big(\bar{\alpha}_0,\bar{\beta}_0,\bar\gamma_0\big) = 0.40\pm0.10 \,.
\end{aligned}
\end{equation}

\begin{figure}[t]
\centering
\large
\scalebox{0.67}{\input{plots/ellipses}}
\caption{Error ellipses for the DR$_3$ fits to BES-III and VES in the plane of the rescaled subtraction constants $(\bar\alpha,\bar\beta)$, corresponding to $68.27\%$ confidence regions.}
\label{fig:ellipses}
\end{figure}
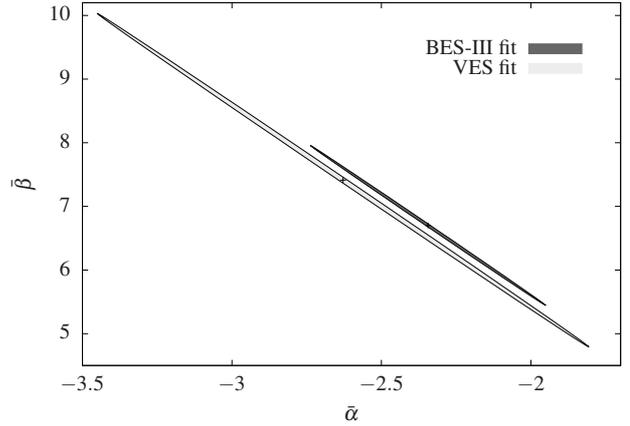

\begin{table}[t!]
\centering
\renewcommand{\arraystretch}{1.5}
\setlength\tabcolsep{0.18cm}
\begin{tabular}{c c c}
\toprule
 & BES-III, DR$_3$ fit & VES, DR$_3$ fit \\
\midrule
$\chi^2/\text{ndof}$ & $459/435\approx1.06$ & $43.1/47\approx0.92$\\
$\alpha$ & $-33\pm4\pm1\,{}^{+5}_{-42}$ & $-36\pm7\pm1\,{}^{+6}_{-56}$\\
$\beta$ & $93\pm12\pm3\,{}^{-15}_{+127}$ & $103\pm24\pm5\,{}^{-19}_{+168}$\\
$\gamma$ & $16\pm2\pm1\,{}^{-3}_{+32}$ & $18\pm4\pm0\,{}^{-4}_{+41}$\\
\bottomrule
\end{tabular}
\renewcommand{\arraystretch}{1.0}
\caption{Fit results for the DR$_3$ subtraction constants for the BES-III and VES data samples. The first error gives the combined uncertainty of the Dalitz-plot data and the partial decay width of $\eta'\to\eta\pi^+\pi^-$, while the second (asymmetric third) error gives the uncertainty due to the $\pi\pi$ ($\pi\eta$) phase input.}
\label{tab:fitetapDR3}
\end{table}

\addtolength\dbltextfloatsep{-1cm}
\begin{table*}[t!]
\centering
\renewcommand{\arraystretch}{1.5}
\setlength\tabcolsep{0.18cm}
\begin{tabular}{c c c c c}
\toprule
 & BES-III, DR$_3$ fit & BES-III, DR$_4$ fit & VES, DR$_3$ fit & VES, DR$_4$ fit \\
\midrule
$\chi^2/\text{ndof}$ & 	$459/435\approx1.06$ & 					$459/434\approx1.06$ & 					$43.1/47\approx0.92$ & 					$42.4/46\approx0.92$ \\
$\alpha_0$ & 			$-11.2\pm1.0\pm0.4\,{}^{+0.7}_{-1.1}$ & 		$-11.6\pm1.1\pm0.1\,{}^{+0.9}_{-2.5}$ & 		$-11.9\pm2.0\pm0.2\,{}^{+1.0}_{-1.0}$ & 		$-11.0\pm2.2\pm0.1\,{}^{+1.1}_{-2.7}$ \\
$\beta_0$ & 			$24\pm3\pm1\,{}^{+2}_{-13}$ & 				$28\pm6\pm3\,{}^{-3}_{+9}$ & 				$24\pm6\pm1\,{}^{+2}_{-14}$ & 				$14\pm12\pm4\,{}^{-4}_{+10}$ \\
$\gamma_0$ & 		$36\pm5\pm5\,{}^{-6}_{+40}$ & 				$25\pm17\pm6\,{}^{-1}_{+2}$ & 			$42\pm10\pm7\,{}^{-7}_{+54}$ & 			$70\pm33\pm7\,{}^{-1}_{+2}$ \\
$\gamma_1$ & 		$5.1\pm0.7\pm0.2\,{}^{-0.8}_{+0.9}$ & 		$5.3\pm0.7\pm0.1\,{}^{-0.9}_{+3.2}$ & 		$6.0\pm1.4\pm0.1\,{}^{-1.1}_{+1.5}$ & 		$5.5\pm1.5\pm0.1\,{}^{-1.1}_{+3.3}$ \\
\bottomrule
\end{tabular}
\renewcommand{\arraystretch}{1.0}
\caption{Fit results for the DR$_4$ subtraction constants for the BES-III and VES data samples, obtained from the three-parameter fits via the sum rule~\eqref{eq:sumruleDR4} and directly from the four-parameter fits. The first error is the fit uncertainty, the second (third) error is the systematic uncertainty due to the $\pi\pi$ ($\pi\eta$) phase input.}
\label{tab:fitetapDR3vsDR4}
\end{table*}

Table~\ref{tab:fitetapDR3} shows the $\chi^2/{\rm ndof}$ and the absolute subtraction constants obtained from the DR$_3$ fits to the sampled \mbox{BES-III} and the VES data sets. The first error is the statistical fit uncertainty. It is dominated by the experimental uncertainty in the Dalitz-plot distribution, while the uncertainty due to the partial decay width is small. The second error is the systematic uncertainty due to the $\pi\pi$ phase input. The very asymmetric third error is due to the $\pi\eta$ phase input with a parameter variation in the range $90° \le \delta_{12} \le 125°$. The upper error corresponds to $\delta_{12} \ge 107.5°$, while the lower error corresponds to $\delta_{12} \le 107.5°$. If the $\pi\eta$ phase variation is restricted to a parameter range of $105° \le \delta_{12} \le 125°$, the large lower error is much reduced and the uncertainty is covered by a symmetric error with the magnitude of the upper error.

The variation of the $\pi\eta$ phase input for $\delta_{12} \le 105°$ has some small effect on the $\chi^2$: for BES-III, the variation is $\chi^2\in[1.05,1.09]$, for VES we find $\chi^2\in[0.90,0.93]$. One might be tempted to minimize the $\chi^2$ with respect to $\delta_{12}$ and try to extract information on the parameter in the $\pi\eta$ phase shift. However, such an attempt is futile. Figure~\ref{fig:delta01} shows that the variation of the phase mostly affects the high-energy region above $1\GeV$. The variation in the parameter region $90°\le\delta_{12}\le105°$ mainly controls how fast the phase reaches $\pi$ in the ``asymptotic'' region. It would certainly be illusionary to extract information on the phase at such high energies from $\eta'\to\eta\pi\pi$ Dalitz-plot data. Hence, the variation of the phase parameter $\delta_{12}$ simply has to be treated as a source of systematic uncertainty.

\addtolength\dbltextfloatsep{1cm}

\bsp
The $\chi^2/{\rm ndof}$ is close to 1 in both fits, even though compared to the phenomenological Dalitz-plot parametrization, the dispersive representation DR$_3$ needs one parameter less to describe the experimental data (disregarding the $C$-parity violating parameter $c$). At first sight, the obtained values for the subtraction constants in the DR$_3$ scheme seem to be compatible between the fits to the two experimental samples. In fact, there is a rather strong tension between the fits to the two experiments, disguised by the strong anti-correlation between $\bar\alpha$ and $\bar\beta$. The error ellipses in the $(\bar\alpha,\bar\beta)$-plane reveal that the two fit results are not compatible with each other; see Fig.~\ref{fig:ellipses}.
\esp

Table~\ref{tab:fitetapDR3vsDR4} shows the $\chi^2/{\rm ndof}$ and a comparison of the absolute subtraction constants in the DR$_4$ scheme, obtained directly from the DR$_4$ fits as well as extracted from the DR$_3$ fits via the sum rule~\eqref{eq:sumruleDR4}. Due to correlations, the large systematic uncertainty from the $\pi\eta$ phase variation, which is visible in all DR$_3$ subtraction constants $\alpha$, $\beta$, and $\gamma$, prominently shows up in the transformed constant $\gamma_0$, while we observe a cancellation of this systematic uncertainty in the other constants, especially in $\alpha_0$ and $\gamma_1$.  The main conclusion is, however, that Table~\ref{tab:fitetapDR3vsDR4} demonstrates the full consistency of the two subtraction schemes with each other already within the statistical (fit) uncertainties alone.

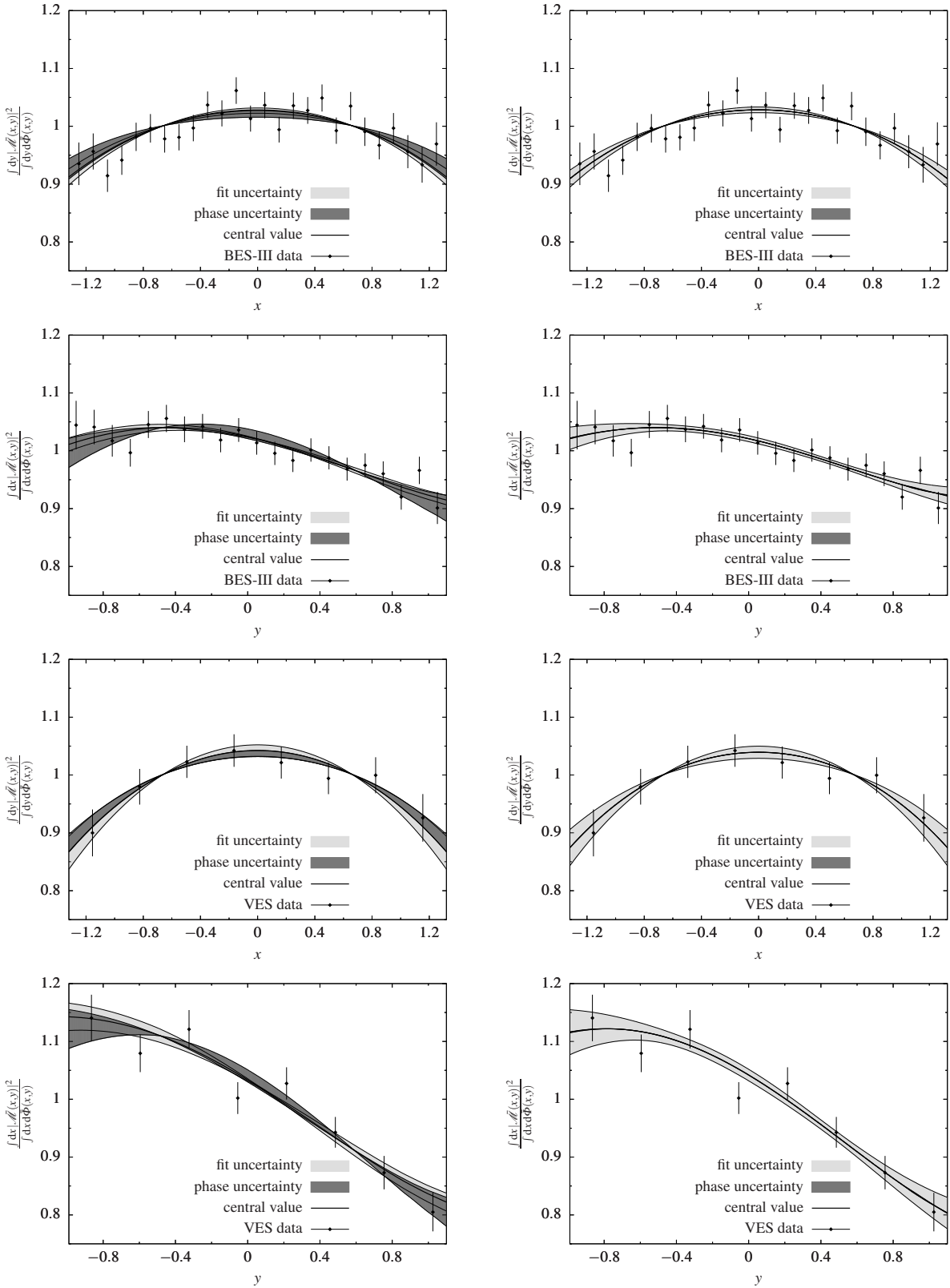
\begin{figure*}[t!]
\centering
\large
\scalebox{0.6}{\input{plots/BES_Xproj_DR3}}\hspace{0.5cm}
\scalebox{0.6}{\input{plots/BES_Xproj_DR4}}\vspace{0.0cm}
\scalebox{0.6}{\input{plots/BES_Yproj_DR3}}\hspace{0.5cm}
\scalebox{0.6}{\input{plots/BES_Yproj_DR4}}\vspace{0.0cm}
\scalebox{0.6}{\input{plots/VES_Xproj_DR3}}\hspace{0.5cm}
\scalebox{0.6}{\input{plots/VES_Xproj_DR4}}\vspace{0.0cm}
\scalebox{0.6}{\input{plots/VES_Yproj_DR3}}\hspace{0.5cm}
\scalebox{0.6}{\input{plots/VES_Yproj_DR4}}
 \caption{Decay spectra for $\eta'\to\eta\pi^{+}\pi^{-}$ integrated over the variable $y$ and divided by the integral over the normalized phase space $\text{d}\bar{\Phi}(x,y)$ (first and third rows), where both are individually normalized as given in Eq.~\eqref{eq:barMnorm}. Analogously for $x\leftrightarrow y$ (second and fourth rows). We show the sampled data sets for the BES-III~\cite{Ablikim:2010kp} (first and second rows) and the VES~\cite{Dorofeev:2006fb} experiments (third and fourth rows). The DR$_3$ (left column) and DR$_4$ fits (right column) are shown. The two error bands in each figure give the uncertainties resulting from the fit to data and originating from the variation of the phase input, respectively.}
\label{fig:etapfit}
\end{figure*}

In Fig.~\ref{fig:etapfit}, we display the decay spectrum integrated over the Dalitz-plot variables $x$ or $y$, respectively. The results of the two subtraction schemes DR$_3$ and DR$_4$ are very similar. The DR$_4$ scheme leads to a much smaller systematic uncertainty due to the phase input at the expense of a larger statistical fit uncertainty.

\begin{table*}[t!]
\centering
\renewcommand{\arraystretch}{1.5}
\begin{tabular}{c c c c c c}
\toprule
 & 					BES-III, DR$_3$ fit & 					BES-III, DR$_4$ fit & 			VES, DR$_3$ fit & 						VES, DR$_4$ fit \\
\midrule
$a$ & 				$-41\pm9\pm1\,{}^{-0}_{+1}$ & 				$-42\pm10\pm1\pm0$ &			$-148\pm18\pm1\,{}^{-1}_{+3}$ &			$-145\pm18\pm1\pm0$ \\
$b$ & 				$-88\pm7\pm10\,{}^{+5}_{-37}$ & 			$-76\pm18\pm0\pm0$ & 			$-82\pm14\pm12\,{}^{+7}_{-51}$ &			$-110\pm34\pm0\pm0$ \\
$d$ & 				$-68\pm11\pm2\,{}^{-0}_{+17}$ & 			$-69\pm11\pm0\pm0$ & 			$-86\pm22\pm1\,{}^{-1}_{+13}$ &			$-85\pm22\pm0\pm1$ \\
\midrule
$\kappa_{03}[y^3]$ & 	$\phantom{+}8\pm1\pm2\,{}^{-1}_{+4}$ & 		$\phantom{+}7\pm2\pm1\pm0$ & 	$\phantom{+}16\pm3\pm3\,{}^{-1}_{+8}$ & 	$\phantom{+}20\pm5\pm2\pm0$ \\
$\kappa_{21}[yx^2]$ & 	$-12\pm2\pm0\pm1$ & 					$-11\pm2\pm0\pm1$ & 			$-9\pm2\pm0\,{}^{+0}_{-1}$ & 				$-10\pm2\pm0\pm1$ \\
$\kappa_{04}[y^4]$ & 	$\phantom{+}3\pm1\pm1\,{}^{-0}_{+1}$ & 		$\phantom{+}3\pm1\pm0\pm0$ & 	$\phantom{+}2\pm1\pm1\,{}^{-0}_{+1}$ & 		$\phantom{+}5\pm2\pm0\pm0$ \\
$\kappa_{22}[y^2x^2]$ & 	$\phantom{+}3\pm1\pm0\,{}^{-0}_{+1}$ & 		$\phantom{+}2\pm1\pm0\pm0$ & 	$\phantom{+}5\pm2\pm1\,{}^{-0}_{+2}$ & 		$\phantom{+}6\pm2\pm0\pm0$ \\
$\kappa_{40}[x^4]$ & 	$\phantom{-}0\pm1\pm0\pm0$ & 			$\phantom{+}0\pm1\pm0\pm0$ & 	$\phantom{+}0\pm1\pm0\pm0$ & 			$\phantom{+}0\pm1\pm0\pm0$ \\
\bottomrule
\end{tabular}
\renewcommand{\arraystretch}{1.0}
\caption{Dalitz-plot parameters obtained from a Taylor expansion of the dispersive amplitude according to Eq.~\eqref{eq:dalitzparametrization}, using the best fitting values of the subtraction constants for the BES-III and VES data samples as input. All values are given in units of $10^{-3}$. 
The values shown here are to be compared with Table~\ref{tab:etapexp}. The first error is the fit uncertainty, the second (third) error is the systematic uncertainty due to the $\pi\pi$ ($\pi\eta$) phase input.}
\label{tab:dalitzfit}
\end{table*}

By expanding the fitted dispersive representations around the center of the Dalitz plot, we extract the Dalitz-plot parameters $a$, $b$, and $d$ listed in Table~\ref{tab:dalitzfit}. The values of the polynomial fit to the Dalitz plot of Table~\ref{tab:etapexp} are well reproduced within the uncertainties. Note that this is a nontrivial observation, as the dispersive amplitude obviously is no polynomial in the Mandelstam variables.

\bsp
By comparing the two subtraction schemes DR$_3$ and DR$_4$, we see that the additional parameter in DR$_4$ has basically no influence on the $\chi^2$ of the fits. From the point of view of the goodness of fit, the additional parameter in DR$_4$ is unnecessary.  In other words, the subtraction constants extracted in the DR$_4$ fit are compatible within errors with the more restrictive high-energy behavior imposed on the DR$_3$ amplitude, and fulfill the corresponding sum rule~\eqref{eq:sumruleDR4}, as can be seen in Table~\ref{tab:fitetapDR3vsDR4}.\footnote{If the DR$_4$ subtraction constants are extracted from the DR$_3$ fit, the integral~\eqref{eq:sumruleDR4integral} leads to a tiny imaginary part in $\gamma_0$ and $\gamma_1$, which we neglect.} However, by comparing the systematic uncertainties, we see that the DR$_3$ representation is rather strongly affected by the uncertainties of the $\pi\eta$ phase shift in the high-energy region, especially the extracted Dalitz-plot parameters. The additional subtraction in DR$_4$ suppresses the influence of the high-energy phase uncertainty significantly. The price to pay is a larger (statistical) fit uncertainty due to the additional fit parameter. Forthcoming data of even higher statistics could reduce this fit uncertainty.
\esp

\section{Predictions of the dispersive representation}\label{sec:predictions}

\bsp
With the subtraction constants of the dispersive representation fitted to data on $\eta'\to\eta\pi^+\pi^-$, we are in the position to make certain additional predictions.
In Sect.~\ref{sec:abd}, we quantify the nontrivial constraint between the leading Dalitz-plot parameters $a$, $b$, and $d$ that exists in the three-parameter scheme DR$_3$. In Sect.~\ref{sec:higherdalitzparam}, we discuss higher terms in the polynomial expansion of the Dalitz-plot distribution.
We study the issue of Adler zeros of the dispersive amplitude in Sect.~\ref{sec:adlerzeros}.
Finally, in view of upcoming high-precision measurements of the neutral decay channel, we specify predictions for the same in Sect.~\ref{sec:neutralchannel}, taking into account the dominant effects of isospin breaking.
\esp

\subsection{The $a$--$b$--$d$ constraint}
\label{sec:abd}

Provided that the dispersive representation DR$_3$ with a more restrictive high-energy behavior and fewer subtractions than DR$_4$ allows a good fit to data, it is possible to formulate a relation between the three parameters $a$, $b$, and $d$, since in this scheme the Dalitz-plot distribution only depends on two parameters. Although this relation is nonlinear and cannot easily be given in closed form, we provide an approximate form of the constraint between $a$, $b$, and $d$ valid in the vicinity of the BES-III DR$_3$ fit values in Table~\ref{tab:dalitzfit}. Defining
\begin{align}
	\Delta a &:= a - a_\text{BES-III} \,, \quad \Delta b := b - b_\text{BES-III} \,, \nn
	\Delta d &:= d - d_\text{BES-III} \,,
\end{align}
we write the $a$--$b$--$d$ constraint expanded to second order as
\begin{equation}
	\label{eq:abdconstraint}
	\Delta d = C_{10} \Delta a + C_{01} \Delta b + C_{20} \Delta a^2 + C_{11} \Delta a \Delta b + C_{02} \Delta b^2
\end{equation}
and find the following results for the coefficients $C_{ij}$:\footnote{The covariance matrices are obtained by combining the correlation matrices with the absolute values of the respective uncertainties.}
\begin{align}
	\label{eq:abdcoeffs}
	\bordermatrix{
		& C_{10} & C_{01} & C_{20} & C_{11} & C_{02} \cr
		C_{10} = \phantom{+}0.21 \,{}^{+0.03}_{-0.09} & 1.00 & {}^{0.91}_{0.98} & {}^{-0.91}_{-0.98} & {}^{\phantom{+}0.92}_{\phantom{+}0.70} & {}^{\phantom{+}0.88}_{-0.41} \cr
		C_{01} = \phantom{+}1.71 \,{}^{+0.43}_{-0.97} & & 1.00 & {}^{-1.00}_{-1.00} & {}^{\phantom{+}0.99}_{\phantom{+}0.80} & {}^{\phantom{+}0.97}_{-0.27} \cr
		C_{20} = -0.43 \,{}^{-0.11}_{+0.24} & & & 1.00 & {}^{-0.99}_{-0.80} & {}^{-0.97}_{\phantom{+}0.27} \cr
		C_{11} = -0.00 \,{}^{-0.00}_{+0.00} & & & & 1.00 & {}^{\phantom{+}0.99}_{\phantom{+}0.35} \cr
		C_{02} = -0.02 \,{}^{-0.01}_{+0.01} & & & & & 1.00
		} \,,
\end{align}
where the uncertainties and correlations are due to the variation of the phase shifts, calculated from covariance matrices based on finite differences and an asymmetry due to the $\pi\eta$ phase variation.

Inserting the differences $\Delta a$ and $\Delta b$ between the VES and \mbox{BES-III} fits from Table~\ref{tab:dalitzfit} into the $a$--$b$--$d$ constraint~\eqref{eq:abdconstraint} and propagating the uncertainties from Eq.~\eqref{eq:abdcoeffs} leads to
\begin{equation}
	\Delta d = - \big(18 \,{}^{+2}_{-7} \big) \times 10^{-3} \, ,
\end{equation}
in agreement with the actual difference $\Delta d$. The systematic uncertainty of the difference $\Delta d$ is reduced compared to the uncertainties of the two determinations of the parameter $d$, since these variations are correlated. Restricting the variation of the $\pi\eta$ phase to $105° \le \delta_{12} \le 125°$ further reduces the lower asymmetric error to a value smaller in magnitude than the upper error.

Equation~\eqref{eq:abdconstraint} could be used in forthcoming experiments to perform a phenomenological fit of the Dalitz-plot distribution, where the number of free parameters is reduced by one. Alternatively, the consistency with this constraint may be checked a posteriori. We emphasize, though, that the relation is based on parameters extracted from fits using the dispersive representation, which has a more physical energy dependence than just a polynomial; as the example of the VES data demonstrates, it is not guaranteed that a direct polynomial fit leads to identical results (cf.\ again Tables~\ref{tab:etapexp}, \ref{tab:dalitzfit}).  In the case of the neutral decay channel $\eta'\to\eta\pi^0\pi^0$, the Dalitz-plot parametrization might differ by an isospin-breaking effect, which should be corrected for before the constraint is applied; see Sect.~\ref{sec:neutralchannel}.

\subsection{Higher order Dalitz-plot parameters}
\label{sec:higherdalitzparam}

From the result of the dispersion relation fitted to data, we can extract not only the Dalitz-plot parameters $a$, $b$, and $d$, but also the coefficients of higher terms in the expansion around the center of the Dalitz plot. We define these coefficients as follows:
\begin{equation}
	|\M_\mathrm{exp}|^2 = | \N_\mathrm{exp} |^2 \sum_{i,j=0}^\infty \kappa_{ij} x^i y^j \, ,
\end{equation}
where $\kappa_{00} = 1$, $\kappa_{01} = a$, $\kappa_{02} = b$, and $\kappa_{20} = d$. $C$-parity implies $\kappa_{ij} = 0$ for odd $i$. The values for the parameters $\kappa_{ij}$ with $i+j \le 4$ are listed in Table~\ref{tab:dalitzfit} for the fits of the dispersion relation to the BES-III and VES data sets. We observe a clear hierarchy
\begin{equation}
	a,b,d > \kappa_{ij}\big|_{i+j=3} > \kappa_{ij}\big|_{i+j=4} \, ,
\end{equation}
with $\kappa_{ij}|_{i+j=4}$ an order of magnitude smaller than the parameters $a$, $b$, and $d$. The results extracted from the DR$_3$ and DR$_4$ schemes are compatible with each other. In the case of DR$_3$, the systematic uncertainties from the variation of the $\pi\eta$ phase shift dominate, while in the case of DR$_4$, the main uncertainties are the statistical fit errors and the systematic uncertainties are suppressed. There are, however, some deviations between the fits to the two different experimental data sets, mainly in $\kappa_{03}$, which are a consequence of the observed tension in the leading Dalitz-plot parameters.

\bsp
If forthcoming experiments reach significantly higher statistics, it might become possible to extract these parameters directly in a phenomenological polynomial fit to data and compare with our predictions.
\esp

\subsection{Adler zeros}\label{sec:adlerzeros}

\begin{sloppypar}
In the limit of one of the pion momenta going to zero, $p_1\to0$ or $p_2\to0$, current algebra predicts two {\em Adler zeros} of the amplitude~\cite{Riazuddin:1971ie,Adler:1964um,Adler:1965ga}. These soft-pion theorems are protected by SU(2)$\times$SU(2) symmetry, hence they only receive corrections of $\order(\mpc^2)$. While the off-shell continuation of the amplitude cannot be defined unambiguously, the Adler theorem implies that the on-shell amplitude is of $\order(\mpc^2)$ at the two soft-pion points
\begin{align}
	\begin{aligned}
		s_1 &= 2 \mpc^2\,, \quad & t_1 &= \metap^2\,, \quad & u_1 &= \meta^2 \,, \\
		s_2 &= 2 \mpc^2\,, \quad & t_2 &= \meta^2\,, \quad & u_2 &= \metap^2 \,.
	\end{aligned}
\end{align}
In the past, claims have been made that the $a_0(980)$ resonance removes the Adler zeros based on the explicit inclusion of a scalar-resonance propagator~\cite{Deshpande:1978iv}. Let us study this issue within our dispersive framework. 

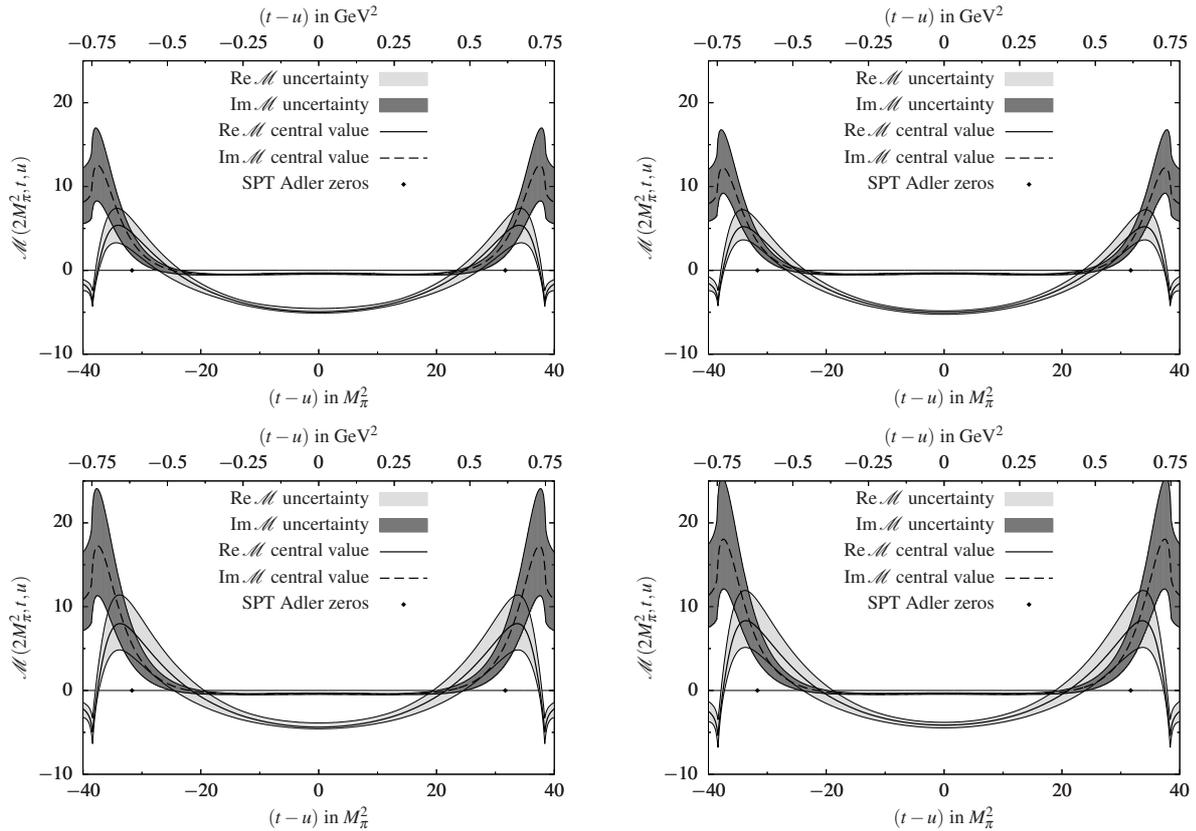
\begin{figure*}[t!]
\centering
\large
\scalebox{0.6}{\input{plots/BES_adler_zero_DR3}}\hspace{0.5cm}
\scalebox{0.6}{\input{plots/BES_adler_zero_DR4}}\vspace{0.2cm}
\scalebox{0.6}{\input{plots/VES_adler_zero_DR3}}\hspace{0.5cm}
\scalebox{0.6}{\input{plots/VES_adler_zero_DR4}}
 \caption{Real and imaginary part of the amplitude along a line of fixed $s=2\mpc^2$. 
 The upper two panels give the fit of the dispersive representation to the BES-III, the lower two panels to the VES data set. The DR$_3$ (left column) and DR$_4$ fits (right column) are shown.}
\label{fig:Adlerzero}
\end{figure*}

In Fig.~\ref{fig:Adlerzero}, we show the result for the dispersive amplitude fitted to data, evaluated along a line of fixed $s=2\mpc^2$. Both subtraction schemes DR$_3$ and DR$_4$ lead to very similar results. We encounter zeros in both the real and imaginary parts of the amplitude at positions close to the soft-pion points, but for slightly smaller values of $|t-u|$. At the resonance positions
\begin{equation}
	|t-u| \approx 2 M_{a_0}^2 - \metap^2 - \meta^2 \, ,
\end{equation}
which are also close but outside the soft-pion points, we observe a large peak in the imaginary part and another zero in the real part. We conclude that the dispersive representation refutes the resonance argument of Ref.~\cite{Deshpande:1978iv} that for $\eta'\to\eta\pi\pi$ the low-energy theorem does not result in an Adler zero of the amplitude. Although the corrections at the soft-pion points are of $\order(\mpc^2 / (\metap^2 - M_{a_0}^2))$, which is not a small quantity, the zeros of the amplitude survive and are just shifted to smaller values of $|t-u|$.
\end{sloppypar}

\subsection{Neutral channel}
\label{sec:neutralchannel}

\bsp
So far, we have analyzed experimental Dalitz-plot data sets for $\eta'\to\eta\pi^+\pi^-$.  To deduce a comparably precise 
prediction for the neutral final state $\eta'\to\eta\pi^0\pi^0$, we have to consider potentially enhanced sources 
of isospin-symmetry violation.
The consideration of isospin breaking, in particular due to the pion mass difference, in Dalitz-plot 
studies is a rather subtle affair, which has recently received some attention in the context of $\eta\to3\pi$ decay
studies~\cite{Ditsche:2008cq,Schneider:2010hs,Colangelo:2016jmc}.  While a correction for phase space alone
is straightforward, it is often less so to construct an amplitude that accordingly has all the thresholds in the right places.
This is particularly true in the context of dispersive analyses, as the ubiquitous phase shifts are typically derived 
from a formalism (the Roy equations) that incorporates isospin symmetry in an essential manner.  Isospin-breaking effects
are bound to affect neutral-pion final states more strongly, as the isospin-symmetric phase shifts use the \emph{charged}
pion mass as their reference scale.  Furthermore, the pion mass difference induces a cusp in $\pi^0\pi^0$ invariant mass spectra
at the $\pi^+\pi^-$ threshold~\cite{Budini:1961bac,Cabibbo:2004gq}, a nonanalyticity that cannot be approximated by a polynomial
Dalitz-plot distribution.  Such a cusp is known to appear more strongly in $\eta'\to\eta\pi^0\pi^0$~\cite{Kubis:2009sb} than,
e.g., in $\eta\to3\pi^0$~\cite{Bissegger:2007yq}.\footnote{At two-loop order, the lower $\pi^0\pi^0$ mass induces an anomalous
threshold in the $\eta'\to\eta\pi^+\pi^-$ amplitude.  We have checked, though, that this does not lead to any enhanced isospin-breaking
effect, using the representation of Ref.~\cite{Kubis:2009sb}.  We thank M.~Mikhasenko for suggesting this check.}

As we want to avoid the complications to solve Khuri--Treiman equations with coupled channels~\cite{Guo:2015kla,Albaladejo:2017hhj}, we once more
follow the strategy proposed in Sect.~\ref{sec:phaseshifts} and construct effective single-channel phase shifts, to be used
as input for the corresponding Omn\`es functions, from the phases of certain scalar form factors. We observe that the cusp
structure of the decay amplitude for $\eta'\to\eta\pi^0\pi^0$ is very similar to that of the neutral-pion scalar form factor $F_0(s)$, 
\begin{align}
F_0(s) = \langle \pi^0(p_1)\pi^0(p_2) | \hat m( \bar uu + \bar dd) | 0 \rangle \,, \quad s = (p_1+p_2)^2\,, \label{eq:defF0}
\end{align}
where $\hat m = (m_u+m_d)/2$ is the average light quark mass, 
in particular given that crossed-channel effects have a negligible influence on its 
strength~\cite{Kubis:2009sb}.  We will therefore employ $\arg F_0(s)$ as the input $\pi^0\pi^0$ $S$-wave phase shift.

The precise construction of the effective $\pi^0\pi^0$ phase shift from the corresponding scalar form factor
is discussed in~\ref{app:isospinbreaking}.
It takes into account the analytic structure near the two-pion thresholds, where isospin breaking is enhanced due
to the proximity of ($S$-wave) threshold cusps, and scales effectively like $\sqrt{M_\pi^2-M_{\pi^0}^2}$, 
where we denote by $\mpc$ the charged and by $\mpn$ the neutral-pion mass.  
Regular, polynomial isospin-breaking effects of order $M_\pi^2-M_{\pi^0}^2$ are still neglected and assumed to be very small.
Similarly, we show there how a simple rescaling can be used to adapt the $\pi^\pm\eta$ phase shift to $\pi^0\eta$ in such a way
as to put all thresholds into the right places.  

\begin{figure}[t!]
\centering
\large
\scalebox{0.66}{\input{plots/BES_Yproj_neutral_DR3}}
\scalebox{0.66}{\input{plots/BES_Yproj_neutral_DR4}}
 \caption{Decay spectrum for $\eta'\to\eta\pi^{0}\pi^{0}$ integrated over the variable $x$ and divided by the integral over the normalized phase space $\text{d}\bar{\Phi}(x,y)$, where both are individually normalized as given in Eq.~\eqref{eq:barMnorm}. The prediction is based on the BES-III fit result for subtraction constants of the charged decay, see Tables~\ref{tab:fitetapDR3} and \ref{tab:fitetapDR3vsDR4}. The DR$_{3}$ (upper panel) and DR$_{4}$ fit results (lower panel) are shown. The two error bands in each figure give the uncertainties resulting from the fit to data and originating from the variation of the phase input, respectively.}
\label{fig:etapYprojneutral}
\end{figure}
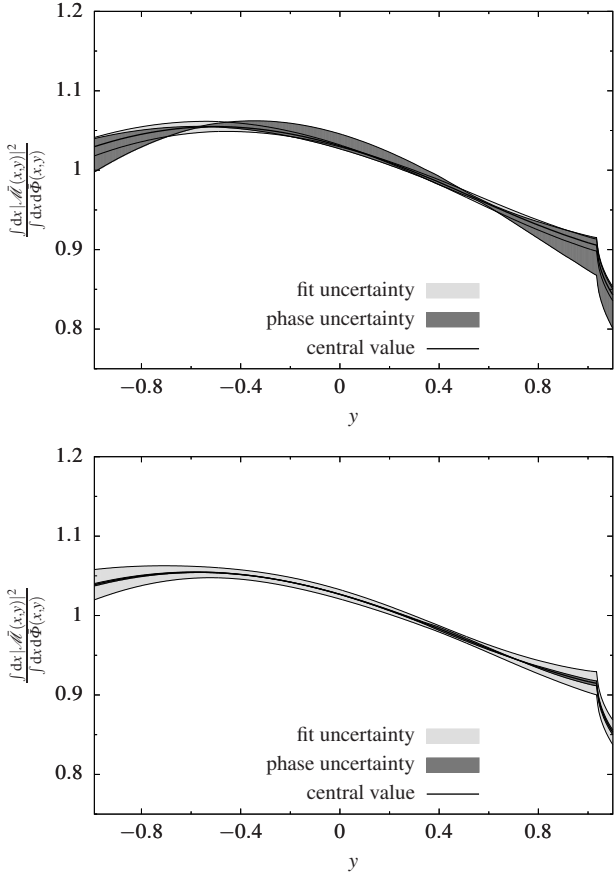

Our prediction for the decay $\eta'\to\eta\pi^0\pi^0$ is therefore based on the subtraction constants as extracted 
from $\eta'\to\eta\pi^+\pi^-$, but with $\pi^0\pi^0$ and $\pi^0\eta$ phase shifts adapted as compared to the 
$\pi^+\pi^-$ and $\pi^\pm\eta$ ones; in this way, the dominant effects of isospin violation due to the charged-to-neutral
pion-mass difference are taken into account. The resulting decay spectrum projected on the $y$ direction is shown in Fig.~\ref{fig:etapYprojneutral}, where the nonanalytic structure of the $\pi^{+}\pi^{-}$ cusp is clearly visible.
\esp

Another rather strong isospin-breaking effect appears in the change of coordinates from the Mandelstam variables to the Dalitz-plot variables $x$ and $y$ if the neutral-pion mass is used in Eq.~\eqref{eq:xyetapdef} for the parametrization of the neutral Dalitz plot. As such this effect has nothing to do with the decay amplitude itself but it affects the Dalitz-plot expansion parameters. We introduce the isospin-breaking parameter
\begin{equation}
	\epsilon_\mathrm{iso} := \frac{(\meta + 2\mpc) \mpn Q_{\eta'}^0}{(\meta + 2\mpn) \mpc Q_{\eta'}} - 1 \approx 4.7\% \, ,
\end{equation}
where $Q_{\eta'}^0 := \metap - \meta - 2\mpn$. Given the phenomenological observation that $1 \gg a, b, d > \kappa_{ij}$ for $i+j\ge3$, we neglect terms of $\order(\epsilon_\mathrm{iso} a^2, \epsilon_\mathrm{iso} ab, \epsilon_\mathrm{iso} ad, \epsilon_\mathrm{iso} \kappa_{ij})$ and of second order in isospin breaking and find the following relation between the Dalitz-plot parameters in the charged (no superscript) and the neutral system (superscript $0$):
\begin{align}
	a^0 &= a + \epsilon_\mathrm{iso} \left( a + 2 b \right) \,, \quad b^0 = b \left( 1  + 2 \epsilon_\mathrm{iso} \right) \,, \nn
	d^0 &=d \bigg( \frac{Q_{\eta'}^0}{Q_{\eta'}} \bigg)^2 \,.
\end{align}
For parameters comparable to the BES-III fit results, this amounts to a sizable correction: we find $a^0 \approx 1.25 a$, $b^0\approx 1.09 b$, $d^0\approx 1.15 d$.

In particular, this correction has to be taken into account if the $a$--$b$--$d$ constraint~\eqref{eq:abdconstraint} formulated in the charged system is employed for neutral Dalitz-plot parameters. For convenience, below we provide the explicit form of the $a$--$b$--$d$ constraint for the neutral system. We apply the isospin correction to the BES-III fit values to define the reference point in the neutral system:
\begin{align}
	\Delta a^0 &:= a^0 - a^0_\text{BES-III} \,, \quad \Delta b^0 := b^0 - b^0_\text{BES-III} \,, \nn
	\Delta d^0 &:= d^0 - d^0_\text{BES-III} \,,
\end{align}
where
\begin{align}
	a^0_\text{BES-III} &= -51 \times 10^{-3} \,, \quad b^0_\text{BES-III} = -96 \times 10^{-3} \,, \nn
	d^0_\text{BES-III} &= -78 \times 10^{-3} \,.
\end{align}
Then the $a$--$b$--$d$ constraint reads again
\begin{align}
	\label{eq:abdconstraintneutral}
	\Delta d^0 &= C_{10}^0 \Delta a^0 + C_{01}^0 \Delta b^0 \nn
		&+ C_{20}^0 (\Delta a^0)^2 + C_{11}^0 \Delta a^0 \Delta b^0 + C_{02}^0 (\Delta b^0)^2 \,,
\end{align}
where the neutral coefficients are given by
\begin{align}
	C_{10}^0 &= \bigg( \frac{Q_{\eta'}^0}{Q_{\eta'}} \bigg)^2 (1 - \epsilon_\mathrm{iso}) C_{10} \, , \nn
	C_{01}^0 &= \bigg( \frac{Q_{\eta'}^0}{Q_{\eta'}} \bigg)^2 \left( C_{01} - 2 \epsilon_\mathrm{iso} (C_{01} + C_{10}) \right) \, , \nn
	C_{20}^0 &= \bigg( \frac{Q_{\eta'}^0}{Q_{\eta'}} \bigg)^2 (1 - 2 \epsilon_\mathrm{iso}) C_{20} \, , \nn
	C_{11}^0 &= \bigg( \frac{Q_{\eta'}^0}{Q_{\eta'}} \bigg)^2 \left( C_{11} - \epsilon_\mathrm{iso} (3C_{11} + 4C_{20}) \right) \, , \nn
	C_{02}^0 &= \bigg( \frac{Q_{\eta'}^0}{Q_{\eta'}} \bigg)^2 \left( C_{02} - 2 \epsilon_\mathrm{iso} (2C_{02} + C_{11}) \right) \, .
\end{align}
With the values for the charged coefficients $C_{ij}$ given in Eq.~\eqref{eq:abdcoeffs}, this results in\footnote{The covariance matrices are obtained by combining the correlation matrices with the absolute values of the respective uncertainties.}
\begin{align}
	\label{eq:abdcoeffsneutral}
	\bordermatrix{
		& C_{10}^0 & C_{01}^0 & C_{20}^0 & C_{11}^0 & C_{02}^0 \cr
		C_{10}^0 = \phantom{+}0.23 \,{}^{+0.03}_{-0.09} & 1.00 & {}^{0.91}_{0.98} & {}^{-0.91}_{-0.98} & {}^{\phantom{+}0.91}_{\phantom{+}0.97} & {}^{\phantom{+}0.87}_{-0.45} \cr
		C_{01}^0 = \phantom{+}1.75 \,{}^{+0.44}_{-0.99} & & 1.00 & {}^{-1.00}_{-1.00} & {}^{\phantom{+}1.00}_{\phantom{+}1.00} & {}^{\phantom{+}0.96}_{-0.32} \cr
		C_{20}^0 = -0.44 \,{}^{-0.11}_{+0.25} & & & 1.00 & {}^{-1.00}_{-1.00} & {}^{-0.96}_{\phantom{+}0.32} \cr
		C_{11}^0 = \phantom{+}0.09 \,{}^{+0.03}_{-0.06} & & & & 1.00 & {}^{\phantom{+}0.97}_{-0.27} \cr
		C_{02}^0 = -0.02 \,{}^{-0.01}_{+0.01} & & & & & 1.00
		} \,.
\end{align}

\section{Comparison to chiral approaches}\label{sec:matching}

As we have seen in Sect.~\ref{sec:subconsts}, our dispersive amplitude allows a good fit to both Dalitz-plot data and the partial decay width. In a next step, we compare the dispersive amplitude with predictions from extensions of chiral perturbation theory; we choose next-to-leading order large-$N_c$ \chpt{} and resonance chiral theory (\rcht{}) for that purpose. The results for the amplitudes in both frameworks, taken from the analysis described in Ref.~\cite{Escribano:2010wt}, are discussed  in Sect.~\ref{sec:chiralamps}. We decompose these amplitudes into forms amenable to a comparison to the dispersion relations and perform the matching in Sect.~\ref{sec:matchingresults}. This allows us to obtain chiral predictions for the subtraction constants and to compare them with the fits to data.

\subsection{Amplitudes from large-\boldmath{$N_c$} \chpt{} and \rcht{}}\label{sec:chiralamps}

\begin{sloppypar}
Large-$N_c$ chiral perturbation theory allows the explicit inclusion of the $\eta'$ meson in an effective-Lagrangian framework. It is founded on the notion that as $N_c\to\infty$, the U(1)$_A$ anomaly
and thus the chiral-limit mass of the $\eta'$ vanishes: the $\eta'$ becomes a Goldstone boson as the U(3)$_L\times$U(3)$_R$ symmetry is spontaneously broken to U(3)$_V$~\cite{Kaiser:2000gs,Gasser:1984gg}.
At leading order (LO) the $\eta'\to\eta\pi\pi$ amplitude is given as~\cite{Cronin:1967jq,Schwinger:1968zz,DiVecchia:1980vpx,Fajfer:1987ij,HerreraSiklody:1999ss,Schechter:1993tc,Singh:1975aq,Escribano:2010wt}
\begin{equation}
 \M_{\rm LO}^{\text{\chpt}}(s,t,u)=\frac{\mpc^2}{6F_\pi^2}\Bigl[2\sqrt{2}\cos(2\theta_P)-\sin(2\theta_P)\Bigr]\,,
\end{equation}
where $\theta_P$ is the $\eta$--$\eta'$ mixing angle that relates the octet and singlet states to the physical $\eta,\eta'$ states at leading order,
and $F_\pi$ is the pion decay constant. At next-to-leading order (NLO) loop contributions are still suppressed in the large-$N_c$ counting, and the full amplitude can be derived
from the NLO Lagrangian~\cite{Escribano:2010wt}, 
\begin{align}\label{eq:NLOetapAmp}
 \M_{\rm NLO}^{\text{\chpt}}&(s,t,u) =c_{qq}\Bigl[\frac{\mpc^2}{2} 
-\frac{2L_5}{F_\pi^2}\big(\metap^2+\meta^2+2\mpc^2\big)\mpc^2 \nn
&+\frac{2(3L_2+L_3)}{F_\pi^2}\big(s^2+t^2+u^2-\metap^4-\meta^4-2\mpc^4\big)\nn
& +\frac{24L_8}{F_\pi^2}\mpc^4+\frac{2}{3}\Lambda_2\mpc^2\Bigr]+c_{sq}\frac{\sqrt{2}}{3}\Lambda_2\mpc^2\,,
\end{align}
where $c_{qq}$ and $c_{sq}$ are functions of the octet and singlet decay constants $F_{8/0}$,
as well as of the \emph{two} mixing angles $\theta_{8/0}$ required in the $\eta$-$\eta'$ mixing scheme
at NLO~\cite{Leutwyler:1997yr,Kaiser:1998ds}:
\begin{align}
	c_{qq} &= \frac{F_0^2 S_0 - 2 F_8^2 S_8 + 2\sqrt{2} F_8 F_0 C_{08}}{3 F_8^2 F_0^2 \cos^2(\theta_8 - \theta_0)} \, , \nn
	c_{sq} &= - \frac{\sqrt{2} F_0^2 S_0 + \sqrt{2} F_8^2 S_8 + F_0 F_8 C_{08}}{3 F_8^2 F_0^2 \cos^2(\theta_8 - \theta_0)} \, , \nn
	S_0 &= \sin(2\theta_0) \,, \quad S_8 = \sin(2\theta_8) \,, \quad C_{08} = \cos(\theta_0 + \theta_8) \, .
\end{align}
Numerically, we use $c_{qq} = (97.6\pm7.1)\GeV^{-2}$ and $c_{sq} = (4.4\pm2.9)\GeV^{-2}$~\cite{Escribano:2010wt,Escribano-private} (compare also Ref.~\cite{Escribano:2015yup}).

For the low-energy constants $L_2$ and $L_3$, we use again the values from Ref.~\cite{Colangelo:2015kha}, given in Eq.~\eqref{eq:kl4lecs}, while for $L_5$ and $L_8$, we use the results of the global BE14 fit~\cite{Bijnens:2014lea}:
\begin{equation}
L_5=1.01(06) \times 10^{-3}\,,~ L_8=0.47(10) \times 10^{-3}\,,
\end{equation}
and finally $\Lambda_2= 0.3$~\cite{Escribano:2010wt}.

The second chiral approach that we consider is resonance chiral theory, which describes the interactions between Goldstone bosons and resonances explicitly~\cite{Ecker:1988te,Ecker:1989yg}. \rcht{} finds its most 
prominent application in the estimate of low-energy constants by means of resonance saturation. It can, however, also be used to directly derive the $\eta'\to\eta\pi\pi$ decay amplitude 
from the \rcht{} Lagrangian~\cite{Escribano:2010wt}.
To properly match it to the dispersive amplitude, it is useful to write it in the form
\begin{align}\label{eq:RChTAmp}
 &\M^\text{\rcht}(s,t,u) =c_{qq}\biggl\{\frac{M_S^2}{M_S^2-s}\bigg(\rho -\frac{c_d^2\Delta}{F_\pi^2M_S^2}
-\rho\frac{s}{M_S^2}\bigg) \nn
&+\bigg[\frac{M_S^2}{M_S^2-s}+\frac{M_S^2}{M_S^2-t}+\frac{M_S^2}{M_S^2-u} \bigg] \frac{M_S^2}{F_\pi^2}\bigg(\frac{\xi}{M_S^4}+\frac{\psi}{M_S^2}+c_d^2\bigg)\biggr\}\,,
\end{align}
where 
\begin{align}
\xi&=(\metap^2+\mpc^2)(\meta^2+\mpc^2)c_d^2-6\mpc^2\setap c_d c_m+4\mpc^4 c_m^2\,,\nn
\psi&=-3\setap c_d^2+4\mpc^2c_mc_d\,, \nn
\rho&= \frac{\mpc^2}{2} - 3\frac{\psi + c_d^2(M_S^2+\setap)}{F_\pi^2} \,.
\end{align}
Here $c_d$ and $c_m$ describe the coupling between the scalar resonances and the Goldstone bosons, and $M_S=0.980\GeV$ is the mass of the scalar multiplet. We will use $c_d=(0.026\pm0.009)\GeV$ and $c_m=(0.080\pm0.021)\GeV$, which fulfill the theoretical constraint $4c_dc_m=F_\pi^2$ rather well~\cite{Escribano:2010wt}.

In the limit of large scalar masses, that is $s$, $t$, $u$, $\mpc^2$, $\meta^2$, $\metap^2\ll M_S^2$, the low-energy expansion 
of the amplitude~\eqref{eq:RChTAmp} agrees with Eq.~\eqref{eq:NLOetapAmp} for~\cite{Ecker:1988te}
\begin{align}
3L_2+L_3=\frac{c_d^2}{2M_S^2}\,,\quad L_5=\frac{c_dc_m}{M_S^2}\,,\quad L_8=\frac{c_m^2}{2M_S^2}\,,\quad \Lambda_{2}=0\,. 
\end{align}
We note that these relations are not at all well fulfilled for the values of the constants we employ,
as cited above: in contrast to vector or axialvector quantities, resonance saturation of low-energy
constants by narrow scalars is problematic at best.
\end{sloppypar}

\subsection{Matching chiral approaches with the dispersion relation}\label{sec:matchingresults}

\begin{sloppypar}
We perform the matching to the dispersion relations as follows: we decompose the chiral amplitudes into single-variable functions and require that the Taylor coefficients of the latter agree between chiral and dispersive representations. This allows us to extract chiral predictions for the subtraction constants. The derivation of the explicit matching equations can be found in~\ref{app:matchingequations}.

In the case of large-$N_c$ \chpt{}, it is not possible to match directly to the three-parameter representation~\eqref{eq:MinteqSWaves}, because the asymptotic behavior of the amplitude violates the condition that was used to fix the ambiguity of the decomposition. Therefore, one has to match the chiral amplitude to the four-parameter representation~\eqref{eq:Minteq}.
In the case of \rcht{}, the situation is different, because the asymptotic behavior allows a matching to the three-parameter representation DR$_3$. Hence, we have two possibilities: either we perform the matching with the DR$_3$ representation and require that the constant and linear terms of the Taylor expansion agree between \rcht{} and dispersive representation, or we can also perform the matching with the DR$_4$ representation and match constant, linear, and quadratic terms in the expansion.
\end{sloppypar}

The results of the matching are shown in Table~\ref{tab:matchingresults} and should be compared to Tables~\ref{tab:fitetapDR3} and \ref{tab:fitetapDR3vsDR4}. In order to compare the chiral predictions with the fits to data, we define the quantity
\begin{equation}
	\label{eq:compatibilityWithFits}
	\Delta_\text{exp}^2 := \sum_{i,j} (t_i \mp t_i^\text{exp}) (C^{-1})_{ij}(t_j \mp t_j^\text{exp}) \,,
\end{equation}
where $t_i$ stands generically for the Taylor coefficients used in the matching equations and $C_{ij}$ is the covariance matrix of $t_i \mp t_i^\text{exp}$, including both statistical and systematic errors. We choose the sign that leads to the smaller value of $\Delta_\text{exp}^2$ (the minus sign for \rcht{} and the plus sign for large-$N_c$ \chpt{})---we stress again that the dispersive fits to data determine the amplitude only up to an overall sign. In Eq.~\eqref{eq:compatibilityWithFits}, we choose to compare the Taylor coefficients instead of the subtraction constants, because their chiral prediction only depends on the model input and is not entangled with Omnès expansion parameters.

\bsp
The analogous quantity for the DR$_3$ fits to BES-III and VES is
\begin{equation}
	\Delta_\text{BES-III,VES}^2 = 22 \, ,
\end{equation}
which quantifies again the tension between the two experiments. The values listed in Table~\ref{tab:matchingresults} show that for both chiral approaches the four-parameter matching involving the quadratic Taylor coefficients does not work at all. The fact that the DR$_3$ matching to \rcht{} gives smaller values for $\Delta_\text{exp}^2$ is explained rather by the larger systematic uncertainties in this setup than a better agreement of the central values.

Given the tension between the two experiments, it is difficult to draw a conclusion concerning the two chiral approaches. We observe mainly two problems in the matching.
\begin{enumerate}
	\item The overall normalization is not well reproduced.
	\item While the matching in both DR$_3$ and DR$_4$ schemes leads to reasonable relative values of $\beta/\alpha$ or $\beta_0/\alpha_0$, the predictions for the relative values of the terms $\gamma/\alpha$ or $\gamma_{0,1}/\alpha_0$ do not work at all.
\end{enumerate}
In the case of large-$N_c$ \chpt{}, the amplitude scales with $(3L_2+L_3)$, up to terms suppressed by $\mpc^2$. On the one hand, the direct insertion of the phenomenological SU$(3)$ LECs~\eqref{eq:kl4lecs} could be problematic. E.g.\ we have not taken into account additional uncertainties due to the scale dependence of the SU$(3)$ LECs, which does not appear at NLO in large-$N_c$ \chpt{}. On the other hand, we cannot exclude that higher-order effects in the chiral and large-$N_c$ expansion (i.e.\ effects only entering at one loop) produce large corrections.
\esp

\begin{table}
\centering
\renewcommand{\arraystretch}{1.4}
	\begin{tabular}{c c | c c c}
	\toprule
					&	\rcht{}, DR$_3$		&					&	\rcht{}, DR$_4$				&	large-$N_c$ \chpt{}	 \\
	\midrule
	$\alpha$			&	$-7\pm4$			&	$\alpha_0$		&	$-6\pm4$					&	$\phantom{+}17\pm13$	\\
	$\beta$			&	$16\pm10$			&	$\beta_0$			&	$12\pm9$					&	$-42\pm32$	\\
	$\gamma$		&	$0.8\pm0.4$		&	$\gamma_0$		&	$24\pm17$				&	$-73\pm57$	\\
					&					&	$\gamma_1$		&	$0.8\pm0.4$				&	$-14\pm12$	\\
	\midrule
	$\Delta^2_\text{BES-III}$	&	$18$			&	&	$145$	&	$451$	\\
	$\Delta^2_\text{VES}$	&	$17$			&	&	$116$	&	$343$	\\
	\bottomrule
	\end{tabular}
\renewcommand{\arraystretch}{1.0}
\caption{Results of the matching between the dispersive amplitude and the large-$N_c$ \chpt{} and \rcht{} representations. In the case of \rcht{}, the uncertainties are due to $c_{qq}$, $c_d$, and $c_m$, while for large-$N_c$ \chpt{}, the errors are due to $c_{qq}$, $c_{sq}$, $L_2$, $L_3$, $L_5$, and $L_8$. The quantity $\Delta^2_\text{exp}$ is defined in Eq.~\eqref{eq:compatibilityWithFits}.}
\label{tab:matchingresults}
\end{table}

\section{Summary and conclusion}

\bsp
In this article we have presented a dispersive analysis of the decay $\eta'\to\eta\pi\pi$. We have derived a set of integral equations on the grounds of unitarity for the corresponding scattering process
and performed an analytic continuation to the physical region of the three-particle decay. 
The integral equations depend on $\pi\pi$ and $\pi\eta$ scattering phase shifts as well as on a set of subtraction constants. The phase shift of $\pi\pi$ scattering is strongly constrained by chiral symmetry
and Roy equations~\cite{Caprini:2011ky}. For the $\pi\eta$ phase shift, the phase of the scalar form factor $F_S^{\eta\pi}$ of Ref.~\cite{Albaladejo:2015aca} is used as input.

Within two different subtraction schemes, the free constants have been fitted to data sets of the Dalitz-plot distribution, sampled from the experimentally measured polynomial Dalitz-plot parametrizations of the VES~\cite{Dorofeev:2006fb} and \mbox{BES-III}~\cite{Ablikim:2010kp} experiments, as well as the partial decay width~\cite{PDG}. The fits to data require a smaller number of free parameters than a polynomial Dalitz-plot parametrization and still exhibit a good $\chi^2$. Therefore, we have been able to derive a constraint between the Dalitz-plot parameters $a$, $b$, and $d$ from one of the two dispersive representations. Furthermore, we have made predictions for higher-order polynomial parameters that have not been measured experimentally so far.  By taking into account the leading isospin-breaking effects, we have also provided predictions for the neutral decay channel. We have further observed that the amplitude exhibits Adler zeros despite the presence of the nearby $a_0(980)$ resonance, which only shifts the position of these zeros somewhat compared to the prediction of the soft-pion theorem.

Matching to large-$N_c$ \chpt{} we find large deviations for the subtraction constants, rendering this approach unfit to be used in an attempt to extract information on $\pi\eta$ scattering. 
When matching to \rcht{}, the deviations are a bit smaller. However, the matching in the three-parameter scheme shows less tension mainly because of the larger systematic uncertainties. Furthermore, the \rcht{} framework does not easily allow for systematic improvements. Therefore, the theoretical prediction of the subtraction constants with chiral models as opposed to fitting them to data currently does not seem to be a viable option.

In the minimally subtracted dispersive amplitude representation, we have observed a rather significant dependence of the subtraction constants on the assumed high-energy behavior of the $\pi\eta$ phase shift input.  More precise experimental data than the one available to us in this study is required to come to definite conclusions about the sensitivity of the $\eta'\to\eta\pi\pi$ decay to low-energy $\pi\eta$ scattering.

The derived amplitudes, compatible with the fundamental principles of analyticity and unitarity, provide ideal tools to analyze forthcoming high-precision Dalitz-plot data, in particular also for the neutral channel, by the A2~collaboration~\cite{A2-private} and BES-III~\cite{Kupscpriv}; see also Ref.~\cite{NuhnDiplom} for a possible measurement at CB-ELSA.
As a further theoretical development,
the fitted dispersive parametrization will be used as an input in a forthcoming analysis of inelasticity effects in $\eta'\to3\pi$~\cite{etaprimeto3pi}.

\esp


\begin{acknowledgements}
\begin{sloppypar}
We thank P.~Adlarson, G.~Colangelo, R.~Escribano, A.~Kup\'s\'c, S.~Lanz,
and S.~Prakhov for useful discussions.
We are grateful to J.~Daub and B.~Moussallam for providing us with the numerical $T$-matrices
of Refs.~\cite{Daub:2015xja,Albaladejo:2015aca}, respectively, 
and to S.~Ropertz for some independent checks.
S.P.S.\ would like to thank the Albert Einstein Center for Fundamental Physics at the University of Bern
for a stimulating stay at an early stage of this project. 
Partial financial support by
the DFG (CRC~16, ``Subnuclear Structure of Matter''),
by DFG and NSFC through funds provided to the Sino--German CRC~110
``Symmetries and the Emergence of Structure in QCD,''
by the project ``Study of Strongly Interacting Matter'' (HadronPhysics3, Grant Agreement No.~283286) 
under the 7th Framework Program of the EU,
by the Bonn--Cologne Graduate School of Physics and Astronomy,
and the DOE (Grant No.\ DE-SC0009919) is gratefully acknowledged.
P.S.\ is supported by a grant of the Swiss National Science Foundation (Project No.\ P300P2\_167751).
\end{sloppypar}
\end{acknowledgements}


\appendix

\section{Decomposition of the amplitude}\label{app:decomposition}

\begin{sloppypar}
We assume that the amplitude for the $\eta'\eta\pi\pi$ four-point function in scattering kinematics is described in terms of Mandelstam variables, $\M(s,t,u)$. The amplitude has an $s$-channel unitarity
cut starting at the threshold $s_0$, the $t$- and $u$-channel cuts start at $t_0=u_0$. The latter appear as left-hand cuts in the $s$-channel. We can
write down a dispersion relation for the amplitude for a fixed value of $t$,
\begin{align}
 \M(s,t,u)&=P_{n-1}^{t}(s,t,u)+\frac{s^n}{2\pi i}\int_{s_0}^\infty \diff s'\frac{\disc\M(s',t,u(s'))}{s'^n(s'-s)} \nn
&+\frac{u^n}{2\pi i}\int_{u_0}^\infty \diff u'\frac{\disc\M(s(u'),t,u')}{u'^n(u'-u)}\,, \label{eq:fixed-t}
\end{align}
where 
$s(u')=3\setap-t-u'=s+u-u'$ and similarly for $u(s')$,
and $P_{n-1}^{t}(s,t,u)$ is a subtraction polynomial of order $n-1$. Its coefficients depend on $t$; it can be written in the form
\begin{equation}
 P_{n-1}^t(s,t,u)=p_0(t)+p_1(t)(s-u)+\ldots\,.
\end{equation}

We now perform a partial-wave expansion of the amplitude in the $s$- and $u$-channel, 
\begin{align}
 \M(s,t,u)&=m_0(s)+m_{\ell\geq2}(s,t,u)\,,\nn
 \M(s,t,u)&=n_0(u)+n_1(u) \,z_u+n_{\ell\geq 2}(s,t,u)\,,
\end{align}
truncating at $\ell=1$. 
Inserting it into the dispersion integrals and using the definition for $z_u$, we find
\begin{align}
& \M(s,t,u)=P_{n-1}^t(s,t,u) \nn &+\frac{s^n}{2\pi i}\int_{s_0}^\infty \diff s'\frac{\disc m_0(s')}{s'^n(s'-s)} 
+\frac{u^n}{2\pi i}\int_{u_0}^\infty \diff u'\frac{\disc n_0(u')}{u'^n(u'-u)}\nn
&+\frac{u^n}{2\pi i}\int_{u_0}^\infty \diff u'\frac{(u'(u-u'+s-t) + \Delta)\disc n_1(u')}{u'^{n+1}\kappa_{\pi\eta}(u')(u'-u)}\,.
\end{align}
We can simplify the integral over the $\pi\eta$ $P$-wave, absorbing parts of it in $P_{n-1}^t(s,t,u)$,
and arrive at
\begin{align}\label{eq:recthmt}
& \M(s,t,u)=P_{n-1}^t(s,t,u)\nn
&+\frac{s^n}{2\pi i}\int_{s_0}^\infty \diff s'\frac{\disc m_0(s')}{s'^n(s'-s)}+\frac{u^n}{2\pi i}\int_{u_0}^\infty \diff u'\frac{\disc n_0(u')}{u'^n(u'-u)}\nn
&+\Bigl[u(s-t) + \Delta\Bigr]\frac{u^{n-2}}{2\pi i}\int_{u_0}^\infty \diff u'\frac{\disc n_1(u')}{u'^{n-1}\kappa_{\pi\eta}(u')(u'-u)}\,.
\end{align}
The same exercise can be performed at fixed $u$, resulting in Eq.~\eqref{eq:recthmt} with $t\leftrightarrow u$.
The term containing $s$-channel $\pi\pi$ scattering is the same at fixed $t$ and fixed $u$;
moreover, at fixed $u$ the latter two integrals of Eq.~\eqref{eq:recthmt}
can be absorbed in $P_{n-1}^u(s,t,u)$ and vice versa for fixed $t$, provided that $n\ge2$. Finally, performing the same steps for fixed $s$ allows us to pin down $P_{n-1}(s,t,u)$ to a polynomial in all three Mandelstam variables. We can thus write down a symmetrized dispersion relation:
\begin{align}\label{eq:recthm}
& \M(s,t,u)=P_{n-1}(s,t,u)+\frac{s^n}{2\pi i}\int_{s_0}^\infty \diff s'\frac{\disc m_0(s')}{s'^n(s'-s)}\nn
&+\frac{u^n}{2\pi i}\int_{u_0}^\infty \diff u'\frac{\disc n_0(u')}{u'^n(u'-u)}+\frac{t^n}{2\pi i}\int_{t_0}^\infty \diff t'\frac{\disc n_0(t')}{t'^n(t'-t)}\nn
&+\Bigl[u(s-t) + \Delta\Bigr]\frac{u^{n-2}}{2\pi i}\int_{u_0}^\infty \diff u'\frac{\disc n_1(u')}{u'^{n-1}\kappa_{\pi\eta}(u')(u'-u)}\nn
&+\Bigl[t(s-u) + \Delta\Bigr]\frac{t^{n-2}}{2\pi i}\int_{t_0}^\infty \diff t'\frac{\disc n_1(t')}{t'^{n-1}\kappa_{\pi\eta}(t')(t'-t)}\,.
\end{align}
This demonstrates the form of the $\eta'\to\eta\pi\pi$ decay amplitude claimed in Eq.~\eqref{eq:decomposition}.
Note again that the above relation is predicated on neglecting discontinuities of 
$\ell\geq 2$ partial waves.
\end{sloppypar}

\section{Construction of $\pi^0\pi^0$ and $\pi^0\eta$ phase shifts}
\label{app:isospinbreaking}

\bsp
The scalar form factors of neutral and charged pions including isospin-breaking effects have been calculated
in the framework of a nonrelativistic effective field theory (NREFT) in Ref.~\cite{Colangelo:2008sm} (see also 
Refs.~\cite{Colangelo:2006va,Gasser:2011ju} for details on the NREFT formalism).  We use a simplified version of 
this result: we only retain the correct thresholds of the two channels ($\pi^0\pi^0$ and $\pi^+\pi^-$), but disregard
isospin-violating corrections in the polynomials that describe the effective-range expansion of the 
scattering partial wave and the low-energy form factor
expansion. In this way, we retain all the nonanalytic effects due to the pion mass difference that scale like
$\sqrt{M_\pi^2-M_{\pi^0}^2}$ near the two-pion thresholds, but neglect regular isospin violation in the form factor 
of order $M_\pi^2-M_{\pi^0}^2$,
which can be calculated in chiral perturbation theory~\cite{Kubis:1999db,DescotesGenon:2012gv}.
In this approximation, the phase of $F_0(s)$ is given by
\begin{equation} \label{eq:argF0}
\arg F_0(s) 
= \arg \bigg[1-i\sigma_0 v_0 - \frac{2}{3}i \frac{(v_0-v_2)(\sigma-\sigma_0)}{1-i\sigma v_2}\bigg]^{-1} \,,
\end{equation}
where $\sigma = \sigma(s)$ as used in the main text, and
$
\sigma_0 = \sqrt{1-{4M_{\pi^0}^2}/{s}}
$. 
The polynomials $v_I = v_I(s)$ for isospin $I=0,2$ are related to the $S$-wave effective-range expansions,
$v_I(s) = a_0^I + \order(s-4M_\pi^2)$, where $a_0^I$ are the $\pi\pi$ $S$-wave scattering lengths.
They can be expressed in terms of the phase shifts of corresponding isospin according to
\begin{equation}
v_I(s) = \frac{1}{\sigma(s) \cot\delta_0^I(s)} \,. \label{eq:vI}
\end{equation}
In order to continue Eqs.~\eqref{eq:argF0} and \eqref{eq:vI} to the region $4M_{\pi^0}^2 \leq s < 4M_\pi^2$, 
we employ an effective-range expansion adapted to the phase shifts used for the $v_I(s)$, 
and the analytic continuation $\sigma \to +i\sqrt{-\sigma^2}$.

\begin{figure}
\centering
\large
  \scalebox{0.66}{\input{plots/isobreak-phase}}\\\vspace{0.2cm}
  \scalebox{0.66}{\input{plots/isobreak-phase-difference}}
  \caption{Top panel: Comparison of $\arg F_0(s)$ (solid line), the phase of the neutral-pion scalar form factor including the effects of different $\pi^0\pi^0$ and $\pi^+\pi^-$ thresholds, to the isospin-symmetric phase $\delta_0^0(s)$ (dashed line).
Bottom panel: $\arg F_0(s)-\delta_0^0(s)$ over a larger range in $s$.} 
\label{fig:iso-phase}
\end{figure}
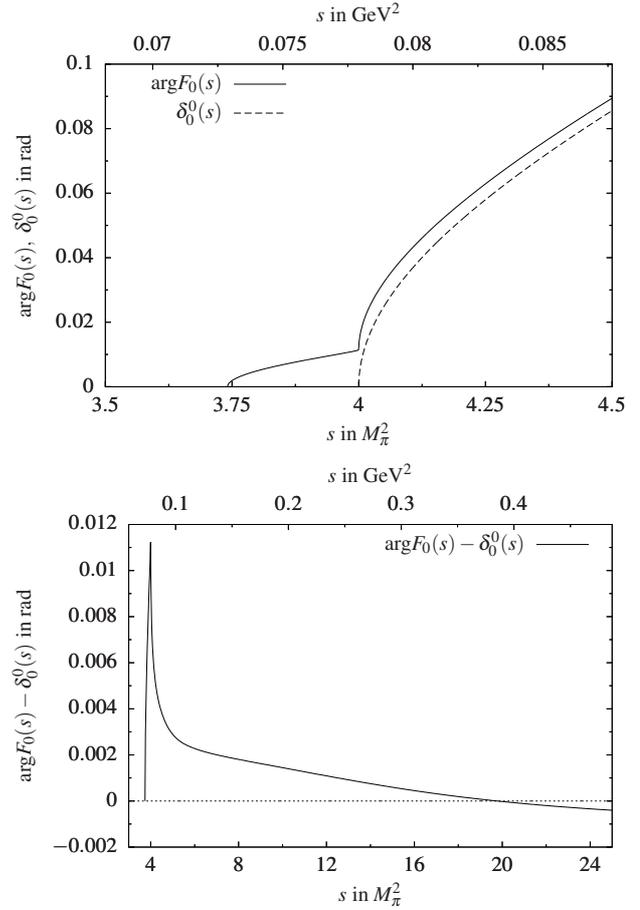

The resulting phase is shown in Fig.~\ref{fig:iso-phase} and compared to $\delta_0^0$, the phase shift in the isospin limit.
We see that $\arg F_0(s)$ starts at $s=4M_{\pi^0}^2$, and has a sharp cusp at $s=4M_\pi^2$, as anticipated.
The difference $\arg F_0(s)-\delta_0^0(s)$ quickly becomes tiny away from the threshold region, 
and in fact crosses $0$ around $s=0.4\GeV^2$.
We neglect the isospin-violating phase difference above this point and use the isospin-symmetric phase shift at higher energies.

We adapt the $\pi\eta$ phase from Ref.~\cite{Albaladejo:2015aca} for $\pi^0\eta$ scattering in a simpler manner.
In this case, there are no different channels coupling/no additional cusps introduced, hence we just need to 
account for the slightly lower threshold.  We achieve this by a linear mapping of the elastic regions
$[(M_{\pi^0}+M_\eta)^2,4M_K^2] \to [(M_\pi+M_\eta)^2,4M_K^2]$, such that the $\pi^0\eta$ scattering phase shift $\tilde\delta_0^1(t)$
is defined as
\begin{align}
\tilde\delta_0^1(t) &= \delta_0^1\big(\tilde t(t)\big) \,, \nn
\tilde t(t) &= 4M_K^2 - \frac{4M_K^2-(M_\pi+M_\eta)^2}{4M_K^2-(M_{\pi^0}+M_\eta)^2} (4M_K^2-t) \,. 
\end{align}
Above the $\bar{K}K$ threshold, we set $\tilde\delta_0^1(t) = \delta_0^1(t)$.  As the $\pi\eta$ phase shift rises only
slowly before the onset of the $a_0(980)$ resonance, the isospin-breaking shift is small compared to the uncertainty 
in the phase shift itself already rather close to threshold. 
\esp

\section{Matching equations}

\label{app:matchingequations}

In the case of \rcht{}, the matching procedure to the dispersive representation is straightforward: an obvious decomposition of the amplitude~\eqref{eq:RChTAmp} is
\begin{align}
	\label{eq:rchtdecomposition}
	\M_0^{0,\text{\rcht}}(s) &= c_{qq} \frac{M_S^2}{M_S^2-s} \begin{aligned}[t]
		& \bigg[ \frac{M_S^2}{F_\pi^2}\bigg(\frac{\xi}{M_S^4} + \frac{\psi}{M_S^2} + c_d^2 \bigg) \nn
		&+ \rho - \frac{c_d^2\Delta}{F_\pi^2M_S^2} - \rho\frac{s}{M_S^2} \bigg] \,, \end{aligned} \nn
	\M_0^{1,\text{\rcht}}(t) &= c_{qq} \frac{M_S^2}{M_S^2-t} \frac{M_S^2}{F_\pi^2}\bigg(\frac{\xi}{M_S^4} + \frac{\psi}{M_S^2} + c_d^2 \bigg) \,,
\end{align}
which is compatible with the required asymptotic behavior. In order also to match the Taylor expansion in~\eqref{eq:MinteqSWaves}, we have to apply a transformation~\eqref{eq:transformation} with $c_2=0$ and
\begin{align}
	c_1 =  2 c_{qq} \frac{M_S^2}{F_\pi^2} A \,, \quad A := \frac{\xi}{M_S^4} + \frac{\psi}{M_S^2} + c_d^2 \,.
\end{align}
Matching the Taylor coefficients to the dispersive representation leads to the subtraction constants
\begin{align}
	\alpha^\text{\rcht} &= c_{qq} \bigg[ \rho + \frac{3M_S^2}{F_\pi^2} \bigg( A- \frac{c_d^2 \Delta}{3M_S^4} \bigg) \bigg] \,, \nn
	\beta^\text{\rcht} &= c_{qq} \begin{aligned}[t]
		&\bigg[ \frac{\metap^2}{F_\pi^2} \bigg( A - \frac{c_d^2\Delta}{M_S^4} \bigg) \nn
		& - \omega_0^0 \bigg( \frac{3\metap^2 M_S^2}{F_\pi^2} \bigg( A - \frac{c_d^2 \Delta}{3M_S^4} \bigg)  + \rho \metap^2 \bigg) \bigg] \,, \end{aligned} \nn
	\gamma^\text{\rcht} &= c_{qq} \frac{\metap^2}{F_\pi^2} A \,.
\end{align}

The \rcht{} amplitude can also be matched to the representation~\eqref{eq:Minteq} with four subtraction constants. In this case, Eq.~\eqref{eq:rchtdecomposition} has to be transformed according to~\eqref{eq:transformation} with
\begin{align}
	c_1 = 2 c_{qq} \frac{M_S^2 + \setap}{F_\pi^2} A \,, \quad c_2 = -c_{qq} \frac{\metap^2 }{F_\pi^2} A \,.
\end{align}
The matching equations for this case are given by
\begin{align}
	\alpha_0^\text{\rcht} &= c_{qq} \bigg[ \rho + \frac{3}{F_\pi^2} \bigg( (M_S^2+\setap) A - \frac{c_d^2 \Delta}{3M_S^2} \bigg) \bigg] \,, \nn
	\beta_0^\text{\rcht} &= -c_{qq} \begin{aligned}[t]
		& \bigg[ \frac{c_d^2\Delta \metap^2}{F_\pi^2 M_S^4}  + \omega_0^0 \frac{3\metap^2}{F_\pi^2} \bigg(  (M_S^2+\setap) A - \frac{c_d^2 \Delta}{3M_S^2} \bigg)  \nn
		& + \metap^2 \omega_0^0 \rho \bigg] \,,  \end{aligned} \nn
	\gamma_0^\text{\rcht} &= c_{qq} \begin{aligned}[t]
		& \bigg[ - \tilde\omega_0^0 \frac{3\metap^4}{2F_\pi^2} \bigg( (M_S^2 + \setap) A - \frac{c_d^2 \Delta}{3 M_S^2} \bigg) \\
		& + \frac{\metap^4}{F_\pi^2 M_S^2} \bigg( A + \frac{c_d^2 \Delta (M_S^2 \omega_0^0 - 1)}{M_S^4} \bigg) - \frac{\metap^4}{2} \tilde\omega_0^0 \rho \bigg] \,, \end{aligned} \nn
	\gamma_1^\text{\rcht} &= c_{qq} \frac{\metap^4}{F_\pi^2 M_S^2} A \,.
\end{align}

\begin{sloppypar}
In the case of NLO large-$N_c$ \chpt{}, an obvious decomposition of the amplitude~\eqref{eq:NLOetapAmp} is
\begin{align}
	\M_{0,\mathrm{NLO}}^{0,\text{\chpt}}(s) &= r_0 + r_2 \frac{s^2}{\metap^4} \,, \quad
	\M_{0,\mathrm{NLO}}^{1,\text{\chpt}}(t) = r_2 \frac{t^2}{\metap^4} \,, \nn
	r_0 &:= c_{qq} \begin{aligned}[t]
		&\Bigl[\frac{\mpc^2}{2} 
			-\frac{2L_5}{F_\pi^2}\big(\metap^2+\meta^2+2\mpc^2\big)\mpc^2 \nn
		&-\frac{2(3L_2+L_3)}{F_\pi^2}\big(\metap^4+\meta^4+2\mpc^4\big)\nn
		& +\frac{24L_8}{F_\pi^2}\mpc^4+\frac{2}{3}\Lambda_2\mpc^2\Bigr]+c_{sq}\frac{\sqrt{2}}{3}\Lambda_2\mpc^2 \,, \end{aligned} \nn
	r_2 &:= c_{qq} \frac{2(3L_2+L_3)}{F_\pi^2} \metap^4 \, . 
\end{align}
In this case, it is not possible to match to the representation~\eqref{eq:MinteqSWaves}, because NLO large-$N_c$ \chpt{} is not compatible with the assumed asymptotic behavior. However, it can be matched directly to the representation~\eqref{eq:Minteq}, as both the asymptotics and the Taylor expansion agree, leading to
\begin{align}
	\alpha_0^\text{\chpt} &= r_0 \,, \quad \beta_0^\text{\chpt} = -r_0 \metap^2 \omega_0^0 \,, \nn
	\gamma_0^\text{\chpt} &= r_2 - \frac{\tilde\omega_0^0}{2} r_0 \metap^4 \,, \quad \gamma_1^\text{\chpt} = r_2 \,.
\end{align}
\end{sloppypar}


\bibliographystyle{utphysmod}
\bibliography{Literature}

\vspace{0cm}

\end{document}

%% file: plots/delta00.tex
\begingroup
  \makeatletter
  \providecommand\color[2][]{%
    \GenericError{(gnuplot) \space\space\space\@spaces}{%
      Package color not loaded in conjunction with
      terminal option `colourtext'%
    }{See the gnuplot documentation for explanation.%
    }{Either use 'blacktext' in gnuplot or load the package
      color.sty in LaTeX.}%
    \renewcommand\color[2][]{}%
  }%
  \providecommand\includegraphics[2][]{%
    \GenericError{(gnuplot) \space\space\space\@spaces}{%
      Package graphicx or graphics not loaded%
    }{See the gnuplot documentation for explanation.%
    }{The gnuplot epslatex terminal needs graphicx.sty or graphics.sty.}%
    \renewcommand\includegraphics[2][]{}%
  }%
  \providecommand\rotatebox[2]{#2}%
  \@ifundefined{ifGPcolor}{%
    \newif\ifGPcolor
    \GPcolorfalse
  }{}%
  \@ifundefined{ifGPblacktext}{%
    \newif\ifGPblacktext
    \GPblacktexttrue
  }{}%
  \let\gplgaddtomacro\g@addto@macro
  \gdef\gplbacktext{}%
  \gdef\gplfronttext{}%
  \makeatother
  \ifGPblacktext
    \def\colorrgb#1{}%
    \def\colorgray#1{}%
  \else
    \ifGPcolor
      \def\colorrgb#1{\color[rgb]{#1}}%
      \def\colorgray#1{\color[gray]{#1}}%
      \expandafter\def\csname LTw\endcsname{\color{white}}%
      \expandafter\def\csname LTb\endcsname{\color{black}}%
      \expandafter\def\csname LTa\endcsname{\color{black}}%
      \expandafter\def\csname LT0\endcsname{\color[rgb]{1,0,0}}%
      \expandafter\def\csname LT1\endcsname{\color[rgb]{0,1,0}}%
      \expandafter\def\csname LT2\endcsname{\color[rgb]{0,0,1}}%
      \expandafter\def\csname LT3\endcsname{\color[rgb]{1,0,1}}%
      \expandafter\def\csname LT4\endcsname{\color[rgb]{0,1,1}}%
      \expandafter\def\csname LT5\endcsname{\color[rgb]{1,1,0}}%
      \expandafter\def\csname LT6\endcsname{\color[rgb]{0,0,0}}%
      \expandafter\def\csname LT7\endcsname{\color[rgb]{1,0.3,0}}%
      \expandafter\def\csname LT8\endcsname{\color[rgb]{0.5,0.5,0.5}}%
    \else
      \def\colorrgb#1{\color{black}}%
      \def\colorgray#1{\color[gray]{#1}}%
      \expandafter\def\csname LTw\endcsname{\color{white}}%
      \expandafter\def\csname LTb\endcsname{\color{black}}%
      \expandafter\def\csname LTa\endcsname{\color{black}}%
      \expandafter\def\csname LT0\endcsname{\color{black}}%
      \expandafter\def\csname LT1\endcsname{\color{black}}%
      \expandafter\def\csname LT2\endcsname{\color{black}}%
      \expandafter\def\csname LT3\endcsname{\color{black}}%
      \expandafter\def\csname LT4\endcsname{\color{black}}%
      \expandafter\def\csname LT5\endcsname{\color{black}}%
      \expandafter\def\csname LT6\endcsname{\color{black}}%
      \expandafter\def\csname LT7\endcsname{\color{black}}%
      \expandafter\def\csname LT8\endcsname{\color{black}}%
    \fi
  \fi
    \setlength{\unitlength}{0.0500bp}%
    \ifx\gptboxheight\undefined%
      \newlength{\gptboxheight}%
      \newlength{\gptboxwidth}%
      \newsavebox{\gptboxtext}%
    \fi%
    \setlength{\fboxrule}{0.5pt}%
    \setlength{\fboxsep}{1pt}%
\begin{picture}(5040.00,4032.00)%
    \gplgaddtomacro\gplbacktext{%
      \csname LTb\endcsname%
      \put(418,704){\makebox(0,0)[r]{\strut{}$0$}}%
      \put(418,1237){\makebox(0,0)[r]{\strut{}$1$}}%
      \put(418,1771){\makebox(0,0)[r]{\strut{}$2$}}%
      \put(418,2304){\makebox(0,0)[r]{\strut{}$3$}}%
      \put(418,2838){\makebox(0,0)[r]{\strut{}$4$}}%
      \put(418,3371){\makebox(0,0)[r]{\strut{}$5$}}%
      \put(550,484){\makebox(0,0){\strut{}$0$}}%
      \put(2255,484){\makebox(0,0){\strut{}$50$}}%
      \put(3961,484){\makebox(0,0){\strut{}$100$}}%
      \put(550,3591){\makebox(0,0){\strut{}$0$}}%
      \put(2301,3591){\makebox(0,0){\strut{}$1$}}%
      \put(4052,3591){\makebox(0,0){\strut{}$2$}}%
    }%
    \gplgaddtomacro\gplfronttext{%
      \csname LTb\endcsname%
      \put(176,2037){\rotatebox{-270}{\makebox(0,0){\strut{}$\delta_0^0(s)$}}}%
      \put(2596,154){\makebox(0,0){\strut{}$s$ in $M_\pi^2$}}%
      \put(2596,3920){\makebox(0,0){\strut{}$s$ in GeV${}^2$}}%
      \csname LTb\endcsname%
      \put(3788,1261){\makebox(0,0)[r]{\strut{}high-energy uncertainty}}%
      \csname LTb\endcsname%
      \put(3788,1041){\makebox(0,0)[r]{\strut{}low-energy uncertainty}}%
      \csname LTb\endcsname%
      \put(3788,821){\makebox(0,0)[r]{\strut{}central value}}%
      \csname LTb\endcsname%
      \put(418,704){\makebox(0,0)[r]{\strut{}$0$}}%
      \put(418,1237){\makebox(0,0)[r]{\strut{}$1$}}%
      \put(418,1771){\makebox(0,0)[r]{\strut{}$2$}}%
      \put(418,2304){\makebox(0,0)[r]{\strut{}$3$}}%
      \put(418,2838){\makebox(0,0)[r]{\strut{}$4$}}%
      \put(418,3371){\makebox(0,0)[r]{\strut{}$5$}}%
      \put(550,484){\makebox(0,0){\strut{}$0$}}%
      \put(2255,484){\makebox(0,0){\strut{}$50$}}%
      \put(3961,484){\makebox(0,0){\strut{}$100$}}%
      \put(550,3591){\makebox(0,0){\strut{}$0$}}%
      \put(2301,3591){\makebox(0,0){\strut{}$1$}}%
      \put(4052,3591){\makebox(0,0){\strut{}$2$}}%
    }%
    \gplbacktext
    \put(0,0){\includegraphics{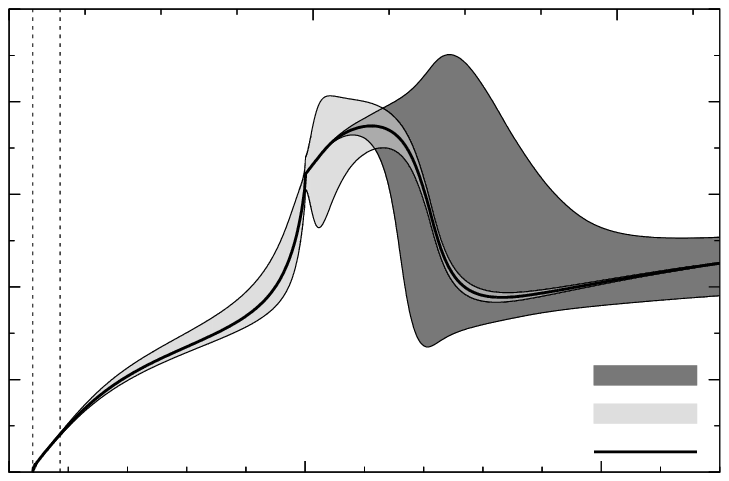}}%
    \gplfronttext
  \end{picture}%
\endgroup

%% file: plots/delta00physreg.tex
\begingroup
  \makeatletter
  \providecommand\color[2][]{%
    \GenericError{(gnuplot) \space\space\space\@spaces}{%
      Package color not loaded in conjunction with
      terminal option `colourtext'%
    }{See the gnuplot documentation for explanation.%
    }{Either use 'blacktext' in gnuplot or load the package
      color.sty in LaTeX.}%
    \renewcommand\color[2][]{}%
  }%
  \providecommand\includegraphics[2][]{%
    \GenericError{(gnuplot) \space\space\space\@spaces}{%
      Package graphicx or graphics not loaded%
    }{See the gnuplot documentation for explanation.%
    }{The gnuplot epslatex terminal needs graphicx.sty or graphics.sty.}%
    \renewcommand\includegraphics[2][]{}%
  }%
  \providecommand\rotatebox[2]{#2}%
  \@ifundefined{ifGPcolor}{%
    \newif\ifGPcolor
    \GPcolorfalse
  }{}%
  \@ifundefined{ifGPblacktext}{%
    \newif\ifGPblacktext
    \GPblacktexttrue
  }{}%
  \let\gplgaddtomacro\g@addto@macro
  \gdef\gplbacktext{}%
  \gdef\gplfronttext{}%
  \makeatother
  \ifGPblacktext
    \def\colorrgb#1{}%
    \def\colorgray#1{}%
  \else
    \ifGPcolor
      \def\colorrgb#1{\color[rgb]{#1}}%
      \def\colorgray#1{\color[gray]{#1}}%
      \expandafter\def\csname LTw\endcsname{\color{white}}%
      \expandafter\def\csname LTb\endcsname{\color{black}}%
      \expandafter\def\csname LTa\endcsname{\color{black}}%
      \expandafter\def\csname LT0\endcsname{\color[rgb]{1,0,0}}%
      \expandafter\def\csname LT1\endcsname{\color[rgb]{0,1,0}}%
      \expandafter\def\csname LT2\endcsname{\color[rgb]{0,0,1}}%
      \expandafter\def\csname LT3\endcsname{\color[rgb]{1,0,1}}%
      \expandafter\def\csname LT4\endcsname{\color[rgb]{0,1,1}}%
      \expandafter\def\csname LT5\endcsname{\color[rgb]{1,1,0}}%
      \expandafter\def\csname LT6\endcsname{\color[rgb]{0,0,0}}%
      \expandafter\def\csname LT7\endcsname{\color[rgb]{1,0.3,0}}%
      \expandafter\def\csname LT8\endcsname{\color[rgb]{0.5,0.5,0.5}}%
    \else
      \def\colorrgb#1{\color{black}}%
      \def\colorgray#1{\color[gray]{#1}}%
      \expandafter\def\csname LTw\endcsname{\color{white}}%
      \expandafter\def\csname LTb\endcsname{\color{black}}%
      \expandafter\def\csname LTa\endcsname{\color{black}}%
      \expandafter\def\csname LT0\endcsname{\color{black}}%
      \expandafter\def\csname LT1\endcsname{\color{black}}%
      \expandafter\def\csname LT2\endcsname{\color{black}}%
      \expandafter\def\csname LT3\endcsname{\color{black}}%
      \expandafter\def\csname LT4\endcsname{\color{black}}%
      \expandafter\def\csname LT5\endcsname{\color{black}}%
      \expandafter\def\csname LT6\endcsname{\color{black}}%
      \expandafter\def\csname LT7\endcsname{\color{black}}%
      \expandafter\def\csname LT8\endcsname{\color{black}}%
    \fi
  \fi
    \setlength{\unitlength}{0.0500bp}%
    \ifx\gptboxheight\undefined%
      \newlength{\gptboxheight}%
      \newlength{\gptboxwidth}%
      \newsavebox{\gptboxtext}%
    \fi%
    \setlength{\fboxrule}{0.5pt}%
    \setlength{\fboxsep}{1pt}%
\begin{picture}(5040.00,4032.00)%
    \gplgaddtomacro\gplbacktext{%
      \csname LTb\endcsname%
      \put(594,704){\makebox(0,0)[r]{\strut{}$0$}}%
      \put(594,3371){\makebox(0,0)[r]{\strut{}$0.5$}}%
      \put(726,484){\makebox(0,0){\strut{}$4$}}%
      \put(1573,484){\makebox(0,0){\strut{}$5$}}%
      \put(2419,484){\makebox(0,0){\strut{}$6$}}%
      \put(3266,484){\makebox(0,0){\strut{}$7$}}%
      \put(4113,484){\makebox(0,0){\strut{}$8$}}%
      \put(1686,3591){\makebox(0,0){\strut{}$0.1$}}%
      \put(3859,3591){\makebox(0,0){\strut{}$0.15$}}%
    }%
    \gplgaddtomacro\gplfronttext{%
      \csname LTb\endcsname%
      \put(352,2037){\rotatebox{-270}{\makebox(0,0){\strut{}$\delta_0^0(s)$}}}%
      \put(2684,154){\makebox(0,0){\strut{}$s$ in $M_\pi^2$}}%
      \put(2684,3920){\makebox(0,0){\strut{}$s$ in GeV${}^2$}}%
      \csname LTb\endcsname%
      \put(3788,1127){\makebox(0,0)[r]{\strut{}low-energy uncertainty}}%
      \csname LTb\endcsname%
      \put(3788,907){\makebox(0,0)[r]{\strut{}central value}}%
      \csname LTb\endcsname%
      \put(594,704){\makebox(0,0)[r]{\strut{}$0$}}%
      \put(594,3371){\makebox(0,0)[r]{\strut{}$0.5$}}%
      \put(726,484){\makebox(0,0){\strut{}$4$}}%
      \put(1573,484){\makebox(0,0){\strut{}$5$}}%
      \put(2419,484){\makebox(0,0){\strut{}$6$}}%
      \put(3266,484){\makebox(0,0){\strut{}$7$}}%
      \put(4113,484){\makebox(0,0){\strut{}$8$}}%
      \put(1686,3591){\makebox(0,0){\strut{}$0.1$}}%
      \put(3859,3591){\makebox(0,0){\strut{}$0.15$}}%
    }%
    \gplbacktext
    \put(0,0){\includegraphics{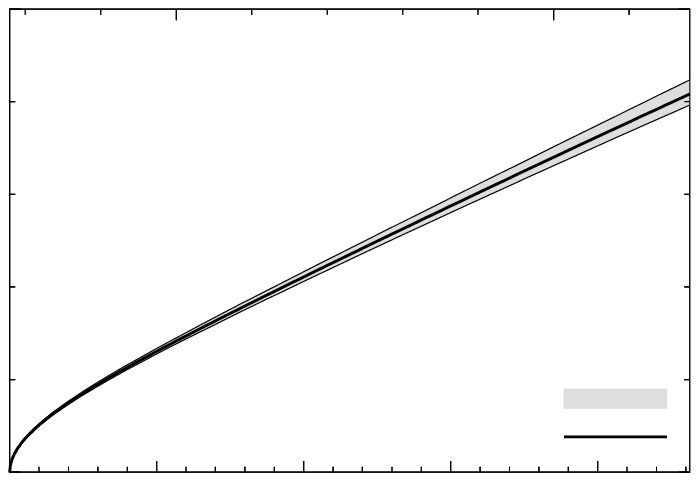}}%
    \gplfronttext
  \end{picture}%
\endgroup

%% file: plots/omega00.tex
\begingroup
  \makeatletter
  \providecommand\color[2][]{%
    \GenericError{(gnuplot) \space\space\space\@spaces}{%
      Package color not loaded in conjunction with
      terminal option `colourtext'%
    }{See the gnuplot documentation for explanation.%
    }{Either use 'blacktext' in gnuplot or load the package
      color.sty in LaTeX.}%
    \renewcommand\color[2][]{}%
  }%
  \providecommand\includegraphics[2][]{%
    \GenericError{(gnuplot) \space\space\space\@spaces}{%
      Package graphicx or graphics not loaded%
    }{See the gnuplot documentation for explanation.%
    }{The gnuplot epslatex terminal needs graphicx.sty or graphics.sty.}%
    \renewcommand\includegraphics[2][]{}%
  }%
  \providecommand\rotatebox[2]{#2}%
  \@ifundefined{ifGPcolor}{%
    \newif\ifGPcolor
    \GPcolorfalse
  }{}%
  \@ifundefined{ifGPblacktext}{%
    \newif\ifGPblacktext
    \GPblacktexttrue
  }{}%
  \let\gplgaddtomacro\g@addto@macro
  \gdef\gplbacktext{}%
  \gdef\gplfronttext{}%
  \makeatother
  \ifGPblacktext
    \def\colorrgb#1{}%
    \def\colorgray#1{}%
  \else
    \ifGPcolor
      \def\colorrgb#1{\color[rgb]{#1}}%
      \def\colorgray#1{\color[gray]{#1}}%
      \expandafter\def\csname LTw\endcsname{\color{white}}%
      \expandafter\def\csname LTb\endcsname{\color{black}}%
      \expandafter\def\csname LTa\endcsname{\color{black}}%
      \expandafter\def\csname LT0\endcsname{\color[rgb]{1,0,0}}%
      \expandafter\def\csname LT1\endcsname{\color[rgb]{0,1,0}}%
      \expandafter\def\csname LT2\endcsname{\color[rgb]{0,0,1}}%
      \expandafter\def\csname LT3\endcsname{\color[rgb]{1,0,1}}%
      \expandafter\def\csname LT4\endcsname{\color[rgb]{0,1,1}}%
      \expandafter\def\csname LT5\endcsname{\color[rgb]{1,1,0}}%
      \expandafter\def\csname LT6\endcsname{\color[rgb]{0,0,0}}%
      \expandafter\def\csname LT7\endcsname{\color[rgb]{1,0.3,0}}%
      \expandafter\def\csname LT8\endcsname{\color[rgb]{0.5,0.5,0.5}}%
    \else
      \def\colorrgb#1{\color{black}}%
      \def\colorgray#1{\color[gray]{#1}}%
      \expandafter\def\csname LTw\endcsname{\color{white}}%
      \expandafter\def\csname LTb\endcsname{\color{black}}%
      \expandafter\def\csname LTa\endcsname{\color{black}}%
      \expandafter\def\csname LT0\endcsname{\color{black}}%
      \expandafter\def\csname LT1\endcsname{\color{black}}%
      \expandafter\def\csname LT2\endcsname{\color{black}}%
      \expandafter\def\csname LT3\endcsname{\color{black}}%
      \expandafter\def\csname LT4\endcsname{\color{black}}%
      \expandafter\def\csname LT5\endcsname{\color{black}}%
      \expandafter\def\csname LT6\endcsname{\color{black}}%
      \expandafter\def\csname LT7\endcsname{\color{black}}%
      \expandafter\def\csname LT8\endcsname{\color{black}}%
    \fi
  \fi
    \setlength{\unitlength}{0.0500bp}%
    \ifx\gptboxheight\undefined%
      \newlength{\gptboxheight}%
      \newlength{\gptboxwidth}%
      \newsavebox{\gptboxtext}%
    \fi%
    \setlength{\fboxrule}{0.5pt}%
    \setlength{\fboxsep}{1pt}%
\begin{picture}(5040.00,4032.00)%
    \gplgaddtomacro\gplbacktext{%
      \csname LTb\endcsname%
      \put(418,704){\makebox(0,0)[r]{\strut{}$0$}}%
      \put(418,1149){\makebox(0,0)[r]{\strut{}$1$}}%
      \put(418,1593){\makebox(0,0)[r]{\strut{}$2$}}%
      \put(418,2038){\makebox(0,0)[r]{\strut{}$3$}}%
      \put(418,2482){\makebox(0,0)[r]{\strut{}$4$}}%
      \put(418,2927){\makebox(0,0)[r]{\strut{}$5$}}%
      \put(418,3371){\makebox(0,0)[r]{\strut{}$6$}}%
      \put(550,484){\makebox(0,0){\strut{}$0$}}%
      \put(2255,484){\makebox(0,0){\strut{}$50$}}%
      \put(3961,484){\makebox(0,0){\strut{}$100$}}%
      \put(550,3591){\makebox(0,0){\strut{}$0$}}%
      \put(2301,3591){\makebox(0,0){\strut{}$1$}}%
      \put(4052,3591){\makebox(0,0){\strut{}$2$}}%
    }%
    \gplgaddtomacro\gplfronttext{%
      \csname LTb\endcsname%
      \put(176,2037){\rotatebox{-270}{\makebox(0,0){\strut{}$|\Omega_0^0(s)|$}}}%
      \put(2596,154){\makebox(0,0){\strut{}$s$ in $M_\pi^2$}}%
      \put(2596,3920){\makebox(0,0){\strut{}$s$ in GeV${}^2$}}%
      \csname LTb\endcsname%
      \put(3788,3128){\makebox(0,0)[r]{\strut{}uncertainty}}%
      \csname LTb\endcsname%
      \put(3788,2908){\makebox(0,0)[r]{\strut{}central value}}%
      \csname LTb\endcsname%
      \put(418,704){\makebox(0,0)[r]{\strut{}$0$}}%
      \put(418,1149){\makebox(0,0)[r]{\strut{}$1$}}%
      \put(418,1593){\makebox(0,0)[r]{\strut{}$2$}}%
      \put(418,2038){\makebox(0,0)[r]{\strut{}$3$}}%
      \put(418,2482){\makebox(0,0)[r]{\strut{}$4$}}%
      \put(418,2927){\makebox(0,0)[r]{\strut{}$5$}}%
      \put(418,3371){\makebox(0,0)[r]{\strut{}$6$}}%
      \put(550,484){\makebox(0,0){\strut{}$0$}}%
      \put(2255,484){\makebox(0,0){\strut{}$50$}}%
      \put(3961,484){\makebox(0,0){\strut{}$100$}}%
      \put(550,3591){\makebox(0,0){\strut{}$0$}}%
      \put(2301,3591){\makebox(0,0){\strut{}$1$}}%
      \put(4052,3591){\makebox(0,0){\strut{}$2$}}%
    }%
    \gplbacktext
    \put(0,0){\includegraphics{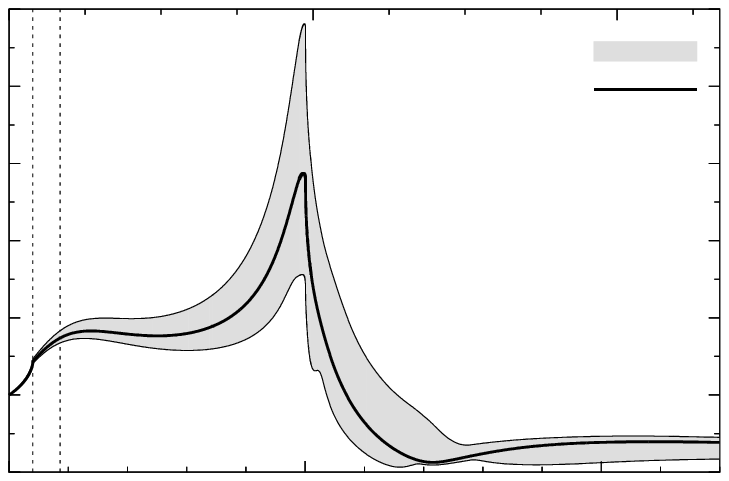}}%
    \gplfronttext
  \end{picture}%
\endgroup

%% file: plots/omega00physreg.tex
\begingroup
  \makeatletter
  \providecommand\color[2][]{%
    \GenericError{(gnuplot) \space\space\space\@spaces}{%
      Package color not loaded in conjunction with
      terminal option `colourtext'%
    }{See the gnuplot documentation for explanation.%
    }{Either use 'blacktext' in gnuplot or load the package
      color.sty in LaTeX.}%
    \renewcommand\color[2][]{}%
  }%
  \providecommand\includegraphics[2][]{%
    \GenericError{(gnuplot) \space\space\space\@spaces}{%
      Package graphicx or graphics not loaded%
    }{See the gnuplot documentation for explanation.%
    }{The gnuplot epslatex terminal needs graphicx.sty or graphics.sty.}%
    \renewcommand\includegraphics[2][]{}%
  }%
  \providecommand\rotatebox[2]{#2}%
  \@ifundefined{ifGPcolor}{%
    \newif\ifGPcolor
    \GPcolorfalse
  }{}%
  \@ifundefined{ifGPblacktext}{%
    \newif\ifGPblacktext
    \GPblacktexttrue
  }{}%
  \let\gplgaddtomacro\g@addto@macro
  \gdef\gplbacktext{}%
  \gdef\gplfronttext{}%
  \makeatother
  \ifGPblacktext
    \def\colorrgb#1{}%
    \def\colorgray#1{}%
  \else
    \ifGPcolor
      \def\colorrgb#1{\color[rgb]{#1}}%
      \def\colorgray#1{\color[gray]{#1}}%
      \expandafter\def\csname LTw\endcsname{\color{white}}%
      \expandafter\def\csname LTb\endcsname{\color{black}}%
      \expandafter\def\csname LTa\endcsname{\color{black}}%
      \expandafter\def\csname LT0\endcsname{\color[rgb]{1,0,0}}%
      \expandafter\def\csname LT1\endcsname{\color[rgb]{0,1,0}}%
      \expandafter\def\csname LT2\endcsname{\color[rgb]{0,0,1}}%
      \expandafter\def\csname LT3\endcsname{\color[rgb]{1,0,1}}%
      \expandafter\def\csname LT4\endcsname{\color[rgb]{0,1,1}}%
      \expandafter\def\csname LT5\endcsname{\color[rgb]{1,1,0}}%
      \expandafter\def\csname LT6\endcsname{\color[rgb]{0,0,0}}%
      \expandafter\def\csname LT7\endcsname{\color[rgb]{1,0.3,0}}%
      \expandafter\def\csname LT8\endcsname{\color[rgb]{0.5,0.5,0.5}}%
    \else
      \def\colorrgb#1{\color{black}}%
      \def\colorgray#1{\color[gray]{#1}}%
      \expandafter\def\csname LTw\endcsname{\color{white}}%
      \expandafter\def\csname LTb\endcsname{\color{black}}%
      \expandafter\def\csname LTa\endcsname{\color{black}}%
      \expandafter\def\csname LT0\endcsname{\color{black}}%
      \expandafter\def\csname LT1\endcsname{\color{black}}%
      \expandafter\def\csname LT2\endcsname{\color{black}}%
      \expandafter\def\csname LT3\endcsname{\color{black}}%
      \expandafter\def\csname LT4\endcsname{\color{black}}%
      \expandafter\def\csname LT5\endcsname{\color{black}}%
      \expandafter\def\csname LT6\endcsname{\color{black}}%
      \expandafter\def\csname LT7\endcsname{\color{black}}%
      \expandafter\def\csname LT8\endcsname{\color{black}}%
    \fi
  \fi
    \setlength{\unitlength}{0.0500bp}%
    \ifx\gptboxheight\undefined%
      \newlength{\gptboxheight}%
      \newlength{\gptboxwidth}%
      \newsavebox{\gptboxtext}%
    \fi%
    \setlength{\fboxrule}{0.5pt}%
    \setlength{\fboxsep}{1pt}%
\begin{picture}(5040.00,4032.00)%
    \gplgaddtomacro\gplbacktext{%
      \csname LTb\endcsname%
      \put(594,704){\makebox(0,0)[r]{\strut{}$1.2$}}%
      \put(594,2990){\makebox(0,0)[r]{\strut{}$1.8$}}%
      \put(726,484){\makebox(0,0){\strut{}$4$}}%
      \put(1573,484){\makebox(0,0){\strut{}$5$}}%
      \put(2419,484){\makebox(0,0){\strut{}$6$}}%
      \put(3266,484){\makebox(0,0){\strut{}$7$}}%
      \put(4113,484){\makebox(0,0){\strut{}$8$}}%
      \put(1686,3591){\makebox(0,0){\strut{}$0.1$}}%
      \put(3859,3591){\makebox(0,0){\strut{}$0.15$}}%
    }%
    \gplgaddtomacro\gplfronttext{%
      \csname LTb\endcsname%
      \put(352,2037){\rotatebox{-270}{\makebox(0,0){\strut{}$|\Omega_0^0(s)|$}}}%
      \put(2684,154){\makebox(0,0){\strut{}$s$ in $M_\pi^2$}}%
      \put(2684,3920){\makebox(0,0){\strut{}$s$ in GeV${}^2$}}%
      \csname LTb\endcsname%
      \put(3788,1127){\makebox(0,0)[r]{\strut{}uncertainty}}%
      \csname LTb\endcsname%
      \put(3788,907){\makebox(0,0)[r]{\strut{}central value}}%
      \csname LTb\endcsname%
      \put(594,704){\makebox(0,0)[r]{\strut{}$1.2$}}%
      \put(594,2990){\makebox(0,0)[r]{\strut{}$1.8$}}%
      \put(726,484){\makebox(0,0){\strut{}$4$}}%
      \put(1573,484){\makebox(0,0){\strut{}$5$}}%
      \put(2419,484){\makebox(0,0){\strut{}$6$}}%
      \put(3266,484){\makebox(0,0){\strut{}$7$}}%
      \put(4113,484){\makebox(0,0){\strut{}$8$}}%
      \put(1686,3591){\makebox(0,0){\strut{}$0.1$}}%
      \put(3859,3591){\makebox(0,0){\strut{}$0.15$}}%
    }%
    \gplbacktext
    \put(0,0){\includegraphics{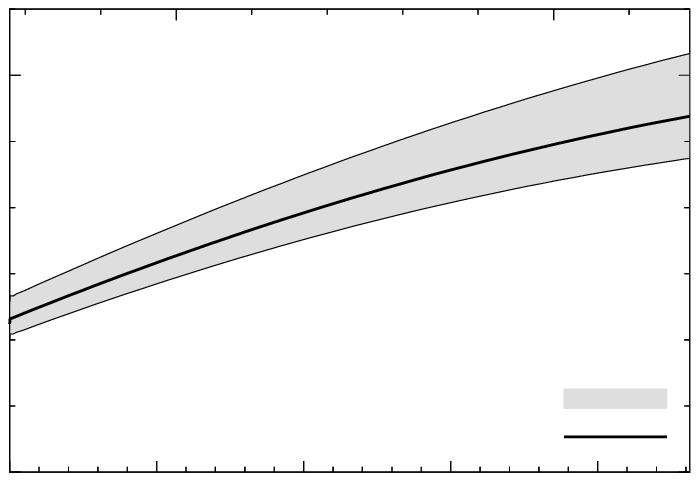}}%
    \gplfronttext
  \end{picture}%
\endgroup

%% file: plots/delta01.tex
\begingroup
  \makeatletter
  \providecommand\color[2][]{%
    \GenericError{(gnuplot) \space\space\space\@spaces}{%
      Package color not loaded in conjunction with
      terminal option `colourtext'%
    }{See the gnuplot documentation for explanation.%
    }{Either use 'blacktext' in gnuplot or load the package
      color.sty in LaTeX.}%
    \renewcommand\color[2][]{}%
  }%
  \providecommand\includegraphics[2][]{%
    \GenericError{(gnuplot) \space\space\space\@spaces}{%
      Package graphicx or graphics not loaded%
    }{See the gnuplot documentation for explanation.%
    }{The gnuplot epslatex terminal needs graphicx.sty or graphics.sty.}%
    \renewcommand\includegraphics[2][]{}%
  }%
  \providecommand\rotatebox[2]{#2}%
  \@ifundefined{ifGPcolor}{%
    \newif\ifGPcolor
    \GPcolorfalse
  }{}%
  \@ifundefined{ifGPblacktext}{%
    \newif\ifGPblacktext
    \GPblacktexttrue
  }{}%
  \let\gplgaddtomacro\g@addto@macro
  \gdef\gplbacktext{}%
  \gdef\gplfronttext{}%
  \makeatother
  \ifGPblacktext
    \def\colorrgb#1{}%
    \def\colorgray#1{}%
  \else
    \ifGPcolor
      \def\colorrgb#1{\color[rgb]{#1}}%
      \def\colorgray#1{\color[gray]{#1}}%
      \expandafter\def\csname LTw\endcsname{\color{white}}%
      \expandafter\def\csname LTb\endcsname{\color{black}}%
      \expandafter\def\csname LTa\endcsname{\color{black}}%
      \expandafter\def\csname LT0\endcsname{\color[rgb]{1,0,0}}%
      \expandafter\def\csname LT1\endcsname{\color[rgb]{0,1,0}}%
      \expandafter\def\csname LT2\endcsname{\color[rgb]{0,0,1}}%
      \expandafter\def\csname LT3\endcsname{\color[rgb]{1,0,1}}%
      \expandafter\def\csname LT4\endcsname{\color[rgb]{0,1,1}}%
      \expandafter\def\csname LT5\endcsname{\color[rgb]{1,1,0}}%
      \expandafter\def\csname LT6\endcsname{\color[rgb]{0,0,0}}%
      \expandafter\def\csname LT7\endcsname{\color[rgb]{1,0.3,0}}%
      \expandafter\def\csname LT8\endcsname{\color[rgb]{0.5,0.5,0.5}}%
    \else
      \def\colorrgb#1{\color{black}}%
      \def\colorgray#1{\color[gray]{#1}}%
      \expandafter\def\csname LTw\endcsname{\color{white}}%
      \expandafter\def\csname LTb\endcsname{\color{black}}%
      \expandafter\def\csname LTa\endcsname{\color{black}}%
      \expandafter\def\csname LT0\endcsname{\color{black}}%
      \expandafter\def\csname LT1\endcsname{\color{black}}%
      \expandafter\def\csname LT2\endcsname{\color{black}}%
      \expandafter\def\csname LT3\endcsname{\color{black}}%
      \expandafter\def\csname LT4\endcsname{\color{black}}%
      \expandafter\def\csname LT5\endcsname{\color{black}}%
      \expandafter\def\csname LT6\endcsname{\color{black}}%
      \expandafter\def\csname LT7\endcsname{\color{black}}%
      \expandafter\def\csname LT8\endcsname{\color{black}}%
    \fi
  \fi
    \setlength{\unitlength}{0.0500bp}%
    \ifx\gptboxheight\undefined%
      \newlength{\gptboxheight}%
      \newlength{\gptboxwidth}%
      \newsavebox{\gptboxtext}%
    \fi%
    \setlength{\fboxrule}{0.5pt}%
    \setlength{\fboxsep}{1pt}%
\begin{picture}(5040.00,4032.00)%
    \gplgaddtomacro\gplbacktext{%
      \csname LTb\endcsname%
      \put(550,704){\makebox(0,0)[r]{\strut{}$0$}}%
      \put(550,1593){\makebox(0,0)[r]{\strut{}$1$}}%
      \put(550,2482){\makebox(0,0)[r]{\strut{}$2$}}%
      \put(550,3371){\makebox(0,0)[r]{\strut{}$3$}}%
      \put(1596,484){\makebox(0,0){\strut{}$50$}}%
      \put(3120,484){\makebox(0,0){\strut{}$100$}}%
      \put(4643,484){\makebox(0,0){\strut{}$150$}}%
      \put(1637,3591){\makebox(0,0){\strut{}$1$}}%
      \put(3201,3591){\makebox(0,0){\strut{}$2$}}%
    }%
    \gplgaddtomacro\gplfronttext{%
      \csname LTb\endcsname%
      \put(176,2037){\rotatebox{-270}{\makebox(0,0){\strut{}$\delta^{1}_{0}(t)$}}}%
      \put(2662,154){\makebox(0,0){\strut{}$t$ in $M_{\pi}^{2}$}}%
      \put(2662,3920){\makebox(0,0){\strut{}$t$ in GeV$^2$}}%
      \csname LTb\endcsname%
      \put(3656,3198){\makebox(0,0)[r]{\strut{}restricted high energy uncertainty}}%
      \csname LTb\endcsname%
      \put(3656,2978){\makebox(0,0)[r]{\strut{}full uncertainty}}%
      \csname LTb\endcsname%
      \put(3656,2758){\makebox(0,0)[r]{\strut{}central value}}%
      \csname LTb\endcsname%
      \put(550,704){\makebox(0,0)[r]{\strut{}$0$}}%
      \put(550,1593){\makebox(0,0)[r]{\strut{}$1$}}%
      \put(550,2482){\makebox(0,0)[r]{\strut{}$2$}}%
      \put(550,3371){\makebox(0,0)[r]{\strut{}$3$}}%
      \put(1596,484){\makebox(0,0){\strut{}$50$}}%
      \put(3120,484){\makebox(0,0){\strut{}$100$}}%
      \put(4643,484){\makebox(0,0){\strut{}$150$}}%
      \put(1637,3591){\makebox(0,0){\strut{}$1$}}%
      \put(3201,3591){\makebox(0,0){\strut{}$2$}}%
    }%
    \gplbacktext
    \put(0,0){\includegraphics{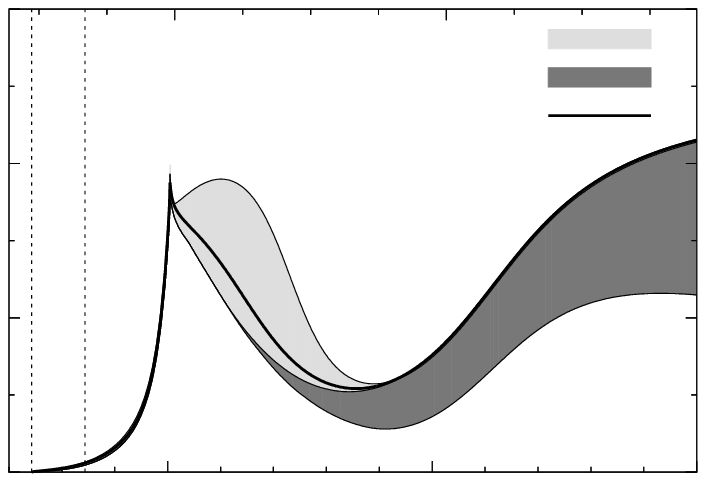}}%
    \gplfronttext
  \end{picture}%
\endgroup

%% file: plots/delta01physreg.tex
\begingroup
  \makeatletter
  \providecommand\color[2][]{%
    \GenericError{(gnuplot) \space\space\space\@spaces}{%
      Package color not loaded in conjunction with
      terminal option `colourtext'%
    }{See the gnuplot documentation for explanation.%
    }{Either use 'blacktext' in gnuplot or load the package
      color.sty in LaTeX.}%
    \renewcommand\color[2][]{}%
  }%
  \providecommand\includegraphics[2][]{%
    \GenericError{(gnuplot) \space\space\space\@spaces}{%
      Package graphicx or graphics not loaded%
    }{See the gnuplot documentation for explanation.%
    }{The gnuplot epslatex terminal needs graphicx.sty or graphics.sty.}%
    \renewcommand\includegraphics[2][]{}%
  }%
  \providecommand\rotatebox[2]{#2}%
  \@ifundefined{ifGPcolor}{%
    \newif\ifGPcolor
    \GPcolorfalse
  }{}%
  \@ifundefined{ifGPblacktext}{%
    \newif\ifGPblacktext
    \GPblacktexttrue
  }{}%
  \let\gplgaddtomacro\g@addto@macro
  \gdef\gplbacktext{}%
  \gdef\gplfronttext{}%
  \makeatother
  \ifGPblacktext
    \def\colorrgb#1{}%
    \def\colorgray#1{}%
  \else
    \ifGPcolor
      \def\colorrgb#1{\color[rgb]{#1}}%
      \def\colorgray#1{\color[gray]{#1}}%
      \expandafter\def\csname LTw\endcsname{\color{white}}%
      \expandafter\def\csname LTb\endcsname{\color{black}}%
      \expandafter\def\csname LTa\endcsname{\color{black}}%
      \expandafter\def\csname LT0\endcsname{\color[rgb]{1,0,0}}%
      \expandafter\def\csname LT1\endcsname{\color[rgb]{0,1,0}}%
      \expandafter\def\csname LT2\endcsname{\color[rgb]{0,0,1}}%
      \expandafter\def\csname LT3\endcsname{\color[rgb]{1,0,1}}%
      \expandafter\def\csname LT4\endcsname{\color[rgb]{0,1,1}}%
      \expandafter\def\csname LT5\endcsname{\color[rgb]{1,1,0}}%
      \expandafter\def\csname LT6\endcsname{\color[rgb]{0,0,0}}%
      \expandafter\def\csname LT7\endcsname{\color[rgb]{1,0.3,0}}%
      \expandafter\def\csname LT8\endcsname{\color[rgb]{0.5,0.5,0.5}}%
    \else
      \def\colorrgb#1{\color{black}}%
      \def\colorgray#1{\color[gray]{#1}}%
      \expandafter\def\csname LTw\endcsname{\color{white}}%
      \expandafter\def\csname LTb\endcsname{\color{black}}%
      \expandafter\def\csname LTa\endcsname{\color{black}}%
      \expandafter\def\csname LT0\endcsname{\color{black}}%
      \expandafter\def\csname LT1\endcsname{\color{black}}%
      \expandafter\def\csname LT2\endcsname{\color{black}}%
      \expandafter\def\csname LT3\endcsname{\color{black}}%
      \expandafter\def\csname LT4\endcsname{\color{black}}%
      \expandafter\def\csname LT5\endcsname{\color{black}}%
      \expandafter\def\csname LT6\endcsname{\color{black}}%
      \expandafter\def\csname LT7\endcsname{\color{black}}%
      \expandafter\def\csname LT8\endcsname{\color{black}}%
    \fi
  \fi
    \setlength{\unitlength}{0.0500bp}%
    \ifx\gptboxheight\undefined%
      \newlength{\gptboxheight}%
      \newlength{\gptboxwidth}%
      \newsavebox{\gptboxtext}%
    \fi%
    \setlength{\fboxrule}{0.5pt}%
    \setlength{\fboxsep}{1pt}%
\begin{picture}(5040.00,4032.00)%
    \gplgaddtomacro\gplbacktext{%
      \csname LTb\endcsname%
      \put(594,704){\makebox(0,0)[r]{\strut{}$0$}}%
      \put(594,3371){\makebox(0,0)[r]{\strut{}$0.1$}}%
      \put(1013,484){\makebox(0,0){\strut{}$25$}}%
      \put(2951,484){\makebox(0,0){\strut{}$30$}}%
      \put(1272,3591){\makebox(0,0){\strut{}$0.5$}}%
      \put(3261,3591){\makebox(0,0){\strut{}$0.6$}}%
    }%
    \gplgaddtomacro\gplfronttext{%
      \csname LTb\endcsname%
      \put(220,2037){\rotatebox{-270}{\makebox(0,0){\strut{}$\delta^{1}_{0}(t)$}}}%
      \put(2684,154){\makebox(0,0){\strut{}$t$ in $M_{\pi}^{2}$}}%
      \put(2684,3920){\makebox(0,0){\strut{}$t$ in GeV$^2$}}%
      \csname LTb\endcsname%
      \put(3656,3198){\makebox(0,0)[r]{\strut{}uncertainty}}%
      \csname LTb\endcsname%
      \put(3656,2978){\makebox(0,0)[r]{\strut{}central value}}%
      \csname LTb\endcsname%
      \put(594,704){\makebox(0,0)[r]{\strut{}$0$}}%
      \put(594,3371){\makebox(0,0)[r]{\strut{}$0.1$}}%
      \put(1013,484){\makebox(0,0){\strut{}$25$}}%
      \put(2951,484){\makebox(0,0){\strut{}$30$}}%
      \put(1272,3591){\makebox(0,0){\strut{}$0.5$}}%
      \put(3261,3591){\makebox(0,0){\strut{}$0.6$}}%
    }%
    \gplbacktext
    \put(0,0){\includegraphics{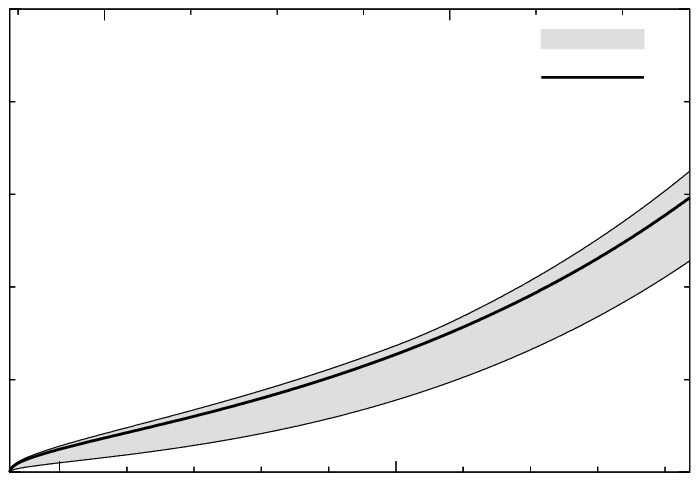}}%
    \gplfronttext
  \end{picture}%
\endgroup

%% file: plots/omega01.tex
\begingroup
  \makeatletter
  \providecommand\color[2][]{%
    \GenericError{(gnuplot) \space\space\space\@spaces}{%
      Package color not loaded in conjunction with
      terminal option `colourtext'%
    }{See the gnuplot documentation for explanation.%
    }{Either use 'blacktext' in gnuplot or load the package
      color.sty in LaTeX.}%
    \renewcommand\color[2][]{}%
  }%
  \providecommand\includegraphics[2][]{%
    \GenericError{(gnuplot) \space\space\space\@spaces}{%
      Package graphicx or graphics not loaded%
    }{See the gnuplot documentation for explanation.%
    }{The gnuplot epslatex terminal needs graphicx.sty or graphics.sty.}%
    \renewcommand\includegraphics[2][]{}%
  }%
  \providecommand\rotatebox[2]{#2}%
  \@ifundefined{ifGPcolor}{%
    \newif\ifGPcolor
    \GPcolorfalse
  }{}%
  \@ifundefined{ifGPblacktext}{%
    \newif\ifGPblacktext
    \GPblacktexttrue
  }{}%
  \let\gplgaddtomacro\g@addto@macro
  \gdef\gplbacktext{}%
  \gdef\gplfronttext{}%
  \makeatother
  \ifGPblacktext
    \def\colorrgb#1{}%
    \def\colorgray#1{}%
  \else
    \ifGPcolor
      \def\colorrgb#1{\color[rgb]{#1}}%
      \def\colorgray#1{\color[gray]{#1}}%
      \expandafter\def\csname LTw\endcsname{\color{white}}%
      \expandafter\def\csname LTb\endcsname{\color{black}}%
      \expandafter\def\csname LTa\endcsname{\color{black}}%
      \expandafter\def\csname LT0\endcsname{\color[rgb]{1,0,0}}%
      \expandafter\def\csname LT1\endcsname{\color[rgb]{0,1,0}}%
      \expandafter\def\csname LT2\endcsname{\color[rgb]{0,0,1}}%
      \expandafter\def\csname LT3\endcsname{\color[rgb]{1,0,1}}%
      \expandafter\def\csname LT4\endcsname{\color[rgb]{0,1,1}}%
      \expandafter\def\csname LT5\endcsname{\color[rgb]{1,1,0}}%
      \expandafter\def\csname LT6\endcsname{\color[rgb]{0,0,0}}%
      \expandafter\def\csname LT7\endcsname{\color[rgb]{1,0.3,0}}%
      \expandafter\def\csname LT8\endcsname{\color[rgb]{0.5,0.5,0.5}}%
    \else
      \def\colorrgb#1{\color{black}}%
      \def\colorgray#1{\color[gray]{#1}}%
      \expandafter\def\csname LTw\endcsname{\color{white}}%
      \expandafter\def\csname LTb\endcsname{\color{black}}%
      \expandafter\def\csname LTa\endcsname{\color{black}}%
      \expandafter\def\csname LT0\endcsname{\color{black}}%
      \expandafter\def\csname LT1\endcsname{\color{black}}%
      \expandafter\def\csname LT2\endcsname{\color{black}}%
      \expandafter\def\csname LT3\endcsname{\color{black}}%
      \expandafter\def\csname LT4\endcsname{\color{black}}%
      \expandafter\def\csname LT5\endcsname{\color{black}}%
      \expandafter\def\csname LT6\endcsname{\color{black}}%
      \expandafter\def\csname LT7\endcsname{\color{black}}%
      \expandafter\def\csname LT8\endcsname{\color{black}}%
    \fi
  \fi
    \setlength{\unitlength}{0.0500bp}%
    \ifx\gptboxheight\undefined%
      \newlength{\gptboxheight}%
      \newlength{\gptboxwidth}%
      \newsavebox{\gptboxtext}%
    \fi%
    \setlength{\fboxrule}{0.5pt}%
    \setlength{\fboxsep}{1pt}%
\begin{picture}(5040.00,4032.00)%
    \gplgaddtomacro\gplbacktext{%
      \csname LTb\endcsname%
      \put(550,704){\makebox(0,0)[r]{\strut{}$0$}}%
      \put(550,1371){\makebox(0,0)[r]{\strut{}$2$}}%
      \put(550,2038){\makebox(0,0)[r]{\strut{}$4$}}%
      \put(550,2704){\makebox(0,0)[r]{\strut{}$6$}}%
      \put(550,3371){\makebox(0,0)[r]{\strut{}$8$}}%
      \put(1596,484){\makebox(0,0){\strut{}$50$}}%
      \put(3120,484){\makebox(0,0){\strut{}$100$}}%
      \put(4643,484){\makebox(0,0){\strut{}$150$}}%
      \put(1637,3591){\makebox(0,0){\strut{}$1$}}%
      \put(3201,3591){\makebox(0,0){\strut{}$2$}}%
    }%
    \gplgaddtomacro\gplfronttext{%
      \csname LTb\endcsname%
      \put(176,2037){\rotatebox{-270}{\makebox(0,0){\strut{}$|\Omega^{1}_{0}(t)|$}}}%
      \put(2662,154){\makebox(0,0){\strut{}$t$ in $M_{\pi}^{2}$}}%
      \put(2662,3920){\makebox(0,0){\strut{}$t$ in GeV$^2$}}%
      \csname LTb\endcsname%
      \put(3656,3198){\makebox(0,0)[r]{\strut{}uncertainty}}%
      \csname LTb\endcsname%
      \put(3656,2978){\makebox(0,0)[r]{\strut{}central value}}%
      \csname LTb\endcsname%
      \put(550,704){\makebox(0,0)[r]{\strut{}$0$}}%
      \put(550,1371){\makebox(0,0)[r]{\strut{}$2$}}%
      \put(550,2038){\makebox(0,0)[r]{\strut{}$4$}}%
      \put(550,2704){\makebox(0,0)[r]{\strut{}$6$}}%
      \put(550,3371){\makebox(0,0)[r]{\strut{}$8$}}%
      \put(1596,484){\makebox(0,0){\strut{}$50$}}%
      \put(3120,484){\makebox(0,0){\strut{}$100$}}%
      \put(4643,484){\makebox(0,0){\strut{}$150$}}%
      \put(1637,3591){\makebox(0,0){\strut{}$1$}}%
      \put(3201,3591){\makebox(0,0){\strut{}$2$}}%
    }%
    \gplbacktext
    \put(0,0){\includegraphics{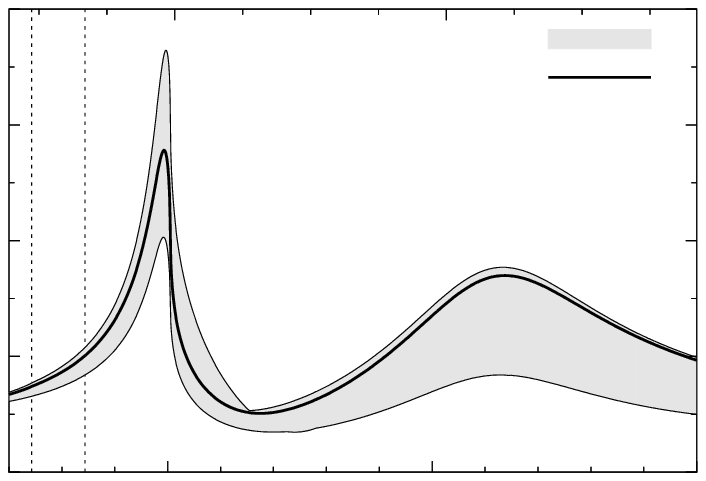}}%
    \gplfronttext
  \end{picture}%
\endgroup

%% file: plots/omega01physreg.tex
\begingroup
  \makeatletter
  \providecommand\color[2][]{%
    \GenericError{(gnuplot) \space\space\space\@spaces}{%
      Package color not loaded in conjunction with
      terminal option `colourtext'%
    }{See the gnuplot documentation for explanation.%
    }{Either use 'blacktext' in gnuplot or load the package
      color.sty in LaTeX.}%
    \renewcommand\color[2][]{}%
  }%
  \providecommand\includegraphics[2][]{%
    \GenericError{(gnuplot) \space\space\space\@spaces}{%
      Package graphicx or graphics not loaded%
    }{See the gnuplot documentation for explanation.%
    }{The gnuplot epslatex terminal needs graphicx.sty or graphics.sty.}%
    \renewcommand\includegraphics[2][]{}%
  }%
  \providecommand\rotatebox[2]{#2}%
  \@ifundefined{ifGPcolor}{%
    \newif\ifGPcolor
    \GPcolorfalse
  }{}%
  \@ifundefined{ifGPblacktext}{%
    \newif\ifGPblacktext
    \GPblacktexttrue
  }{}%
  \let\gplgaddtomacro\g@addto@macro
  \gdef\gplbacktext{}%
  \gdef\gplfronttext{}%
  \makeatother
  \ifGPblacktext
    \def\colorrgb#1{}%
    \def\colorgray#1{}%
  \else
    \ifGPcolor
      \def\colorrgb#1{\color[rgb]{#1}}%
      \def\colorgray#1{\color[gray]{#1}}%
      \expandafter\def\csname LTw\endcsname{\color{white}}%
      \expandafter\def\csname LTb\endcsname{\color{black}}%
      \expandafter\def\csname LTa\endcsname{\color{black}}%
      \expandafter\def\csname LT0\endcsname{\color[rgb]{1,0,0}}%
      \expandafter\def\csname LT1\endcsname{\color[rgb]{0,1,0}}%
      \expandafter\def\csname LT2\endcsname{\color[rgb]{0,0,1}}%
      \expandafter\def\csname LT3\endcsname{\color[rgb]{1,0,1}}%
      \expandafter\def\csname LT4\endcsname{\color[rgb]{0,1,1}}%
      \expandafter\def\csname LT5\endcsname{\color[rgb]{1,1,0}}%
      \expandafter\def\csname LT6\endcsname{\color[rgb]{0,0,0}}%
      \expandafter\def\csname LT7\endcsname{\color[rgb]{1,0.3,0}}%
      \expandafter\def\csname LT8\endcsname{\color[rgb]{0.5,0.5,0.5}}%
    \else
      \def\colorrgb#1{\color{black}}%
      \def\colorgray#1{\color[gray]{#1}}%
      \expandafter\def\csname LTw\endcsname{\color{white}}%
      \expandafter\def\csname LTb\endcsname{\color{black}}%
      \expandafter\def\csname LTa\endcsname{\color{black}}%
      \expandafter\def\csname LT0\endcsname{\color{black}}%
      \expandafter\def\csname LT1\endcsname{\color{black}}%
      \expandafter\def\csname LT2\endcsname{\color{black}}%
      \expandafter\def\csname LT3\endcsname{\color{black}}%
      \expandafter\def\csname LT4\endcsname{\color{black}}%
      \expandafter\def\csname LT5\endcsname{\color{black}}%
      \expandafter\def\csname LT6\endcsname{\color{black}}%
      \expandafter\def\csname LT7\endcsname{\color{black}}%
      \expandafter\def\csname LT8\endcsname{\color{black}}%
    \fi
  \fi
    \setlength{\unitlength}{0.0500bp}%
    \ifx\gptboxheight\undefined%
      \newlength{\gptboxheight}%
      \newlength{\gptboxwidth}%
      \newsavebox{\gptboxtext}%
    \fi%
    \setlength{\fboxrule}{0.5pt}%
    \setlength{\fboxsep}{1pt}%
\begin{picture}(5040.00,4032.00)%
    \gplgaddtomacro\gplbacktext{%
      \csname LTb\endcsname%
      \put(550,704){\makebox(0,0)[r]{\strut{}$1$}}%
      \put(550,2038){\makebox(0,0)[r]{\strut{}$2$}}%
      \put(550,3371){\makebox(0,0)[r]{\strut{}$3$}}%
      \put(973,484){\makebox(0,0){\strut{}$25$}}%
      \put(2932,484){\makebox(0,0){\strut{}$30$}}%
      \put(1234,3591){\makebox(0,0){\strut{}$0.5$}}%
      \put(3246,3591){\makebox(0,0){\strut{}$0.6$}}%
    }%
    \gplgaddtomacro\gplfronttext{%
      \csname LTb\endcsname%
      \put(176,2037){\rotatebox{-270}{\makebox(0,0){\strut{}$|\Omega^{1}_{0}(t)|$}}}%
      \put(2662,154){\makebox(0,0){\strut{}$t$ in $M_{\pi}^{2}$}}%
      \put(2662,3920){\makebox(0,0){\strut{}$t$ in GeV$^2$}}%
      \csname LTb\endcsname%
      \put(3656,3198){\makebox(0,0)[r]{\strut{}uncertainty}}%
      \csname LTb\endcsname%
      \put(3656,2978){\makebox(0,0)[r]{\strut{}central value}}%
      \csname LTb\endcsname%
      \put(550,704){\makebox(0,0)[r]{\strut{}$1$}}%
      \put(550,2038){\makebox(0,0)[r]{\strut{}$2$}}%
      \put(550,3371){\makebox(0,0)[r]{\strut{}$3$}}%
      \put(973,484){\makebox(0,0){\strut{}$25$}}%
      \put(2932,484){\makebox(0,0){\strut{}$30$}}%
      \put(1234,3591){\makebox(0,0){\strut{}$0.5$}}%
      \put(3246,3591){\makebox(0,0){\strut{}$0.6$}}%
    }%
    \gplbacktext
    \put(0,0){\includegraphics{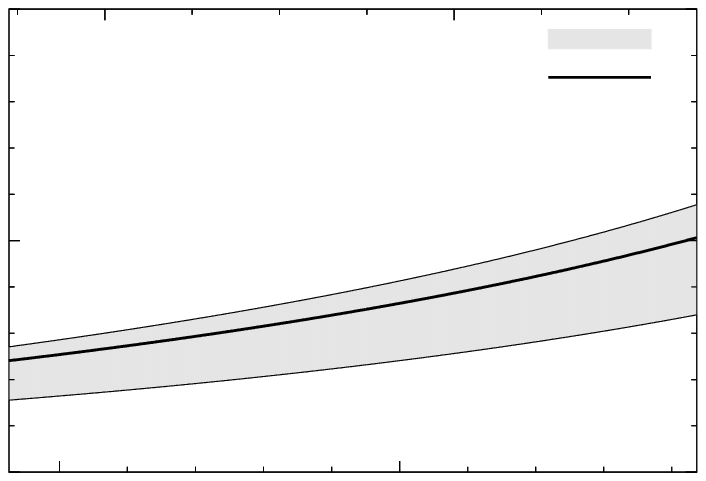}}%
    \gplfronttext
  \end{picture}%
\endgroup

%% file: plots/ellipses.tex
\begingroup
  \makeatletter
  \providecommand\color[2][]{%
    \GenericError{(gnuplot) \space\space\space\@spaces}{%
      Package color not loaded in conjunction with
      terminal option `colourtext'%
    }{See the gnuplot documentation for explanation.%
    }{Either use 'blacktext' in gnuplot or load the package
      color.sty in LaTeX.}%
    \renewcommand\color[2][]{}%
  }%
  \providecommand\includegraphics[2][]{%
    \GenericError{(gnuplot) \space\space\space\@spaces}{%
      Package graphicx or graphics not loaded%
    }{See the gnuplot documentation for explanation.%
    }{The gnuplot epslatex terminal needs graphicx.sty or graphics.sty.}%
    \renewcommand\includegraphics[2][]{}%
  }%
  \providecommand\rotatebox[2]{#2}%
  \@ifundefined{ifGPcolor}{%
    \newif\ifGPcolor
    \GPcolorfalse
  }{}%
  \@ifundefined{ifGPblacktext}{%
    \newif\ifGPblacktext
    \GPblacktexttrue
  }{}%
  \let\gplgaddtomacro\g@addto@macro
  \gdef\gplbacktext{}%
  \gdef\gplfronttext{}%
  \makeatother
  \ifGPblacktext
    \def\colorrgb#1{}%
    \def\colorgray#1{}%
  \else
    \ifGPcolor
      \def\colorrgb#1{\color[rgb]{#1}}%
      \def\colorgray#1{\color[gray]{#1}}%
      \expandafter\def\csname LTw\endcsname{\color{white}}%
      \expandafter\def\csname LTb\endcsname{\color{black}}%
      \expandafter\def\csname LTa\endcsname{\color{black}}%
      \expandafter\def\csname LT0\endcsname{\color[rgb]{1,0,0}}%
      \expandafter\def\csname LT1\endcsname{\color[rgb]{0,1,0}}%
      \expandafter\def\csname LT2\endcsname{\color[rgb]{0,0,1}}%
      \expandafter\def\csname LT3\endcsname{\color[rgb]{1,0,1}}%
      \expandafter\def\csname LT4\endcsname{\color[rgb]{0,1,1}}%
      \expandafter\def\csname LT5\endcsname{\color[rgb]{1,1,0}}%
      \expandafter\def\csname LT6\endcsname{\color[rgb]{0,0,0}}%
      \expandafter\def\csname LT7\endcsname{\color[rgb]{1,0.3,0}}%
      \expandafter\def\csname LT8\endcsname{\color[rgb]{0.5,0.5,0.5}}%
    \else
      \def\colorrgb#1{\color{black}}%
      \def\colorgray#1{\color[gray]{#1}}%
      \expandafter\def\csname LTw\endcsname{\color{white}}%
      \expandafter\def\csname LTb\endcsname{\color{black}}%
      \expandafter\def\csname LTa\endcsname{\color{black}}%
      \expandafter\def\csname LT0\endcsname{\color{black}}%
      \expandafter\def\csname LT1\endcsname{\color{black}}%
      \expandafter\def\csname LT2\endcsname{\color{black}}%
      \expandafter\def\csname LT3\endcsname{\color{black}}%
      \expandafter\def\csname LT4\endcsname{\color{black}}%
      \expandafter\def\csname LT5\endcsname{\color{black}}%
      \expandafter\def\csname LT6\endcsname{\color{black}}%
      \expandafter\def\csname LT7\endcsname{\color{black}}%
      \expandafter\def\csname LT8\endcsname{\color{black}}%
    \fi
  \fi
    \setlength{\unitlength}{0.0500bp}%
    \ifx\gptboxheight\undefined%
      \newlength{\gptboxheight}%
      \newlength{\gptboxwidth}%
      \newsavebox{\gptboxtext}%
    \fi%
    \setlength{\fboxrule}{0.5pt}%
    \setlength{\fboxsep}{1pt}%
\begin{picture}(7200.00,5040.00)%
    \gplgaddtomacro\gplbacktext{%
      \csname LTb\endcsname%
      \put(682,1061){\makebox(0,0)[r]{\strut{}$5$}}%
      \put(682,1775){\makebox(0,0)[r]{\strut{}$6$}}%
      \put(682,2490){\makebox(0,0)[r]{\strut{}$7$}}%
      \put(682,3204){\makebox(0,0)[r]{\strut{}$8$}}%
      \put(682,3918){\makebox(0,0)[r]{\strut{}$9$}}%
      \put(682,4632){\makebox(0,0)[r]{\strut{}$10$}}%
      \put(814,484){\makebox(0,0){\strut{}$-3.5$}}%
      \put(2478,484){\makebox(0,0){\strut{}$-3$}}%
      \put(4141,484){\makebox(0,0){\strut{}$-2.5$}}%
      \put(5805,484){\makebox(0,0){\strut{}$-2$}}%
    }%
    \gplgaddtomacro\gplfronttext{%
      \csname LTb\endcsname%
      \put(176,2739){\rotatebox{-270}{\makebox(0,0){\strut{}$\bar\beta$}}}%
      \put(3808,154){\makebox(0,0){\strut{}$\bar\alpha$}}%
      \csname LTb\endcsname%
      \put(5649,4258){\makebox(0,0)[r]{\strut{}BES-III fit}}%
      \csname LTb\endcsname%
      \put(5649,4038){\makebox(0,0)[r]{\strut{}VES fit}}%
    }%
    \gplbacktext
    \put(0,0){\includegraphics{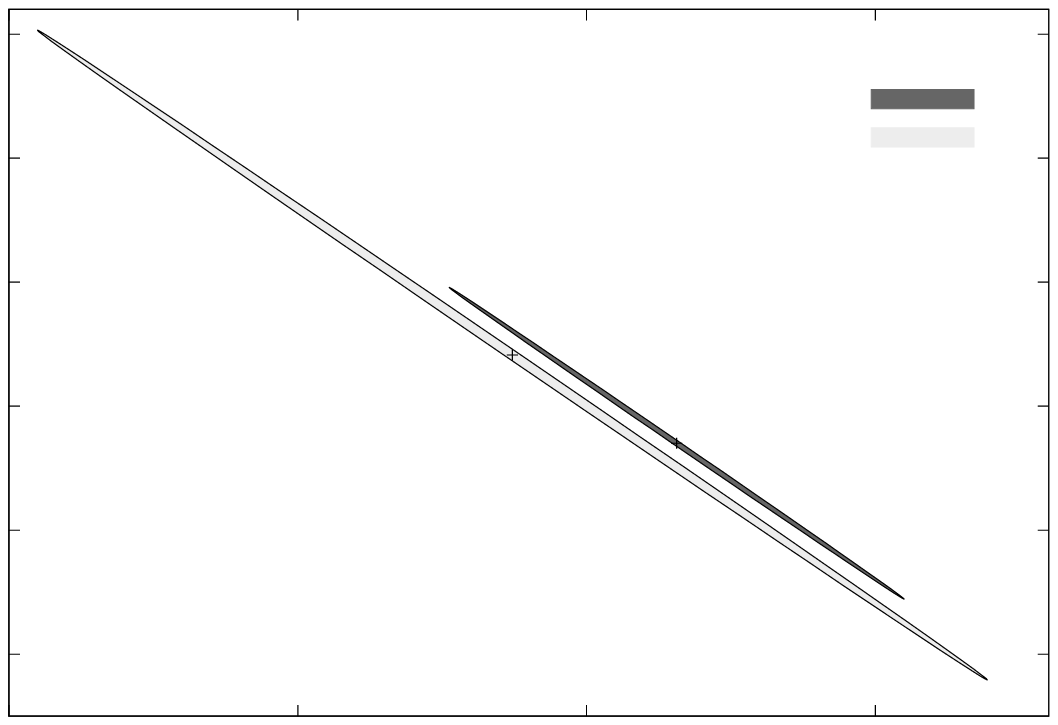}}%
    \gplfronttext
  \end{picture}%
\endgroup

%% file: plots/BES_Xproj_DR3.tex
\begingroup
  \makeatletter
  \providecommand\color[2][]{%
    \GenericError{(gnuplot) \space\space\space\@spaces}{%
      Package color not loaded in conjunction with
      terminal option `colourtext'%
    }{See the gnuplot documentation for explanation.%
    }{Either use 'blacktext' in gnuplot or load the package
      color.sty in LaTeX.}%
    \renewcommand\color[2][]{}%
  }%
  \providecommand\includegraphics[2][]{%
    \GenericError{(gnuplot) \space\space\space\@spaces}{%
      Package graphicx or graphics not loaded%
    }{See the gnuplot documentation for explanation.%
    }{The gnuplot epslatex terminal needs graphicx.sty or graphics.sty.}%
    \renewcommand\includegraphics[2][]{}%
  }%
  \providecommand\rotatebox[2]{#2}%
  \@ifundefined{ifGPcolor}{%
    \newif\ifGPcolor
    \GPcolorfalse
  }{}%
  \@ifundefined{ifGPblacktext}{%
    \newif\ifGPblacktext
    \GPblacktexttrue
  }{}%
  \let\gplgaddtomacro\g@addto@macro
  \gdef\gplbacktext{}%
  \gdef\gplfronttext{}%
  \makeatother
  \ifGPblacktext
    \def\colorrgb#1{}%
    \def\colorgray#1{}%
  \else
    \ifGPcolor
      \def\colorrgb#1{\color[rgb]{#1}}%
      \def\colorgray#1{\color[gray]{#1}}%
      \expandafter\def\csname LTw\endcsname{\color{white}}%
      \expandafter\def\csname LTb\endcsname{\color{black}}%
      \expandafter\def\csname LTa\endcsname{\color{black}}%
      \expandafter\def\csname LT0\endcsname{\color[rgb]{1,0,0}}%
      \expandafter\def\csname LT1\endcsname{\color[rgb]{0,1,0}}%
      \expandafter\def\csname LT2\endcsname{\color[rgb]{0,0,1}}%
      \expandafter\def\csname LT3\endcsname{\color[rgb]{1,0,1}}%
      \expandafter\def\csname LT4\endcsname{\color[rgb]{0,1,1}}%
      \expandafter\def\csname LT5\endcsname{\color[rgb]{1,1,0}}%
      \expandafter\def\csname LT6\endcsname{\color[rgb]{0,0,0}}%
      \expandafter\def\csname LT7\endcsname{\color[rgb]{1,0.3,0}}%
      \expandafter\def\csname LT8\endcsname{\color[rgb]{0.5,0.5,0.5}}%
    \else
      \def\colorrgb#1{\color{black}}%
      \def\colorgray#1{\color[gray]{#1}}%
      \expandafter\def\csname LTw\endcsname{\color{white}}%
      \expandafter\def\csname LTb\endcsname{\color{black}}%
      \expandafter\def\csname LTa\endcsname{\color{black}}%
      \expandafter\def\csname LT0\endcsname{\color{black}}%
      \expandafter\def\csname LT1\endcsname{\color{black}}%
      \expandafter\def\csname LT2\endcsname{\color{black}}%
      \expandafter\def\csname LT3\endcsname{\color{black}}%
      \expandafter\def\csname LT4\endcsname{\color{black}}%
      \expandafter\def\csname LT5\endcsname{\color{black}}%
      \expandafter\def\csname LT6\endcsname{\color{black}}%
      \expandafter\def\csname LT7\endcsname{\color{black}}%
      \expandafter\def\csname LT8\endcsname{\color{black}}%
    \fi
  \fi
    \setlength{\unitlength}{0.0500bp}%
    \ifx\gptboxheight\undefined%
      \newlength{\gptboxheight}%
      \newlength{\gptboxwidth}%
      \newsavebox{\gptboxtext}%
    \fi%
    \setlength{\fboxrule}{0.5pt}%
    \setlength{\fboxsep}{1pt}%
\begin{picture}(7200.00,5040.00)%
    \gplgaddtomacro\gplbacktext{%
      \csname LTb\endcsname%
      \put(814,1156){\makebox(0,0)[r]{\strut{}$0.8$}}%
      \put(814,2061){\makebox(0,0)[r]{\strut{}$0.9$}}%
      \put(814,2966){\makebox(0,0)[r]{\strut{}$1$}}%
      \put(814,3870){\makebox(0,0)[r]{\strut{}$1.1$}}%
      \put(814,4775){\makebox(0,0)[r]{\strut{}$1.2$}}%
      \put(1212,484){\makebox(0,0){\strut{}$-1.2$}}%
      \put(2100,484){\makebox(0,0){\strut{}$-0.8$}}%
      \put(2987,484){\makebox(0,0){\strut{}$-0.4$}}%
      \put(3874,484){\makebox(0,0){\strut{}$0$}}%
      \put(4762,484){\makebox(0,0){\strut{}$0.4$}}%
      \put(5649,484){\makebox(0,0){\strut{}$0.8$}}%
      \put(6537,484){\makebox(0,0){\strut{}$1.2$}}%
    }%
    \gplgaddtomacro\gplfronttext{%
      \csname LTb\endcsname%
      \put(176,2739){\rotatebox{-270}{\makebox(0,0){\strut{}$\frac{\int \diff y\,|\bar\M(x,y)|^{2}}{\int\diff y\,\diff\bar\Phi(x,y)}$}}}%
      \put(3874,154){\makebox(0,0){\strut{}$x$}}%
      \csname LTb\endcsname%
      \put(4569,1922){\makebox(0,0)[r]{\strut{}fit uncertainty}}%
      \csname LTb\endcsname%
      \put(4569,1592){\makebox(0,0)[r]{\strut{}phase uncertainty}}%
      \csname LTb\endcsname%
      \put(4569,1262){\makebox(0,0)[r]{\strut{}central value}}%
      \csname LTb\endcsname%
      \put(4569,932){\makebox(0,0)[r]{\strut{}BES-III data}}%
      \csname LTb\endcsname%
      \put(814,1156){\makebox(0,0)[r]{\strut{}$0.8$}}%
      \put(814,2061){\makebox(0,0)[r]{\strut{}$0.9$}}%
      \put(814,2966){\makebox(0,0)[r]{\strut{}$1$}}%
      \put(814,3870){\makebox(0,0)[r]{\strut{}$1.1$}}%
      \put(814,4775){\makebox(0,0)[r]{\strut{}$1.2$}}%
      \put(1212,484){\makebox(0,0){\strut{}$-1.2$}}%
      \put(2100,484){\makebox(0,0){\strut{}$-0.8$}}%
      \put(2987,484){\makebox(0,0){\strut{}$-0.4$}}%
      \put(3874,484){\makebox(0,0){\strut{}$0$}}%
      \put(4762,484){\makebox(0,0){\strut{}$0.4$}}%
      \put(5649,484){\makebox(0,0){\strut{}$0.8$}}%
      \put(6537,484){\makebox(0,0){\strut{}$1.2$}}%
    }%
    \gplbacktext
    \put(0,0){\includegraphics{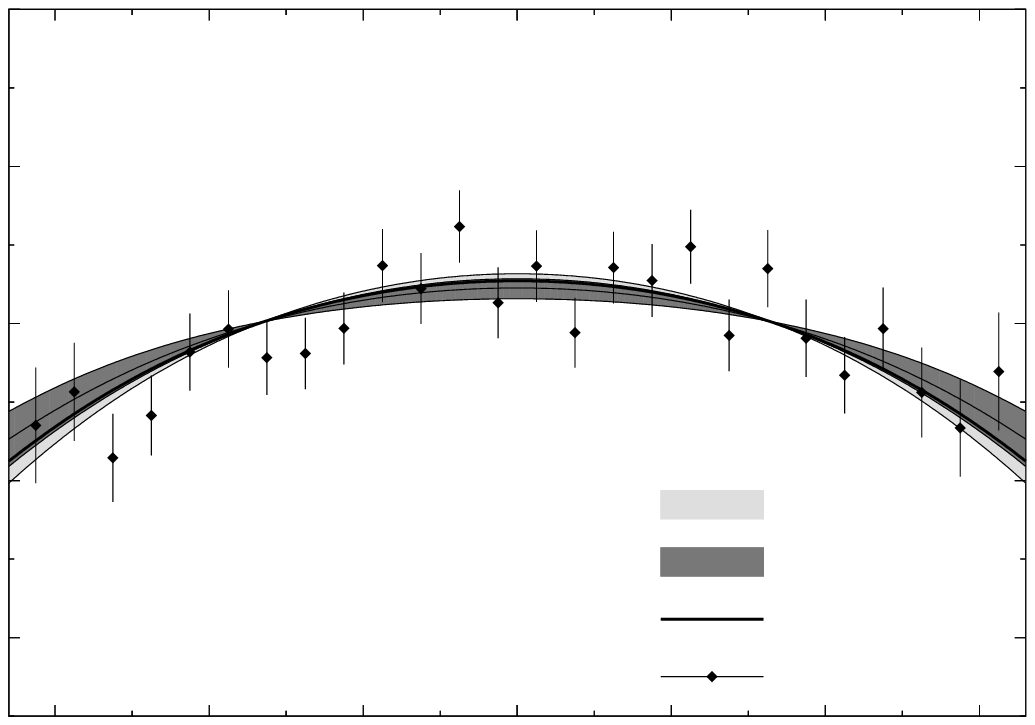}}%
    \gplfronttext
  \end{picture}%
\endgroup

%% file: plots/BES_Xproj_DR4.tex
\begingroup
  \makeatletter
  \providecommand\color[2][]{%
    \GenericError{(gnuplot) \space\space\space\@spaces}{%
      Package color not loaded in conjunction with
      terminal option `colourtext'%
    }{See the gnuplot documentation for explanation.%
    }{Either use 'blacktext' in gnuplot or load the package
      color.sty in LaTeX.}%
    \renewcommand\color[2][]{}%
  }%
  \providecommand\includegraphics[2][]{%
    \GenericError{(gnuplot) \space\space\space\@spaces}{%
      Package graphicx or graphics not loaded%
    }{See the gnuplot documentation for explanation.%
    }{The gnuplot epslatex terminal needs graphicx.sty or graphics.sty.}%
    \renewcommand\includegraphics[2][]{}%
  }%
  \providecommand\rotatebox[2]{#2}%
  \@ifundefined{ifGPcolor}{%
    \newif\ifGPcolor
    \GPcolorfalse
  }{}%
  \@ifundefined{ifGPblacktext}{%
    \newif\ifGPblacktext
    \GPblacktexttrue
  }{}%
  \let\gplgaddtomacro\g@addto@macro
  \gdef\gplbacktext{}%
  \gdef\gplfronttext{}%
  \makeatother
  \ifGPblacktext
    \def\colorrgb#1{}%
    \def\colorgray#1{}%
  \else
    \ifGPcolor
      \def\colorrgb#1{\color[rgb]{#1}}%
      \def\colorgray#1{\color[gray]{#1}}%
      \expandafter\def\csname LTw\endcsname{\color{white}}%
      \expandafter\def\csname LTb\endcsname{\color{black}}%
      \expandafter\def\csname LTa\endcsname{\color{black}}%
      \expandafter\def\csname LT0\endcsname{\color[rgb]{1,0,0}}%
      \expandafter\def\csname LT1\endcsname{\color[rgb]{0,1,0}}%
      \expandafter\def\csname LT2\endcsname{\color[rgb]{0,0,1}}%
      \expandafter\def\csname LT3\endcsname{\color[rgb]{1,0,1}}%
      \expandafter\def\csname LT4\endcsname{\color[rgb]{0,1,1}}%
      \expandafter\def\csname LT5\endcsname{\color[rgb]{1,1,0}}%
      \expandafter\def\csname LT6\endcsname{\color[rgb]{0,0,0}}%
      \expandafter\def\csname LT7\endcsname{\color[rgb]{1,0.3,0}}%
      \expandafter\def\csname LT8\endcsname{\color[rgb]{0.5,0.5,0.5}}%
    \else
      \def\colorrgb#1{\color{black}}%
      \def\colorgray#1{\color[gray]{#1}}%
      \expandafter\def\csname LTw\endcsname{\color{white}}%
      \expandafter\def\csname LTb\endcsname{\color{black}}%
      \expandafter\def\csname LTa\endcsname{\color{black}}%
      \expandafter\def\csname LT0\endcsname{\color{black}}%
      \expandafter\def\csname LT1\endcsname{\color{black}}%
      \expandafter\def\csname LT2\endcsname{\color{black}}%
      \expandafter\def\csname LT3\endcsname{\color{black}}%
      \expandafter\def\csname LT4\endcsname{\color{black}}%
      \expandafter\def\csname LT5\endcsname{\color{black}}%
      \expandafter\def\csname LT6\endcsname{\color{black}}%
      \expandafter\def\csname LT7\endcsname{\color{black}}%
      \expandafter\def\csname LT8\endcsname{\color{black}}%
    \fi
  \fi
    \setlength{\unitlength}{0.0500bp}%
    \ifx\gptboxheight\undefined%
      \newlength{\gptboxheight}%
      \newlength{\gptboxwidth}%
      \newsavebox{\gptboxtext}%
    \fi%
    \setlength{\fboxrule}{0.5pt}%
    \setlength{\fboxsep}{1pt}%
\begin{picture}(7200.00,5040.00)%
    \gplgaddtomacro\gplbacktext{%
      \csname LTb\endcsname%
      \put(814,1156){\makebox(0,0)[r]{\strut{}$0.8$}}%
      \put(814,2061){\makebox(0,0)[r]{\strut{}$0.9$}}%
      \put(814,2966){\makebox(0,0)[r]{\strut{}$1$}}%
      \put(814,3870){\makebox(0,0)[r]{\strut{}$1.1$}}%
      \put(814,4775){\makebox(0,0)[r]{\strut{}$1.2$}}%
      \put(1212,484){\makebox(0,0){\strut{}$-1.2$}}%
      \put(2100,484){\makebox(0,0){\strut{}$-0.8$}}%
      \put(2987,484){\makebox(0,0){\strut{}$-0.4$}}%
      \put(3874,484){\makebox(0,0){\strut{}$0$}}%
      \put(4762,484){\makebox(0,0){\strut{}$0.4$}}%
      \put(5649,484){\makebox(0,0){\strut{}$0.8$}}%
      \put(6537,484){\makebox(0,0){\strut{}$1.2$}}%
    }%
    \gplgaddtomacro\gplfronttext{%
      \csname LTb\endcsname%
      \put(176,2739){\rotatebox{-270}{\makebox(0,0){\strut{}$\frac{\int \diff y\,|\bar\M(x,y)|^{2}}{\int\diff y\,\diff\bar\Phi(x,y)}$}}}%
      \put(3874,154){\makebox(0,0){\strut{}$x$}}%
      \csname LTb\endcsname%
      \put(4569,1922){\makebox(0,0)[r]{\strut{}fit uncertainty}}%
      \csname LTb\endcsname%
      \put(4569,1592){\makebox(0,0)[r]{\strut{}phase uncertainty}}%
      \csname LTb\endcsname%
      \put(4569,1262){\makebox(0,0)[r]{\strut{}central value}}%
      \csname LTb\endcsname%
      \put(4569,932){\makebox(0,0)[r]{\strut{}BES-III data}}%
      \csname LTb\endcsname%
      \put(814,1156){\makebox(0,0)[r]{\strut{}$0.8$}}%
      \put(814,2061){\makebox(0,0)[r]{\strut{}$0.9$}}%
      \put(814,2966){\makebox(0,0)[r]{\strut{}$1$}}%
      \put(814,3870){\makebox(0,0)[r]{\strut{}$1.1$}}%
      \put(814,4775){\makebox(0,0)[r]{\strut{}$1.2$}}%
      \put(1212,484){\makebox(0,0){\strut{}$-1.2$}}%
      \put(2100,484){\makebox(0,0){\strut{}$-0.8$}}%
      \put(2987,484){\makebox(0,0){\strut{}$-0.4$}}%
      \put(3874,484){\makebox(0,0){\strut{}$0$}}%
      \put(4762,484){\makebox(0,0){\strut{}$0.4$}}%
      \put(5649,484){\makebox(0,0){\strut{}$0.8$}}%
      \put(6537,484){\makebox(0,0){\strut{}$1.2$}}%
    }%
    \gplbacktext
    \put(0,0){\includegraphics{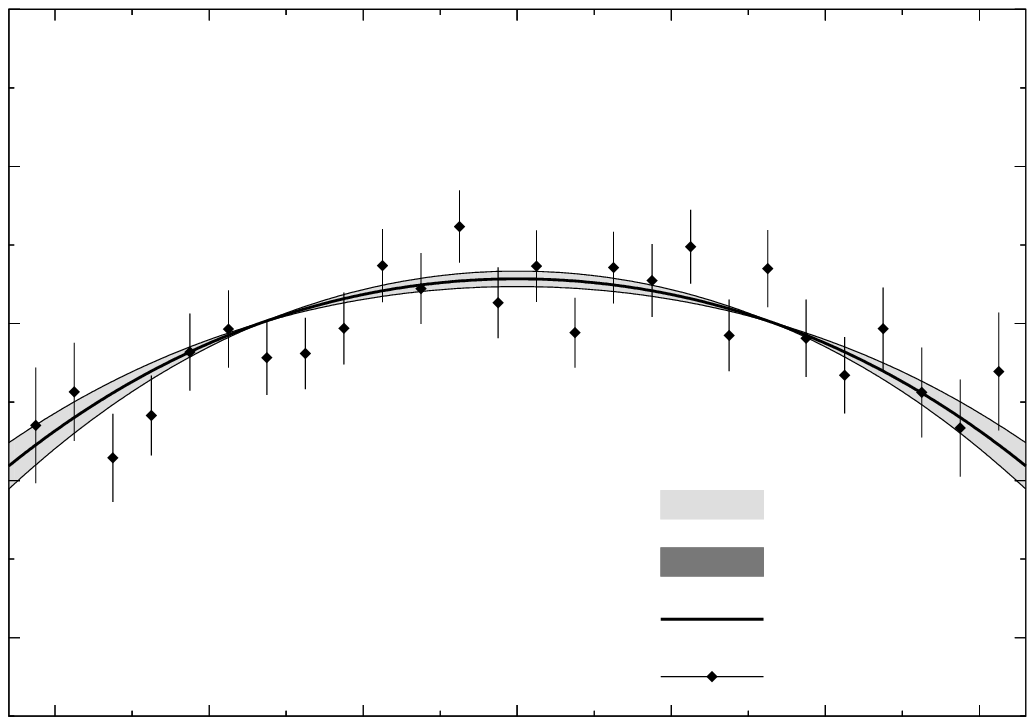}}%
    \gplfronttext
  \end{picture}%
\endgroup

%% file: plots/BES_Yproj_DR3.tex
\begingroup
  \makeatletter
  \providecommand\color[2][]{%
    \GenericError{(gnuplot) \space\space\space\@spaces}{%
      Package color not loaded in conjunction with
      terminal option `colourtext'%
    }{See the gnuplot documentation for explanation.%
    }{Either use 'blacktext' in gnuplot or load the package
      color.sty in LaTeX.}%
    \renewcommand\color[2][]{}%
  }%
  \providecommand\includegraphics[2][]{%
    \GenericError{(gnuplot) \space\space\space\@spaces}{%
      Package graphicx or graphics not loaded%
    }{See the gnuplot documentation for explanation.%
    }{The gnuplot epslatex terminal needs graphicx.sty or graphics.sty.}%
    \renewcommand\includegraphics[2][]{}%
  }%
  \providecommand\rotatebox[2]{#2}%
  \@ifundefined{ifGPcolor}{%
    \newif\ifGPcolor
    \GPcolorfalse
  }{}%
  \@ifundefined{ifGPblacktext}{%
    \newif\ifGPblacktext
    \GPblacktexttrue
  }{}%
  \let\gplgaddtomacro\g@addto@macro
  \gdef\gplbacktext{}%
  \gdef\gplfronttext{}%
  \makeatother
  \ifGPblacktext
    \def\colorrgb#1{}%
    \def\colorgray#1{}%
  \else
    \ifGPcolor
      \def\colorrgb#1{\color[rgb]{#1}}%
      \def\colorgray#1{\color[gray]{#1}}%
      \expandafter\def\csname LTw\endcsname{\color{white}}%
      \expandafter\def\csname LTb\endcsname{\color{black}}%
      \expandafter\def\csname LTa\endcsname{\color{black}}%
      \expandafter\def\csname LT0\endcsname{\color[rgb]{1,0,0}}%
      \expandafter\def\csname LT1\endcsname{\color[rgb]{0,1,0}}%
      \expandafter\def\csname LT2\endcsname{\color[rgb]{0,0,1}}%
      \expandafter\def\csname LT3\endcsname{\color[rgb]{1,0,1}}%
      \expandafter\def\csname LT4\endcsname{\color[rgb]{0,1,1}}%
      \expandafter\def\csname LT5\endcsname{\color[rgb]{1,1,0}}%
      \expandafter\def\csname LT6\endcsname{\color[rgb]{0,0,0}}%
      \expandafter\def\csname LT7\endcsname{\color[rgb]{1,0.3,0}}%
      \expandafter\def\csname LT8\endcsname{\color[rgb]{0.5,0.5,0.5}}%
    \else
      \def\colorrgb#1{\color{black}}%
      \def\colorgray#1{\color[gray]{#1}}%
      \expandafter\def\csname LTw\endcsname{\color{white}}%
      \expandafter\def\csname LTb\endcsname{\color{black}}%
      \expandafter\def\csname LTa\endcsname{\color{black}}%
      \expandafter\def\csname LT0\endcsname{\color{black}}%
      \expandafter\def\csname LT1\endcsname{\color{black}}%
      \expandafter\def\csname LT2\endcsname{\color{black}}%
      \expandafter\def\csname LT3\endcsname{\color{black}}%
      \expandafter\def\csname LT4\endcsname{\color{black}}%
      \expandafter\def\csname LT5\endcsname{\color{black}}%
      \expandafter\def\csname LT6\endcsname{\color{black}}%
      \expandafter\def\csname LT7\endcsname{\color{black}}%
      \expandafter\def\csname LT8\endcsname{\color{black}}%
    \fi
  \fi
    \setlength{\unitlength}{0.0500bp}%
    \ifx\gptboxheight\undefined%
      \newlength{\gptboxheight}%
      \newlength{\gptboxwidth}%
      \newsavebox{\gptboxtext}%
    \fi%
    \setlength{\fboxrule}{0.5pt}%
    \setlength{\fboxsep}{1pt}%
\begin{picture}(7200.00,5040.00)%
    \gplgaddtomacro\gplbacktext{%
      \csname LTb\endcsname%
      \put(814,1156){\makebox(0,0)[r]{\strut{}$0.8$}}%
      \put(814,2061){\makebox(0,0)[r]{\strut{}$0.9$}}%
      \put(814,2966){\makebox(0,0)[r]{\strut{}$1$}}%
      \put(814,3870){\makebox(0,0)[r]{\strut{}$1.1$}}%
      \put(814,4775){\makebox(0,0)[r]{\strut{}$1.2$}}%
      \put(1478,484){\makebox(0,0){\strut{}$-0.8$}}%
      \put(2599,484){\makebox(0,0){\strut{}$-0.4$}}%
      \put(3720,484){\makebox(0,0){\strut{}$0$}}%
      \put(4841,484){\makebox(0,0){\strut{}$0.4$}}%
      \put(5962,484){\makebox(0,0){\strut{}$0.8$}}%
    }%
    \gplgaddtomacro\gplfronttext{%
      \csname LTb\endcsname%
      \put(176,2739){\rotatebox{-270}{\makebox(0,0){\strut{}$\frac{\int \diff x\,|\bar\M(x,y)|^{2}}{\int\diff x\,\diff\bar\Phi(x,y)}$}}}%
      \put(3874,154){\makebox(0,0){\strut{}$y$}}%
      \csname LTb\endcsname%
      \put(4569,1922){\makebox(0,0)[r]{\strut{}fit uncertainty}}%
      \csname LTb\endcsname%
      \put(4569,1592){\makebox(0,0)[r]{\strut{}phase uncertainty}}%
      \csname LTb\endcsname%
      \put(4569,1262){\makebox(0,0)[r]{\strut{}central value}}%
      \csname LTb\endcsname%
      \put(4569,932){\makebox(0,0)[r]{\strut{}BES-III data}}%
      \csname LTb\endcsname%
      \put(814,1156){\makebox(0,0)[r]{\strut{}$0.8$}}%
      \put(814,2061){\makebox(0,0)[r]{\strut{}$0.9$}}%
      \put(814,2966){\makebox(0,0)[r]{\strut{}$1$}}%
      \put(814,3870){\makebox(0,0)[r]{\strut{}$1.1$}}%
      \put(814,4775){\makebox(0,0)[r]{\strut{}$1.2$}}%
      \put(1478,484){\makebox(0,0){\strut{}$-0.8$}}%
      \put(2599,484){\makebox(0,0){\strut{}$-0.4$}}%
      \put(3720,484){\makebox(0,0){\strut{}$0$}}%
      \put(4841,484){\makebox(0,0){\strut{}$0.4$}}%
      \put(5962,484){\makebox(0,0){\strut{}$0.8$}}%
    }%
    \gplbacktext
    \put(0,0){\includegraphics{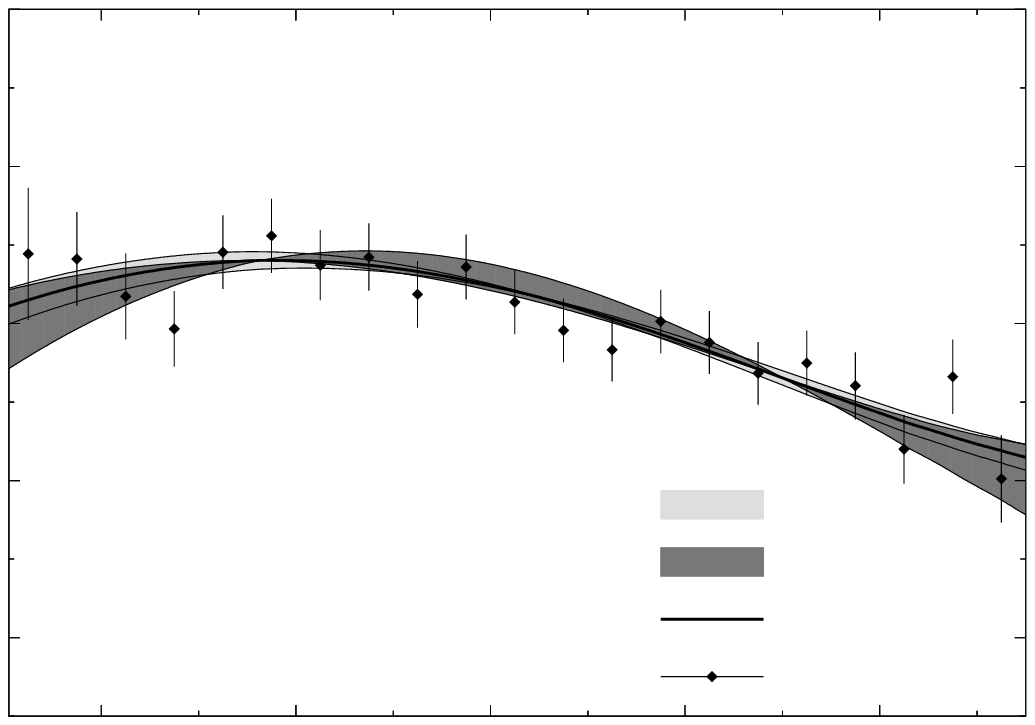}}%
    \gplfronttext
  \end{picture}%
\endgroup

%% file: plots/BES_Yproj_DR4.tex
\begingroup
  \makeatletter
  \providecommand\color[2][]{%
    \GenericError{(gnuplot) \space\space\space\@spaces}{%
      Package color not loaded in conjunction with
      terminal option `colourtext'%
    }{See the gnuplot documentation for explanation.%
    }{Either use 'blacktext' in gnuplot or load the package
      color.sty in LaTeX.}%
    \renewcommand\color[2][]{}%
  }%
  \providecommand\includegraphics[2][]{%
    \GenericError{(gnuplot) \space\space\space\@spaces}{%
      Package graphicx or graphics not loaded%
    }{See the gnuplot documentation for explanation.%
    }{The gnuplot epslatex terminal needs graphicx.sty or graphics.sty.}%
    \renewcommand\includegraphics[2][]{}%
  }%
  \providecommand\rotatebox[2]{#2}%
  \@ifundefined{ifGPcolor}{%
    \newif\ifGPcolor
    \GPcolorfalse
  }{}%
  \@ifundefined{ifGPblacktext}{%
    \newif\ifGPblacktext
    \GPblacktexttrue
  }{}%
  \let\gplgaddtomacro\g@addto@macro
  \gdef\gplbacktext{}%
  \gdef\gplfronttext{}%
  \makeatother
  \ifGPblacktext
    \def\colorrgb#1{}%
    \def\colorgray#1{}%
  \else
    \ifGPcolor
      \def\colorrgb#1{\color[rgb]{#1}}%
      \def\colorgray#1{\color[gray]{#1}}%
      \expandafter\def\csname LTw\endcsname{\color{white}}%
      \expandafter\def\csname LTb\endcsname{\color{black}}%
      \expandafter\def\csname LTa\endcsname{\color{black}}%
      \expandafter\def\csname LT0\endcsname{\color[rgb]{1,0,0}}%
      \expandafter\def\csname LT1\endcsname{\color[rgb]{0,1,0}}%
      \expandafter\def\csname LT2\endcsname{\color[rgb]{0,0,1}}%
      \expandafter\def\csname LT3\endcsname{\color[rgb]{1,0,1}}%
      \expandafter\def\csname LT4\endcsname{\color[rgb]{0,1,1}}%
      \expandafter\def\csname LT5\endcsname{\color[rgb]{1,1,0}}%
      \expandafter\def\csname LT6\endcsname{\color[rgb]{0,0,0}}%
      \expandafter\def\csname LT7\endcsname{\color[rgb]{1,0.3,0}}%
      \expandafter\def\csname LT8\endcsname{\color[rgb]{0.5,0.5,0.5}}%
    \else
      \def\colorrgb#1{\color{black}}%
      \def\colorgray#1{\color[gray]{#1}}%
      \expandafter\def\csname LTw\endcsname{\color{white}}%
      \expandafter\def\csname LTb\endcsname{\color{black}}%
      \expandafter\def\csname LTa\endcsname{\color{black}}%
      \expandafter\def\csname LT0\endcsname{\color{black}}%
      \expandafter\def\csname LT1\endcsname{\color{black}}%
      \expandafter\def\csname LT2\endcsname{\color{black}}%
      \expandafter\def\csname LT3\endcsname{\color{black}}%
      \expandafter\def\csname LT4\endcsname{\color{black}}%
      \expandafter\def\csname LT5\endcsname{\color{black}}%
      \expandafter\def\csname LT6\endcsname{\color{black}}%
      \expandafter\def\csname LT7\endcsname{\color{black}}%
      \expandafter\def\csname LT8\endcsname{\color{black}}%
    \fi
  \fi
    \setlength{\unitlength}{0.0500bp}%
    \ifx\gptboxheight\undefined%
      \newlength{\gptboxheight}%
      \newlength{\gptboxwidth}%
      \newsavebox{\gptboxtext}%
    \fi%
    \setlength{\fboxrule}{0.5pt}%
    \setlength{\fboxsep}{1pt}%
\begin{picture}(7200.00,5040.00)%
    \gplgaddtomacro\gplbacktext{%
      \csname LTb\endcsname%
      \put(814,1156){\makebox(0,0)[r]{\strut{}$0.8$}}%
      \put(814,2061){\makebox(0,0)[r]{\strut{}$0.9$}}%
      \put(814,2966){\makebox(0,0)[r]{\strut{}$1$}}%
      \put(814,3870){\makebox(0,0)[r]{\strut{}$1.1$}}%
      \put(814,4775){\makebox(0,0)[r]{\strut{}$1.2$}}%
      \put(1478,484){\makebox(0,0){\strut{}$-0.8$}}%
      \put(2599,484){\makebox(0,0){\strut{}$-0.4$}}%
      \put(3720,484){\makebox(0,0){\strut{}$0$}}%
      \put(4841,484){\makebox(0,0){\strut{}$0.4$}}%
      \put(5962,484){\makebox(0,0){\strut{}$0.8$}}%
    }%
    \gplgaddtomacro\gplfronttext{%
      \csname LTb\endcsname%
      \put(176,2739){\rotatebox{-270}{\makebox(0,0){\strut{}$\frac{\int \diff x\,|\bar\M(x,y)|^{2}}{\int\diff x\,\diff\bar\Phi(x,y)}$}}}%
      \put(3874,154){\makebox(0,0){\strut{}$y$}}%
      \csname LTb\endcsname%
      \put(4569,1922){\makebox(0,0)[r]{\strut{}fit uncertainty}}%
      \csname LTb\endcsname%
      \put(4569,1592){\makebox(0,0)[r]{\strut{}phase uncertainty}}%
      \csname LTb\endcsname%
      \put(4569,1262){\makebox(0,0)[r]{\strut{}central value}}%
      \csname LTb\endcsname%
      \put(4569,932){\makebox(0,0)[r]{\strut{}BES-III data}}%
      \csname LTb\endcsname%
      \put(814,1156){\makebox(0,0)[r]{\strut{}$0.8$}}%
      \put(814,2061){\makebox(0,0)[r]{\strut{}$0.9$}}%
      \put(814,2966){\makebox(0,0)[r]{\strut{}$1$}}%
      \put(814,3870){\makebox(0,0)[r]{\strut{}$1.1$}}%
      \put(814,4775){\makebox(0,0)[r]{\strut{}$1.2$}}%
      \put(1478,484){\makebox(0,0){\strut{}$-0.8$}}%
      \put(2599,484){\makebox(0,0){\strut{}$-0.4$}}%
      \put(3720,484){\makebox(0,0){\strut{}$0$}}%
      \put(4841,484){\makebox(0,0){\strut{}$0.4$}}%
      \put(5962,484){\makebox(0,0){\strut{}$0.8$}}%
    }%
    \gplbacktext
    \put(0,0){\includegraphics{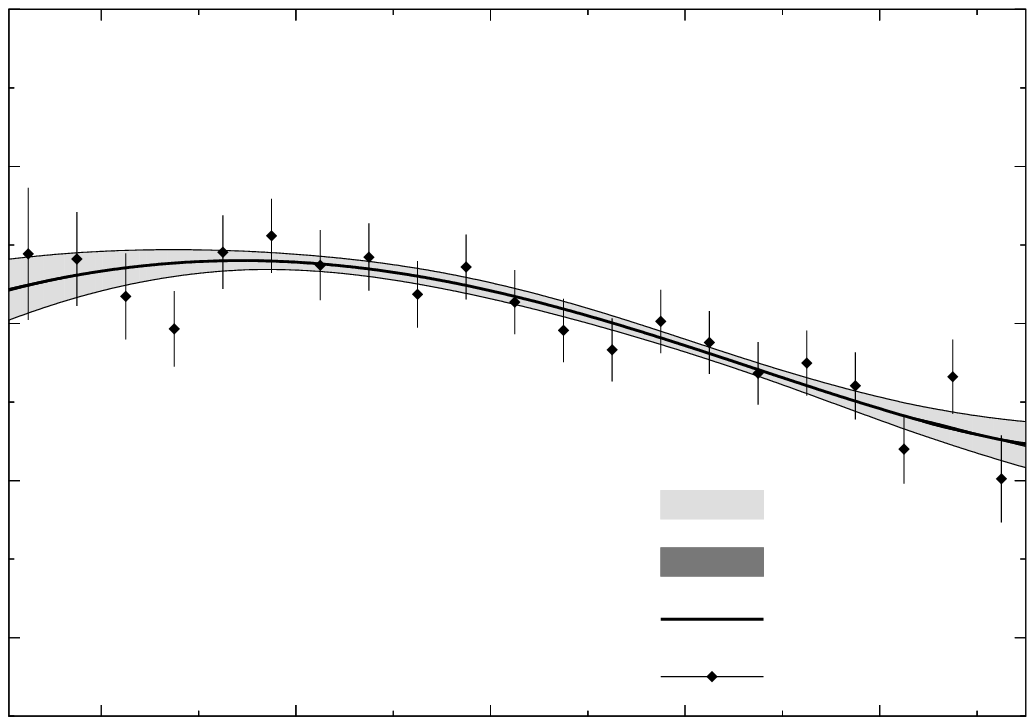}}%
    \gplfronttext
  \end{picture}%
\endgroup

%% file: plots/VES_Xproj_DR3.tex
\begingroup
  \makeatletter
  \providecommand\color[2][]{%
    \GenericError{(gnuplot) \space\space\space\@spaces}{%
      Package color not loaded in conjunction with
      terminal option `colourtext'%
    }{See the gnuplot documentation for explanation.%
    }{Either use 'blacktext' in gnuplot or load the package
      color.sty in LaTeX.}%
    \renewcommand\color[2][]{}%
  }%
  \providecommand\includegraphics[2][]{%
    \GenericError{(gnuplot) \space\space\space\@spaces}{%
      Package graphicx or graphics not loaded%
    }{See the gnuplot documentation for explanation.%
    }{The gnuplot epslatex terminal needs graphicx.sty or graphics.sty.}%
    \renewcommand\includegraphics[2][]{}%
  }%
  \providecommand\rotatebox[2]{#2}%
  \@ifundefined{ifGPcolor}{%
    \newif\ifGPcolor
    \GPcolorfalse
  }{}%
  \@ifundefined{ifGPblacktext}{%
    \newif\ifGPblacktext
    \GPblacktexttrue
  }{}%
  \let\gplgaddtomacro\g@addto@macro
  \gdef\gplbacktext{}%
  \gdef\gplfronttext{}%
  \makeatother
  \ifGPblacktext
    \def\colorrgb#1{}%
    \def\colorgray#1{}%
  \else
    \ifGPcolor
      \def\colorrgb#1{\color[rgb]{#1}}%
      \def\colorgray#1{\color[gray]{#1}}%
      \expandafter\def\csname LTw\endcsname{\color{white}}%
      \expandafter\def\csname LTb\endcsname{\color{black}}%
      \expandafter\def\csname LTa\endcsname{\color{black}}%
      \expandafter\def\csname LT0\endcsname{\color[rgb]{1,0,0}}%
      \expandafter\def\csname LT1\endcsname{\color[rgb]{0,1,0}}%
      \expandafter\def\csname LT2\endcsname{\color[rgb]{0,0,1}}%
      \expandafter\def\csname LT3\endcsname{\color[rgb]{1,0,1}}%
      \expandafter\def\csname LT4\endcsname{\color[rgb]{0,1,1}}%
      \expandafter\def\csname LT5\endcsname{\color[rgb]{1,1,0}}%
      \expandafter\def\csname LT6\endcsname{\color[rgb]{0,0,0}}%
      \expandafter\def\csname LT7\endcsname{\color[rgb]{1,0.3,0}}%
      \expandafter\def\csname LT8\endcsname{\color[rgb]{0.5,0.5,0.5}}%
    \else
      \def\colorrgb#1{\color{black}}%
      \def\colorgray#1{\color[gray]{#1}}%
      \expandafter\def\csname LTw\endcsname{\color{white}}%
      \expandafter\def\csname LTb\endcsname{\color{black}}%
      \expandafter\def\csname LTa\endcsname{\color{black}}%
      \expandafter\def\csname LT0\endcsname{\color{black}}%
      \expandafter\def\csname LT1\endcsname{\color{black}}%
      \expandafter\def\csname LT2\endcsname{\color{black}}%
      \expandafter\def\csname LT3\endcsname{\color{black}}%
      \expandafter\def\csname LT4\endcsname{\color{black}}%
      \expandafter\def\csname LT5\endcsname{\color{black}}%
      \expandafter\def\csname LT6\endcsname{\color{black}}%
      \expandafter\def\csname LT7\endcsname{\color{black}}%
      \expandafter\def\csname LT8\endcsname{\color{black}}%
    \fi
  \fi
    \setlength{\unitlength}{0.0500bp}%
    \ifx\gptboxheight\undefined%
      \newlength{\gptboxheight}%
      \newlength{\gptboxwidth}%
      \newsavebox{\gptboxtext}%
    \fi%
    \setlength{\fboxrule}{0.5pt}%
    \setlength{\fboxsep}{1pt}%
\begin{picture}(7200.00,5040.00)%
    \gplgaddtomacro\gplbacktext{%
      \csname LTb\endcsname%
      \put(814,1156){\makebox(0,0)[r]{\strut{}$0.8$}}%
      \put(814,2061){\makebox(0,0)[r]{\strut{}$0.9$}}%
      \put(814,2966){\makebox(0,0)[r]{\strut{}$1$}}%
      \put(814,3870){\makebox(0,0)[r]{\strut{}$1.1$}}%
      \put(814,4775){\makebox(0,0)[r]{\strut{}$1.2$}}%
      \put(1212,484){\makebox(0,0){\strut{}$-1.2$}}%
      \put(2100,484){\makebox(0,0){\strut{}$-0.8$}}%
      \put(2987,484){\makebox(0,0){\strut{}$-0.4$}}%
      \put(3874,484){\makebox(0,0){\strut{}$0$}}%
      \put(4762,484){\makebox(0,0){\strut{}$0.4$}}%
      \put(5649,484){\makebox(0,0){\strut{}$0.8$}}%
      \put(6537,484){\makebox(0,0){\strut{}$1.2$}}%
    }%
    \gplgaddtomacro\gplfronttext{%
      \csname LTb\endcsname%
      \put(176,2739){\rotatebox{-270}{\makebox(0,0){\strut{}$\frac{\int \diff y\,|\bar\M(x,y)|^{2}}{\int\diff y\,\diff\bar\Phi(x,y)}$}}}%
      \put(3874,154){\makebox(0,0){\strut{}$x$}}%
      \csname LTb\endcsname%
      \put(4569,1922){\makebox(0,0)[r]{\strut{}fit uncertainty}}%
      \csname LTb\endcsname%
      \put(4569,1592){\makebox(0,0)[r]{\strut{}phase uncertainty}}%
      \csname LTb\endcsname%
      \put(4569,1262){\makebox(0,0)[r]{\strut{}central value}}%
      \csname LTb\endcsname%
      \put(4569,932){\makebox(0,0)[r]{\strut{}VES data}}%
      \csname LTb\endcsname%
      \put(814,1156){\makebox(0,0)[r]{\strut{}$0.8$}}%
      \put(814,2061){\makebox(0,0)[r]{\strut{}$0.9$}}%
      \put(814,2966){\makebox(0,0)[r]{\strut{}$1$}}%
      \put(814,3870){\makebox(0,0)[r]{\strut{}$1.1$}}%
      \put(814,4775){\makebox(0,0)[r]{\strut{}$1.2$}}%
      \put(1212,484){\makebox(0,0){\strut{}$-1.2$}}%
      \put(2100,484){\makebox(0,0){\strut{}$-0.8$}}%
      \put(2987,484){\makebox(0,0){\strut{}$-0.4$}}%
      \put(3874,484){\makebox(0,0){\strut{}$0$}}%
      \put(4762,484){\makebox(0,0){\strut{}$0.4$}}%
      \put(5649,484){\makebox(0,0){\strut{}$0.8$}}%
      \put(6537,484){\makebox(0,0){\strut{}$1.2$}}%
    }%
    \gplbacktext
    \put(0,0){\includegraphics{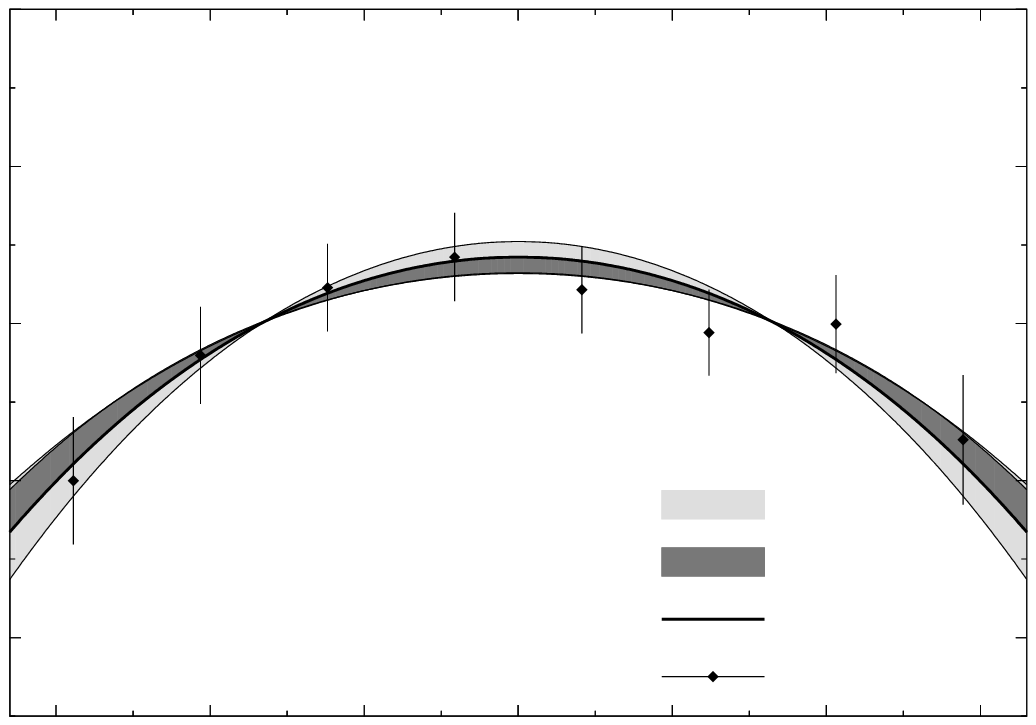}}%
    \gplfronttext
  \end{picture}%
\endgroup

%% file: plots/VES_Xproj_DR4.tex
\begingroup
  \makeatletter
  \providecommand\color[2][]{%
    \GenericError{(gnuplot) \space\space\space\@spaces}{%
      Package color not loaded in conjunction with
      terminal option `colourtext'%
    }{See the gnuplot documentation for explanation.%
    }{Either use 'blacktext' in gnuplot or load the package
      color.sty in LaTeX.}%
    \renewcommand\color[2][]{}%
  }%
  \providecommand\includegraphics[2][]{%
    \GenericError{(gnuplot) \space\space\space\@spaces}{%
      Package graphicx or graphics not loaded%
    }{See the gnuplot documentation for explanation.%
    }{The gnuplot epslatex terminal needs graphicx.sty or graphics.sty.}%
    \renewcommand\includegraphics[2][]{}%
  }%
  \providecommand\rotatebox[2]{#2}%
  \@ifundefined{ifGPcolor}{%
    \newif\ifGPcolor
    \GPcolorfalse
  }{}%
  \@ifundefined{ifGPblacktext}{%
    \newif\ifGPblacktext
    \GPblacktexttrue
  }{}%
  \let\gplgaddtomacro\g@addto@macro
  \gdef\gplbacktext{}%
  \gdef\gplfronttext{}%
  \makeatother
  \ifGPblacktext
    \def\colorrgb#1{}%
    \def\colorgray#1{}%
  \else
    \ifGPcolor
      \def\colorrgb#1{\color[rgb]{#1}}%
      \def\colorgray#1{\color[gray]{#1}}%
      \expandafter\def\csname LTw\endcsname{\color{white}}%
      \expandafter\def\csname LTb\endcsname{\color{black}}%
      \expandafter\def\csname LTa\endcsname{\color{black}}%
      \expandafter\def\csname LT0\endcsname{\color[rgb]{1,0,0}}%
      \expandafter\def\csname LT1\endcsname{\color[rgb]{0,1,0}}%
      \expandafter\def\csname LT2\endcsname{\color[rgb]{0,0,1}}%
      \expandafter\def\csname LT3\endcsname{\color[rgb]{1,0,1}}%
      \expandafter\def\csname LT4\endcsname{\color[rgb]{0,1,1}}%
      \expandafter\def\csname LT5\endcsname{\color[rgb]{1,1,0}}%
      \expandafter\def\csname LT6\endcsname{\color[rgb]{0,0,0}}%
      \expandafter\def\csname LT7\endcsname{\color[rgb]{1,0.3,0}}%
      \expandafter\def\csname LT8\endcsname{\color[rgb]{0.5,0.5,0.5}}%
    \else
      \def\colorrgb#1{\color{black}}%
      \def\colorgray#1{\color[gray]{#1}}%
      \expandafter\def\csname LTw\endcsname{\color{white}}%
      \expandafter\def\csname LTb\endcsname{\color{black}}%
      \expandafter\def\csname LTa\endcsname{\color{black}}%
      \expandafter\def\csname LT0\endcsname{\color{black}}%
      \expandafter\def\csname LT1\endcsname{\color{black}}%
      \expandafter\def\csname LT2\endcsname{\color{black}}%
      \expandafter\def\csname LT3\endcsname{\color{black}}%
      \expandafter\def\csname LT4\endcsname{\color{black}}%
      \expandafter\def\csname LT5\endcsname{\color{black}}%
      \expandafter\def\csname LT6\endcsname{\color{black}}%
      \expandafter\def\csname LT7\endcsname{\color{black}}%
      \expandafter\def\csname LT8\endcsname{\color{black}}%
    \fi
  \fi
    \setlength{\unitlength}{0.0500bp}%
    \ifx\gptboxheight\undefined%
      \newlength{\gptboxheight}%
      \newlength{\gptboxwidth}%
      \newsavebox{\gptboxtext}%
    \fi%
    \setlength{\fboxrule}{0.5pt}%
    \setlength{\fboxsep}{1pt}%
\begin{picture}(7200.00,5040.00)%
    \gplgaddtomacro\gplbacktext{%
      \csname LTb\endcsname%
      \put(814,1156){\makebox(0,0)[r]{\strut{}$0.8$}}%
      \put(814,2061){\makebox(0,0)[r]{\strut{}$0.9$}}%
      \put(814,2966){\makebox(0,0)[r]{\strut{}$1$}}%
      \put(814,3870){\makebox(0,0)[r]{\strut{}$1.1$}}%
      \put(814,4775){\makebox(0,0)[r]{\strut{}$1.2$}}%
      \put(1212,484){\makebox(0,0){\strut{}$-1.2$}}%
      \put(2100,484){\makebox(0,0){\strut{}$-0.8$}}%
      \put(2987,484){\makebox(0,0){\strut{}$-0.4$}}%
      \put(3874,484){\makebox(0,0){\strut{}$0$}}%
      \put(4762,484){\makebox(0,0){\strut{}$0.4$}}%
      \put(5649,484){\makebox(0,0){\strut{}$0.8$}}%
      \put(6537,484){\makebox(0,0){\strut{}$1.2$}}%
    }%
    \gplgaddtomacro\gplfronttext{%
      \csname LTb\endcsname%
      \put(176,2739){\rotatebox{-270}{\makebox(0,0){\strut{}$\frac{\int \diff y\,|\bar\M(x,y)|^{2}}{\int\diff y\,\diff\bar\Phi(x,y)}$}}}%
      \put(3874,154){\makebox(0,0){\strut{}$x$}}%
      \csname LTb\endcsname%
      \put(4569,1922){\makebox(0,0)[r]{\strut{}fit uncertainty}}%
      \csname LTb\endcsname%
      \put(4569,1592){\makebox(0,0)[r]{\strut{}phase uncertainty}}%
      \csname LTb\endcsname%
      \put(4569,1262){\makebox(0,0)[r]{\strut{}central value}}%
      \csname LTb\endcsname%
      \put(4569,932){\makebox(0,0)[r]{\strut{}VES data}}%
      \csname LTb\endcsname%
      \put(814,1156){\makebox(0,0)[r]{\strut{}$0.8$}}%
      \put(814,2061){\makebox(0,0)[r]{\strut{}$0.9$}}%
      \put(814,2966){\makebox(0,0)[r]{\strut{}$1$}}%
      \put(814,3870){\makebox(0,0)[r]{\strut{}$1.1$}}%
      \put(814,4775){\makebox(0,0)[r]{\strut{}$1.2$}}%
      \put(1212,484){\makebox(0,0){\strut{}$-1.2$}}%
      \put(2100,484){\makebox(0,0){\strut{}$-0.8$}}%
      \put(2987,484){\makebox(0,0){\strut{}$-0.4$}}%
      \put(3874,484){\makebox(0,0){\strut{}$0$}}%
      \put(4762,484){\makebox(0,0){\strut{}$0.4$}}%
      \put(5649,484){\makebox(0,0){\strut{}$0.8$}}%
      \put(6537,484){\makebox(0,0){\strut{}$1.2$}}%
    }%
    \gplbacktext
    \put(0,0){\includegraphics{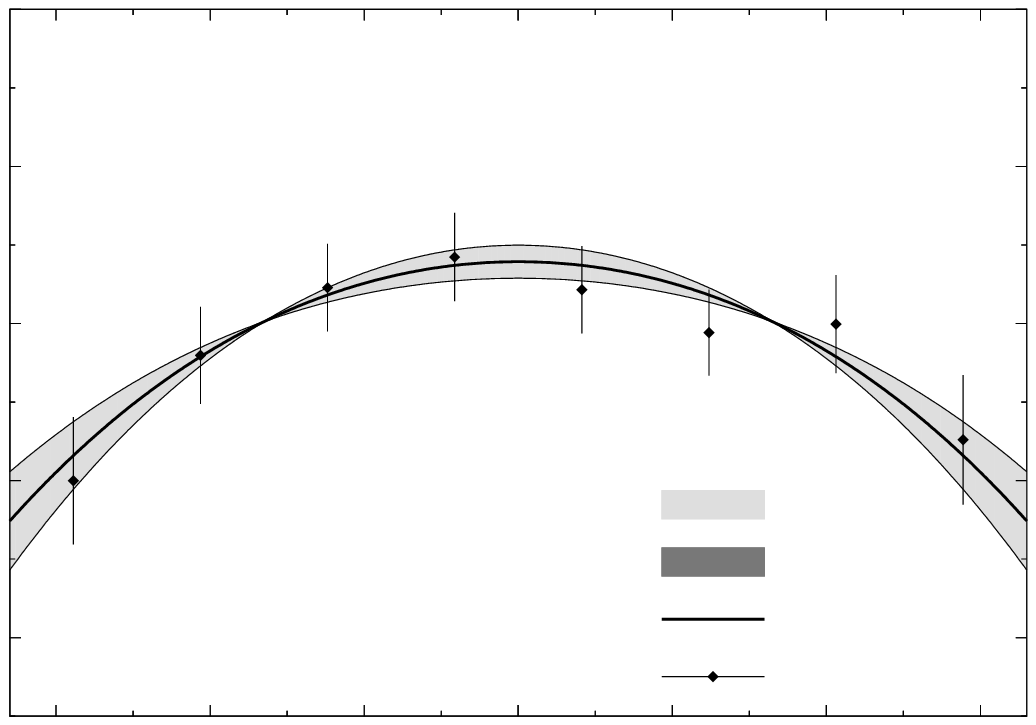}}%
    \gplfronttext
  \end{picture}%
\endgroup

%% file: plots/VES_Yproj_DR3.tex
\begingroup
  \makeatletter
  \providecommand\color[2][]{%
    \GenericError{(gnuplot) \space\space\space\@spaces}{%
      Package color not loaded in conjunction with
      terminal option `colourtext'%
    }{See the gnuplot documentation for explanation.%
    }{Either use 'blacktext' in gnuplot or load the package
      color.sty in LaTeX.}%
    \renewcommand\color[2][]{}%
  }%
  \providecommand\includegraphics[2][]{%
    \GenericError{(gnuplot) \space\space\space\@spaces}{%
      Package graphicx or graphics not loaded%
    }{See the gnuplot documentation for explanation.%
    }{The gnuplot epslatex terminal needs graphicx.sty or graphics.sty.}%
    \renewcommand\includegraphics[2][]{}%
  }%
  \providecommand\rotatebox[2]{#2}%
  \@ifundefined{ifGPcolor}{%
    \newif\ifGPcolor
    \GPcolorfalse
  }{}%
  \@ifundefined{ifGPblacktext}{%
    \newif\ifGPblacktext
    \GPblacktexttrue
  }{}%
  \let\gplgaddtomacro\g@addto@macro
  \gdef\gplbacktext{}%
  \gdef\gplfronttext{}%
  \makeatother
  \ifGPblacktext
    \def\colorrgb#1{}%
    \def\colorgray#1{}%
  \else
    \ifGPcolor
      \def\colorrgb#1{\color[rgb]{#1}}%
      \def\colorgray#1{\color[gray]{#1}}%
      \expandafter\def\csname LTw\endcsname{\color{white}}%
      \expandafter\def\csname LTb\endcsname{\color{black}}%
      \expandafter\def\csname LTa\endcsname{\color{black}}%
      \expandafter\def\csname LT0\endcsname{\color[rgb]{1,0,0}}%
      \expandafter\def\csname LT1\endcsname{\color[rgb]{0,1,0}}%
      \expandafter\def\csname LT2\endcsname{\color[rgb]{0,0,1}}%
      \expandafter\def\csname LT3\endcsname{\color[rgb]{1,0,1}}%
      \expandafter\def\csname LT4\endcsname{\color[rgb]{0,1,1}}%
      \expandafter\def\csname LT5\endcsname{\color[rgb]{1,1,0}}%
      \expandafter\def\csname LT6\endcsname{\color[rgb]{0,0,0}}%
      \expandafter\def\csname LT7\endcsname{\color[rgb]{1,0.3,0}}%
      \expandafter\def\csname LT8\endcsname{\color[rgb]{0.5,0.5,0.5}}%
    \else
      \def\colorrgb#1{\color{black}}%
      \def\colorgray#1{\color[gray]{#1}}%
      \expandafter\def\csname LTw\endcsname{\color{white}}%
      \expandafter\def\csname LTb\endcsname{\color{black}}%
      \expandafter\def\csname LTa\endcsname{\color{black}}%
      \expandafter\def\csname LT0\endcsname{\color{black}}%
      \expandafter\def\csname LT1\endcsname{\color{black}}%
      \expandafter\def\csname LT2\endcsname{\color{black}}%
      \expandafter\def\csname LT3\endcsname{\color{black}}%
      \expandafter\def\csname LT4\endcsname{\color{black}}%
      \expandafter\def\csname LT5\endcsname{\color{black}}%
      \expandafter\def\csname LT6\endcsname{\color{black}}%
      \expandafter\def\csname LT7\endcsname{\color{black}}%
      \expandafter\def\csname LT8\endcsname{\color{black}}%
    \fi
  \fi
    \setlength{\unitlength}{0.0500bp}%
    \ifx\gptboxheight\undefined%
      \newlength{\gptboxheight}%
      \newlength{\gptboxwidth}%
      \newsavebox{\gptboxtext}%
    \fi%
    \setlength{\fboxrule}{0.5pt}%
    \setlength{\fboxsep}{1pt}%
\begin{picture}(7200.00,5040.00)%
    \gplgaddtomacro\gplbacktext{%
      \csname LTb\endcsname%
      \put(814,1156){\makebox(0,0)[r]{\strut{}$0.8$}}%
      \put(814,2061){\makebox(0,0)[r]{\strut{}$0.9$}}%
      \put(814,2966){\makebox(0,0)[r]{\strut{}$1$}}%
      \put(814,3870){\makebox(0,0)[r]{\strut{}$1.1$}}%
      \put(814,4775){\makebox(0,0)[r]{\strut{}$1.2$}}%
      \put(1478,484){\makebox(0,0){\strut{}$-0.8$}}%
      \put(2599,484){\makebox(0,0){\strut{}$-0.4$}}%
      \put(3720,484){\makebox(0,0){\strut{}$0$}}%
      \put(4841,484){\makebox(0,0){\strut{}$0.4$}}%
      \put(5962,484){\makebox(0,0){\strut{}$0.8$}}%
    }%
    \gplgaddtomacro\gplfronttext{%
      \csname LTb\endcsname%
      \put(176,2739){\rotatebox{-270}{\makebox(0,0){\strut{}$\frac{\int \diff x\,|\bar\M(x,y)|^{2}}{\int\diff x\,\diff\bar\Phi(x,y)}$}}}%
      \put(3874,154){\makebox(0,0){\strut{}$y$}}%
      \csname LTb\endcsname%
      \put(4569,1922){\makebox(0,0)[r]{\strut{}fit uncertainty}}%
      \csname LTb\endcsname%
      \put(4569,1592){\makebox(0,0)[r]{\strut{}phase uncertainty}}%
      \csname LTb\endcsname%
      \put(4569,1262){\makebox(0,0)[r]{\strut{}central value}}%
      \csname LTb\endcsname%
      \put(4569,932){\makebox(0,0)[r]{\strut{}VES data}}%
      \csname LTb\endcsname%
      \put(814,1156){\makebox(0,0)[r]{\strut{}$0.8$}}%
      \put(814,2061){\makebox(0,0)[r]{\strut{}$0.9$}}%
      \put(814,2966){\makebox(0,0)[r]{\strut{}$1$}}%
      \put(814,3870){\makebox(0,0)[r]{\strut{}$1.1$}}%
      \put(814,4775){\makebox(0,0)[r]{\strut{}$1.2$}}%
      \put(1478,484){\makebox(0,0){\strut{}$-0.8$}}%
      \put(2599,484){\makebox(0,0){\strut{}$-0.4$}}%
      \put(3720,484){\makebox(0,0){\strut{}$0$}}%
      \put(4841,484){\makebox(0,0){\strut{}$0.4$}}%
      \put(5962,484){\makebox(0,0){\strut{}$0.8$}}%
    }%
    \gplbacktext
    \put(0,0){\includegraphics{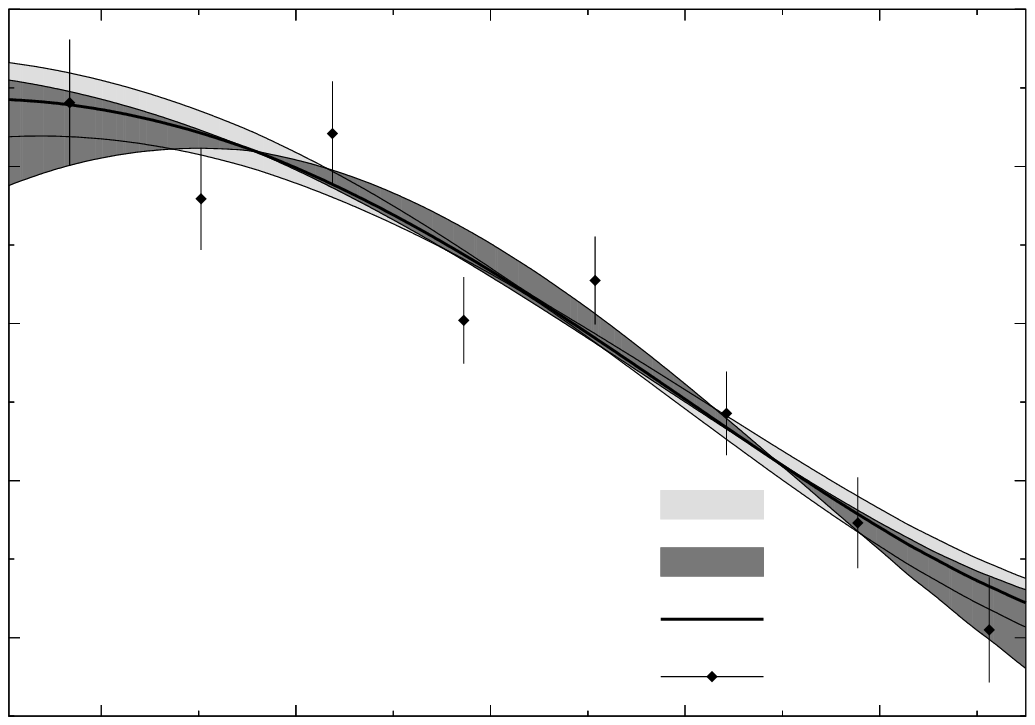}}%
    \gplfronttext
  \end{picture}%
\endgroup

%% file: plots/VES_Yproj_DR4.tex
\begingroup
  \makeatletter
  \providecommand\color[2][]{%
    \GenericError{(gnuplot) \space\space\space\@spaces}{%
      Package color not loaded in conjunction with
      terminal option `colourtext'%
    }{See the gnuplot documentation for explanation.%
    }{Either use 'blacktext' in gnuplot or load the package
      color.sty in LaTeX.}%
    \renewcommand\color[2][]{}%
  }%
  \providecommand\includegraphics[2][]{%
    \GenericError{(gnuplot) \space\space\space\@spaces}{%
      Package graphicx or graphics not loaded%
    }{See the gnuplot documentation for explanation.%
    }{The gnuplot epslatex terminal needs graphicx.sty or graphics.sty.}%
    \renewcommand\includegraphics[2][]{}%
  }%
  \providecommand\rotatebox[2]{#2}%
  \@ifundefined{ifGPcolor}{%
    \newif\ifGPcolor
    \GPcolorfalse
  }{}%
  \@ifundefined{ifGPblacktext}{%
    \newif\ifGPblacktext
    \GPblacktexttrue
  }{}%
  \let\gplgaddtomacro\g@addto@macro
  \gdef\gplbacktext{}%
  \gdef\gplfronttext{}%
  \makeatother
  \ifGPblacktext
    \def\colorrgb#1{}%
    \def\colorgray#1{}%
  \else
    \ifGPcolor
      \def\colorrgb#1{\color[rgb]{#1}}%
      \def\colorgray#1{\color[gray]{#1}}%
      \expandafter\def\csname LTw\endcsname{\color{white}}%
      \expandafter\def\csname LTb\endcsname{\color{black}}%
      \expandafter\def\csname LTa\endcsname{\color{black}}%
      \expandafter\def\csname LT0\endcsname{\color[rgb]{1,0,0}}%
      \expandafter\def\csname LT1\endcsname{\color[rgb]{0,1,0}}%
      \expandafter\def\csname LT2\endcsname{\color[rgb]{0,0,1}}%
      \expandafter\def\csname LT3\endcsname{\color[rgb]{1,0,1}}%
      \expandafter\def\csname LT4\endcsname{\color[rgb]{0,1,1}}%
      \expandafter\def\csname LT5\endcsname{\color[rgb]{1,1,0}}%
      \expandafter\def\csname LT6\endcsname{\color[rgb]{0,0,0}}%
      \expandafter\def\csname LT7\endcsname{\color[rgb]{1,0.3,0}}%
      \expandafter\def\csname LT8\endcsname{\color[rgb]{0.5,0.5,0.5}}%
    \else
      \def\colorrgb#1{\color{black}}%
      \def\colorgray#1{\color[gray]{#1}}%
      \expandafter\def\csname LTw\endcsname{\color{white}}%
      \expandafter\def\csname LTb\endcsname{\color{black}}%
      \expandafter\def\csname LTa\endcsname{\color{black}}%
      \expandafter\def\csname LT0\endcsname{\color{black}}%
      \expandafter\def\csname LT1\endcsname{\color{black}}%
      \expandafter\def\csname LT2\endcsname{\color{black}}%
      \expandafter\def\csname LT3\endcsname{\color{black}}%
      \expandafter\def\csname LT4\endcsname{\color{black}}%
      \expandafter\def\csname LT5\endcsname{\color{black}}%
      \expandafter\def\csname LT6\endcsname{\color{black}}%
      \expandafter\def\csname LT7\endcsname{\color{black}}%
      \expandafter\def\csname LT8\endcsname{\color{black}}%
    \fi
  \fi
    \setlength{\unitlength}{0.0500bp}%
    \ifx\gptboxheight\undefined%
      \newlength{\gptboxheight}%
      \newlength{\gptboxwidth}%
      \newsavebox{\gptboxtext}%
    \fi%
    \setlength{\fboxrule}{0.5pt}%
    \setlength{\fboxsep}{1pt}%
\begin{picture}(7200.00,5040.00)%
    \gplgaddtomacro\gplbacktext{%
      \csname LTb\endcsname%
      \put(814,1156){\makebox(0,0)[r]{\strut{}$0.8$}}%
      \put(814,2061){\makebox(0,0)[r]{\strut{}$0.9$}}%
      \put(814,2966){\makebox(0,0)[r]{\strut{}$1$}}%
      \put(814,3870){\makebox(0,0)[r]{\strut{}$1.1$}}%
      \put(814,4775){\makebox(0,0)[r]{\strut{}$1.2$}}%
      \put(1478,484){\makebox(0,0){\strut{}$-0.8$}}%
      \put(2599,484){\makebox(0,0){\strut{}$-0.4$}}%
      \put(3720,484){\makebox(0,0){\strut{}$0$}}%
      \put(4841,484){\makebox(0,0){\strut{}$0.4$}}%
      \put(5962,484){\makebox(0,0){\strut{}$0.8$}}%
    }%
    \gplgaddtomacro\gplfronttext{%
      \csname LTb\endcsname%
      \put(176,2739){\rotatebox{-270}{\makebox(0,0){\strut{}$\frac{\int \diff x\,|\bar\M(x,y)|^{2}}{\int\diff x\,\diff\bar\Phi(x,y)}$}}}%
      \put(3874,154){\makebox(0,0){\strut{}$y$}}%
      \csname LTb\endcsname%
      \put(4569,1922){\makebox(0,0)[r]{\strut{}fit uncertainty}}%
      \csname LTb\endcsname%
      \put(4569,1592){\makebox(0,0)[r]{\strut{}phase uncertainty}}%
      \csname LTb\endcsname%
      \put(4569,1262){\makebox(0,0)[r]{\strut{}central value}}%
      \csname LTb\endcsname%
      \put(4569,932){\makebox(0,0)[r]{\strut{}VES data}}%
      \csname LTb\endcsname%
      \put(814,1156){\makebox(0,0)[r]{\strut{}$0.8$}}%
      \put(814,2061){\makebox(0,0)[r]{\strut{}$0.9$}}%
      \put(814,2966){\makebox(0,0)[r]{\strut{}$1$}}%
      \put(814,3870){\makebox(0,0)[r]{\strut{}$1.1$}}%
      \put(814,4775){\makebox(0,0)[r]{\strut{}$1.2$}}%
      \put(1478,484){\makebox(0,0){\strut{}$-0.8$}}%
      \put(2599,484){\makebox(0,0){\strut{}$-0.4$}}%
      \put(3720,484){\makebox(0,0){\strut{}$0$}}%
      \put(4841,484){\makebox(0,0){\strut{}$0.4$}}%
      \put(5962,484){\makebox(0,0){\strut{}$0.8$}}%
    }%
    \gplbacktext
    \put(0,0){\includegraphics{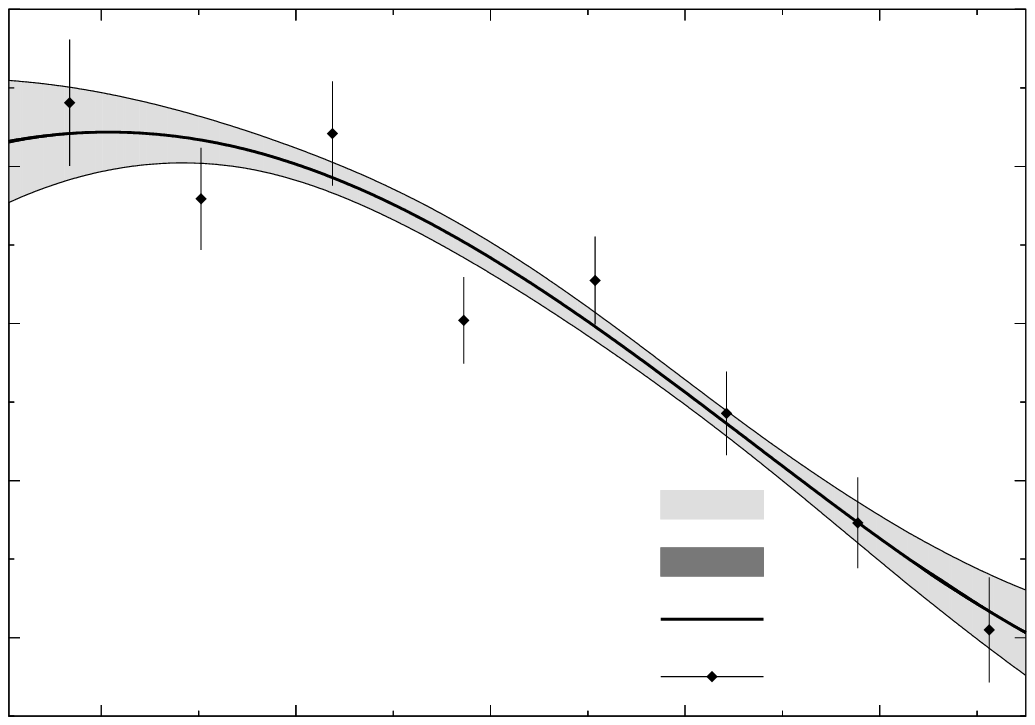}}%
    \gplfronttext
  \end{picture}%
\endgroup

%% file: plots/BES_adler_zero_DR3.tex
\begingroup
  \makeatletter
  \providecommand\color[2][]{%
    \GenericError{(gnuplot) \space\space\space\@spaces}{%
      Package color not loaded in conjunction with
      terminal option `colourtext'%
    }{See the gnuplot documentation for explanation.%
    }{Either use 'blacktext' in gnuplot or load the package
      color.sty in LaTeX.}%
    \renewcommand\color[2][]{}%
  }%
  \providecommand\includegraphics[2][]{%
    \GenericError{(gnuplot) \space\space\space\@spaces}{%
      Package graphicx or graphics not loaded%
    }{See the gnuplot documentation for explanation.%
    }{The gnuplot epslatex terminal needs graphicx.sty or graphics.sty.}%
    \renewcommand\includegraphics[2][]{}%
  }%
  \providecommand\rotatebox[2]{#2}%
  \@ifundefined{ifGPcolor}{%
    \newif\ifGPcolor
    \GPcolorfalse
  }{}%
  \@ifundefined{ifGPblacktext}{%
    \newif\ifGPblacktext
    \GPblacktexttrue
  }{}%
  \let\gplgaddtomacro\g@addto@macro
  \gdef\gplbacktext{}%
  \gdef\gplfronttext{}%
  \makeatother
  \ifGPblacktext
    \def\colorrgb#1{}%
    \def\colorgray#1{}%
  \else
    \ifGPcolor
      \def\colorrgb#1{\color[rgb]{#1}}%
      \def\colorgray#1{\color[gray]{#1}}%
      \expandafter\def\csname LTw\endcsname{\color{white}}%
      \expandafter\def\csname LTb\endcsname{\color{black}}%
      \expandafter\def\csname LTa\endcsname{\color{black}}%
      \expandafter\def\csname LT0\endcsname{\color[rgb]{1,0,0}}%
      \expandafter\def\csname LT1\endcsname{\color[rgb]{0,1,0}}%
      \expandafter\def\csname LT2\endcsname{\color[rgb]{0,0,1}}%
      \expandafter\def\csname LT3\endcsname{\color[rgb]{1,0,1}}%
      \expandafter\def\csname LT4\endcsname{\color[rgb]{0,1,1}}%
      \expandafter\def\csname LT5\endcsname{\color[rgb]{1,1,0}}%
      \expandafter\def\csname LT6\endcsname{\color[rgb]{0,0,0}}%
      \expandafter\def\csname LT7\endcsname{\color[rgb]{1,0.3,0}}%
      \expandafter\def\csname LT8\endcsname{\color[rgb]{0.5,0.5,0.5}}%
    \else
      \def\colorrgb#1{\color{black}}%
      \def\colorgray#1{\color[gray]{#1}}%
      \expandafter\def\csname LTw\endcsname{\color{white}}%
      \expandafter\def\csname LTb\endcsname{\color{black}}%
      \expandafter\def\csname LTa\endcsname{\color{black}}%
      \expandafter\def\csname LT0\endcsname{\color{black}}%
      \expandafter\def\csname LT1\endcsname{\color{black}}%
      \expandafter\def\csname LT2\endcsname{\color{black}}%
      \expandafter\def\csname LT3\endcsname{\color{black}}%
      \expandafter\def\csname LT4\endcsname{\color{black}}%
      \expandafter\def\csname LT5\endcsname{\color{black}}%
      \expandafter\def\csname LT6\endcsname{\color{black}}%
      \expandafter\def\csname LT7\endcsname{\color{black}}%
      \expandafter\def\csname LT8\endcsname{\color{black}}%
    \fi
  \fi
    \setlength{\unitlength}{0.0500bp}%
    \ifx\gptboxheight\undefined%
      \newlength{\gptboxheight}%
      \newlength{\gptboxwidth}%
      \newsavebox{\gptboxtext}%
    \fi%
    \setlength{\fboxrule}{0.5pt}%
    \setlength{\fboxsep}{1pt}%
\begin{picture}(7200.00,5040.00)%
    \gplgaddtomacro\gplbacktext{%
      \csname LTb\endcsname%
      \put(814,704){\makebox(0,0)[r]{\strut{}$-10$}}%
      \put(814,1754){\makebox(0,0)[r]{\strut{}$0$}}%
      \put(814,2804){\makebox(0,0)[r]{\strut{}$10$}}%
      \put(814,3854){\makebox(0,0)[r]{\strut{}$20$}}%
      \put(946,484){\makebox(0,0){\strut{}$-40$}}%
      \put(2410,484){\makebox(0,0){\strut{}$-20$}}%
      \put(3875,484){\makebox(0,0){\strut{}$0$}}%
      \put(5339,484){\makebox(0,0){\strut{}$20$}}%
      \put(6803,484){\makebox(0,0){\strut{}$40$}}%
      \put(1056,4599){\makebox(0,0){\strut{}$-0.75$}}%
      \put(1995,4599){\makebox(0,0){\strut{}$-0.5$}}%
      \put(2935,4599){\makebox(0,0){\strut{}$-0.25$}}%
      \put(3875,4599){\makebox(0,0){\strut{}$0$}}%
      \put(4814,4599){\makebox(0,0){\strut{}$0.25$}}%
      \put(5754,4599){\makebox(0,0){\strut{}$0.5$}}%
      \put(6693,4599){\makebox(0,0){\strut{}$0.75$}}%
    }%
    \gplgaddtomacro\gplfronttext{%
      \csname LTb\endcsname%
      \put(176,2541){\rotatebox{-270}{\makebox(0,0){\strut{}$\M(2M_{\pi}^{2},t,u)$}}}%
      \put(3874,154){\makebox(0,0){\strut{}$(t-u)$ in $M_{\pi}^{2}$}}%
      \put(3874,4928){\makebox(0,0){\strut{}$(t-u)$ in GeV$^{2}$}}%
      \csname LTb\endcsname%
      \put(4503,4151){\makebox(0,0)[r]{\strut{}$\Re\M$ uncertainty}}%
      \csname LTb\endcsname%
      \put(4503,3821){\makebox(0,0)[r]{\strut{}$\Im\M$ uncertainty}}%
      \csname LTb\endcsname%
      \put(4503,3491){\makebox(0,0)[r]{\strut{}$\Re\M$ central value}}%
      \csname LTb\endcsname%
      \put(4503,3161){\makebox(0,0)[r]{\strut{}$\Im\M$ central value}}%
      \csname LTb\endcsname%
      \put(4503,2831){\makebox(0,0)[r]{\strut{}SPT Adler zeros}}%
      \csname LTb\endcsname%
      \put(814,704){\makebox(0,0)[r]{\strut{}$-10$}}%
      \put(814,1754){\makebox(0,0)[r]{\strut{}$0$}}%
      \put(814,2804){\makebox(0,0)[r]{\strut{}$10$}}%
      \put(814,3854){\makebox(0,0)[r]{\strut{}$20$}}%
      \put(946,484){\makebox(0,0){\strut{}$-40$}}%
      \put(2410,484){\makebox(0,0){\strut{}$-20$}}%
      \put(3875,484){\makebox(0,0){\strut{}$0$}}%
      \put(5339,484){\makebox(0,0){\strut{}$20$}}%
      \put(6803,484){\makebox(0,0){\strut{}$40$}}%
      \put(1056,4599){\makebox(0,0){\strut{}$-0.75$}}%
      \put(1995,4599){\makebox(0,0){\strut{}$-0.5$}}%
      \put(2935,4599){\makebox(0,0){\strut{}$-0.25$}}%
      \put(3875,4599){\makebox(0,0){\strut{}$0$}}%
      \put(4814,4599){\makebox(0,0){\strut{}$0.25$}}%
      \put(5754,4599){\makebox(0,0){\strut{}$0.5$}}%
      \put(6693,4599){\makebox(0,0){\strut{}$0.75$}}%
    }%
    \gplbacktext
    \put(0,0){\includegraphics{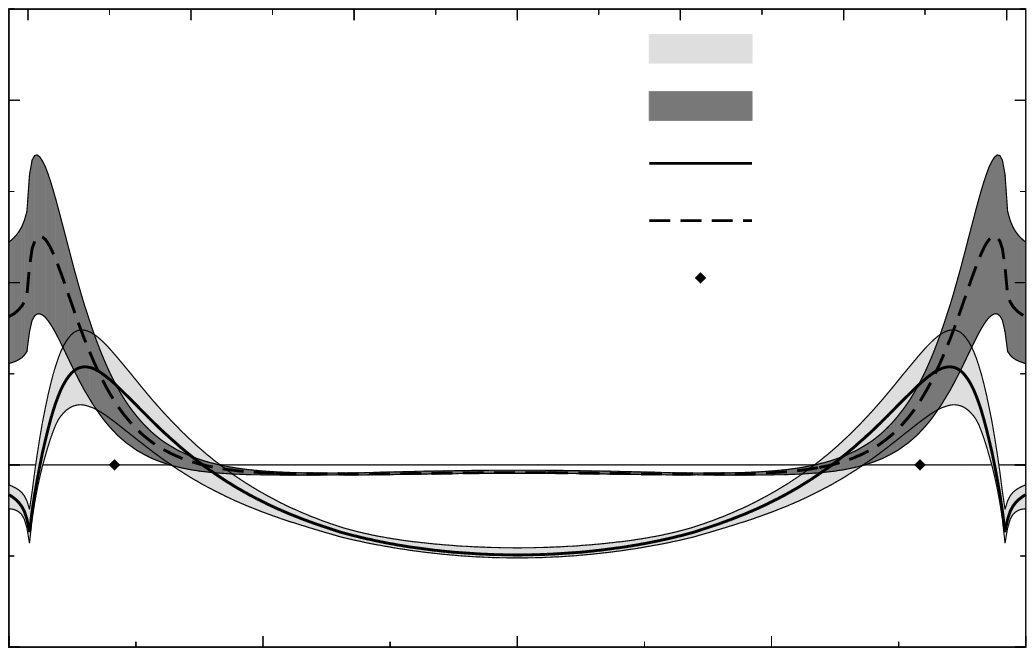}}%
    \gplfronttext
  \end{picture}%
\endgroup

%% file: plots/BES_adler_zero_DR4.tex
\begingroup
  \makeatletter
  \providecommand\color[2][]{%
    \GenericError{(gnuplot) \space\space\space\@spaces}{%
      Package color not loaded in conjunction with
      terminal option `colourtext'%
    }{See the gnuplot documentation for explanation.%
    }{Either use 'blacktext' in gnuplot or load the package
      color.sty in LaTeX.}%
    \renewcommand\color[2][]{}%
  }%
  \providecommand\includegraphics[2][]{%
    \GenericError{(gnuplot) \space\space\space\@spaces}{%
      Package graphicx or graphics not loaded%
    }{See the gnuplot documentation for explanation.%
    }{The gnuplot epslatex terminal needs graphicx.sty or graphics.sty.}%
    \renewcommand\includegraphics[2][]{}%
  }%
  \providecommand\rotatebox[2]{#2}%
  \@ifundefined{ifGPcolor}{%
    \newif\ifGPcolor
    \GPcolorfalse
  }{}%
  \@ifundefined{ifGPblacktext}{%
    \newif\ifGPblacktext
    \GPblacktexttrue
  }{}%
  \let\gplgaddtomacro\g@addto@macro
  \gdef\gplbacktext{}%
  \gdef\gplfronttext{}%
  \makeatother
  \ifGPblacktext
    \def\colorrgb#1{}%
    \def\colorgray#1{}%
  \else
    \ifGPcolor
      \def\colorrgb#1{\color[rgb]{#1}}%
      \def\colorgray#1{\color[gray]{#1}}%
      \expandafter\def\csname LTw\endcsname{\color{white}}%
      \expandafter\def\csname LTb\endcsname{\color{black}}%
      \expandafter\def\csname LTa\endcsname{\color{black}}%
      \expandafter\def\csname LT0\endcsname{\color[rgb]{1,0,0}}%
      \expandafter\def\csname LT1\endcsname{\color[rgb]{0,1,0}}%
      \expandafter\def\csname LT2\endcsname{\color[rgb]{0,0,1}}%
      \expandafter\def\csname LT3\endcsname{\color[rgb]{1,0,1}}%
      \expandafter\def\csname LT4\endcsname{\color[rgb]{0,1,1}}%
      \expandafter\def\csname LT5\endcsname{\color[rgb]{1,1,0}}%
      \expandafter\def\csname LT6\endcsname{\color[rgb]{0,0,0}}%
      \expandafter\def\csname LT7\endcsname{\color[rgb]{1,0.3,0}}%
      \expandafter\def\csname LT8\endcsname{\color[rgb]{0.5,0.5,0.5}}%
    \else
      \def\colorrgb#1{\color{black}}%
      \def\colorgray#1{\color[gray]{#1}}%
      \expandafter\def\csname LTw\endcsname{\color{white}}%
      \expandafter\def\csname LTb\endcsname{\color{black}}%
      \expandafter\def\csname LTa\endcsname{\color{black}}%
      \expandafter\def\csname LT0\endcsname{\color{black}}%
      \expandafter\def\csname LT1\endcsname{\color{black}}%
      \expandafter\def\csname LT2\endcsname{\color{black}}%
      \expandafter\def\csname LT3\endcsname{\color{black}}%
      \expandafter\def\csname LT4\endcsname{\color{black}}%
      \expandafter\def\csname LT5\endcsname{\color{black}}%
      \expandafter\def\csname LT6\endcsname{\color{black}}%
      \expandafter\def\csname LT7\endcsname{\color{black}}%
      \expandafter\def\csname LT8\endcsname{\color{black}}%
    \fi
  \fi
    \setlength{\unitlength}{0.0500bp}%
    \ifx\gptboxheight\undefined%
      \newlength{\gptboxheight}%
      \newlength{\gptboxwidth}%
      \newsavebox{\gptboxtext}%
    \fi%
    \setlength{\fboxrule}{0.5pt}%
    \setlength{\fboxsep}{1pt}%
\begin{picture}(7200.00,5040.00)%
    \gplgaddtomacro\gplbacktext{%
      \csname LTb\endcsname%
      \put(814,704){\makebox(0,0)[r]{\strut{}$-10$}}%
      \put(814,1754){\makebox(0,0)[r]{\strut{}$0$}}%
      \put(814,2804){\makebox(0,0)[r]{\strut{}$10$}}%
      \put(814,3854){\makebox(0,0)[r]{\strut{}$20$}}%
      \put(946,484){\makebox(0,0){\strut{}$-40$}}%
      \put(2410,484){\makebox(0,0){\strut{}$-20$}}%
      \put(3875,484){\makebox(0,0){\strut{}$0$}}%
      \put(5339,484){\makebox(0,0){\strut{}$20$}}%
      \put(6803,484){\makebox(0,0){\strut{}$40$}}%
      \put(1056,4599){\makebox(0,0){\strut{}$-0.75$}}%
      \put(1995,4599){\makebox(0,0){\strut{}$-0.5$}}%
      \put(2935,4599){\makebox(0,0){\strut{}$-0.25$}}%
      \put(3875,4599){\makebox(0,0){\strut{}$0$}}%
      \put(4814,4599){\makebox(0,0){\strut{}$0.25$}}%
      \put(5754,4599){\makebox(0,0){\strut{}$0.5$}}%
      \put(6693,4599){\makebox(0,0){\strut{}$0.75$}}%
    }%
    \gplgaddtomacro\gplfronttext{%
      \csname LTb\endcsname%
      \put(176,2541){\rotatebox{-270}{\makebox(0,0){\strut{}$\M(2M_{\pi}^{2},t,u)$}}}%
      \put(3874,154){\makebox(0,0){\strut{}$(t-u)$ in $M_{\pi}^{2}$}}%
      \put(3874,4928){\makebox(0,0){\strut{}$(t-u)$ in GeV$^{2}$}}%
      \csname LTb\endcsname%
      \put(4503,4151){\makebox(0,0)[r]{\strut{}$\Re\M$ uncertainty}}%
      \csname LTb\endcsname%
      \put(4503,3821){\makebox(0,0)[r]{\strut{}$\Im\M$ uncertainty}}%
      \csname LTb\endcsname%
      \put(4503,3491){\makebox(0,0)[r]{\strut{}$\Re\M$ central value}}%
      \csname LTb\endcsname%
      \put(4503,3161){\makebox(0,0)[r]{\strut{}$\Im\M$ central value}}%
      \csname LTb\endcsname%
      \put(4503,2831){\makebox(0,0)[r]{\strut{}SPT Adler zeros}}%
      \csname LTb\endcsname%
      \put(814,704){\makebox(0,0)[r]{\strut{}$-10$}}%
      \put(814,1754){\makebox(0,0)[r]{\strut{}$0$}}%
      \put(814,2804){\makebox(0,0)[r]{\strut{}$10$}}%
      \put(814,3854){\makebox(0,0)[r]{\strut{}$20$}}%
      \put(946,484){\makebox(0,0){\strut{}$-40$}}%
      \put(2410,484){\makebox(0,0){\strut{}$-20$}}%
      \put(3875,484){\makebox(0,0){\strut{}$0$}}%
      \put(5339,484){\makebox(0,0){\strut{}$20$}}%
      \put(6803,484){\makebox(0,0){\strut{}$40$}}%
      \put(1056,4599){\makebox(0,0){\strut{}$-0.75$}}%
      \put(1995,4599){\makebox(0,0){\strut{}$-0.5$}}%
      \put(2935,4599){\makebox(0,0){\strut{}$-0.25$}}%
      \put(3875,4599){\makebox(0,0){\strut{}$0$}}%
      \put(4814,4599){\makebox(0,0){\strut{}$0.25$}}%
      \put(5754,4599){\makebox(0,0){\strut{}$0.5$}}%
      \put(6693,4599){\makebox(0,0){\strut{}$0.75$}}%
    }%
    \gplbacktext
    \put(0,0){\includegraphics{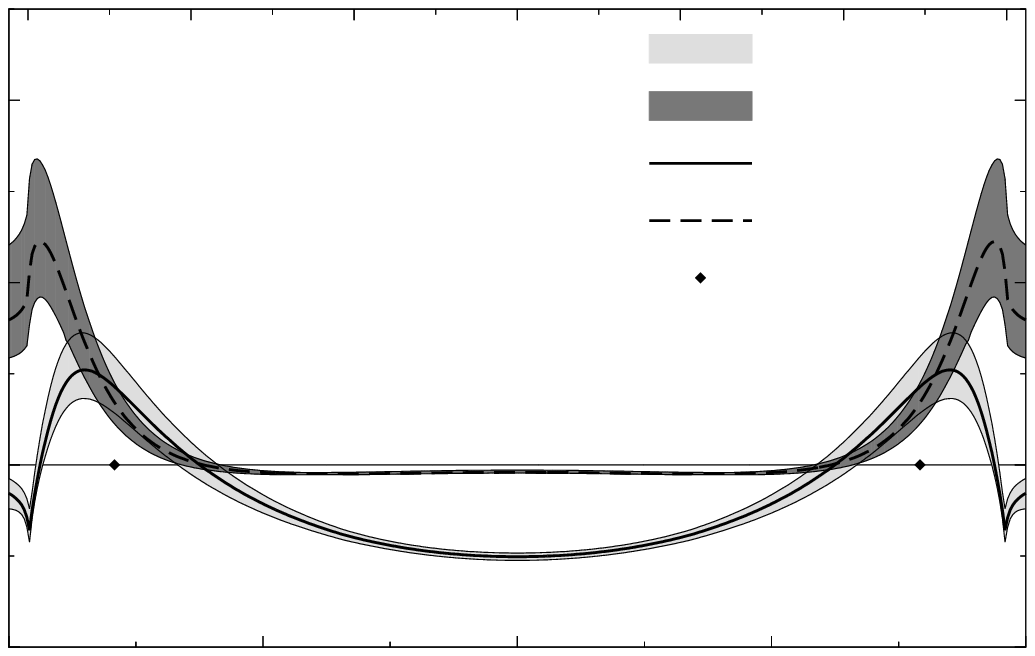}}%
    \gplfronttext
  \end{picture}%
\endgroup

%% file: plots/VES_adler_zero_DR3.tex
\begingroup
  \makeatletter
  \providecommand\color[2][]{%
    \GenericError{(gnuplot) \space\space\space\@spaces}{%
      Package color not loaded in conjunction with
      terminal option `colourtext'%
    }{See the gnuplot documentation for explanation.%
    }{Either use 'blacktext' in gnuplot or load the package
      color.sty in LaTeX.}%
    \renewcommand\color[2][]{}%
  }%
  \providecommand\includegraphics[2][]{%
    \GenericError{(gnuplot) \space\space\space\@spaces}{%
      Package graphicx or graphics not loaded%
    }{See the gnuplot documentation for explanation.%
    }{The gnuplot epslatex terminal needs graphicx.sty or graphics.sty.}%
    \renewcommand\includegraphics[2][]{}%
  }%
  \providecommand\rotatebox[2]{#2}%
  \@ifundefined{ifGPcolor}{%
    \newif\ifGPcolor
    \GPcolorfalse
  }{}%
  \@ifundefined{ifGPblacktext}{%
    \newif\ifGPblacktext
    \GPblacktexttrue
  }{}%
  \let\gplgaddtomacro\g@addto@macro
  \gdef\gplbacktext{}%
  \gdef\gplfronttext{}%
  \makeatother
  \ifGPblacktext
    \def\colorrgb#1{}%
    \def\colorgray#1{}%
  \else
    \ifGPcolor
      \def\colorrgb#1{\color[rgb]{#1}}%
      \def\colorgray#1{\color[gray]{#1}}%
      \expandafter\def\csname LTw\endcsname{\color{white}}%
      \expandafter\def\csname LTb\endcsname{\color{black}}%
      \expandafter\def\csname LTa\endcsname{\color{black}}%
      \expandafter\def\csname LT0\endcsname{\color[rgb]{1,0,0}}%
      \expandafter\def\csname LT1\endcsname{\color[rgb]{0,1,0}}%
      \expandafter\def\csname LT2\endcsname{\color[rgb]{0,0,1}}%
      \expandafter\def\csname LT3\endcsname{\color[rgb]{1,0,1}}%
      \expandafter\def\csname LT4\endcsname{\color[rgb]{0,1,1}}%
      \expandafter\def\csname LT5\endcsname{\color[rgb]{1,1,0}}%
      \expandafter\def\csname LT6\endcsname{\color[rgb]{0,0,0}}%
      \expandafter\def\csname LT7\endcsname{\color[rgb]{1,0.3,0}}%
      \expandafter\def\csname LT8\endcsname{\color[rgb]{0.5,0.5,0.5}}%
    \else
      \def\colorrgb#1{\color{black}}%
      \def\colorgray#1{\color[gray]{#1}}%
      \expandafter\def\csname LTw\endcsname{\color{white}}%
      \expandafter\def\csname LTb\endcsname{\color{black}}%
      \expandafter\def\csname LTa\endcsname{\color{black}}%
      \expandafter\def\csname LT0\endcsname{\color{black}}%
      \expandafter\def\csname LT1\endcsname{\color{black}}%
      \expandafter\def\csname LT2\endcsname{\color{black}}%
      \expandafter\def\csname LT3\endcsname{\color{black}}%
      \expandafter\def\csname LT4\endcsname{\color{black}}%
      \expandafter\def\csname LT5\endcsname{\color{black}}%
      \expandafter\def\csname LT6\endcsname{\color{black}}%
      \expandafter\def\csname LT7\endcsname{\color{black}}%
      \expandafter\def\csname LT8\endcsname{\color{black}}%
    \fi
  \fi
    \setlength{\unitlength}{0.0500bp}%
    \ifx\gptboxheight\undefined%
      \newlength{\gptboxheight}%
      \newlength{\gptboxwidth}%
      \newsavebox{\gptboxtext}%
    \fi%
    \setlength{\fboxrule}{0.5pt}%
    \setlength{\fboxsep}{1pt}%
\begin{picture}(7200.00,5040.00)%
    \gplgaddtomacro\gplbacktext{%
      \csname LTb\endcsname%
      \put(814,704){\makebox(0,0)[r]{\strut{}$-10$}}%
      \put(814,1754){\makebox(0,0)[r]{\strut{}$0$}}%
      \put(814,2804){\makebox(0,0)[r]{\strut{}$10$}}%
      \put(814,3854){\makebox(0,0)[r]{\strut{}$20$}}%
      \put(946,484){\makebox(0,0){\strut{}$-40$}}%
      \put(2410,484){\makebox(0,0){\strut{}$-20$}}%
      \put(3875,484){\makebox(0,0){\strut{}$0$}}%
      \put(5339,484){\makebox(0,0){\strut{}$20$}}%
      \put(6803,484){\makebox(0,0){\strut{}$40$}}%
      \put(1056,4599){\makebox(0,0){\strut{}$-0.75$}}%
      \put(1995,4599){\makebox(0,0){\strut{}$-0.5$}}%
      \put(2935,4599){\makebox(0,0){\strut{}$-0.25$}}%
      \put(3875,4599){\makebox(0,0){\strut{}$0$}}%
      \put(4814,4599){\makebox(0,0){\strut{}$0.25$}}%
      \put(5754,4599){\makebox(0,0){\strut{}$0.5$}}%
      \put(6693,4599){\makebox(0,0){\strut{}$0.75$}}%
    }%
    \gplgaddtomacro\gplfronttext{%
      \csname LTb\endcsname%
      \put(176,2541){\rotatebox{-270}{\makebox(0,0){\strut{}$\M(2M_{\pi}^{2},t,u)$}}}%
      \put(3874,154){\makebox(0,0){\strut{}$(t-u)$ in $M_{\pi}^{2}$}}%
      \put(3874,4928){\makebox(0,0){\strut{}$(t-u)$ in GeV$^{2}$}}%
      \csname LTb\endcsname%
      \put(4503,4151){\makebox(0,0)[r]{\strut{}$\Re\M$ uncertainty}}%
      \csname LTb\endcsname%
      \put(4503,3821){\makebox(0,0)[r]{\strut{}$\Im\M$ uncertainty}}%
      \csname LTb\endcsname%
      \put(4503,3491){\makebox(0,0)[r]{\strut{}$\Re\M$ central value}}%
      \csname LTb\endcsname%
      \put(4503,3161){\makebox(0,0)[r]{\strut{}$\Im\M$ central value}}%
      \csname LTb\endcsname%
      \put(4503,2831){\makebox(0,0)[r]{\strut{}SPT Adler zeros}}%
      \csname LTb\endcsname%
      \put(814,704){\makebox(0,0)[r]{\strut{}$-10$}}%
      \put(814,1754){\makebox(0,0)[r]{\strut{}$0$}}%
      \put(814,2804){\makebox(0,0)[r]{\strut{}$10$}}%
      \put(814,3854){\makebox(0,0)[r]{\strut{}$20$}}%
      \put(946,484){\makebox(0,0){\strut{}$-40$}}%
      \put(2410,484){\makebox(0,0){\strut{}$-20$}}%
      \put(3875,484){\makebox(0,0){\strut{}$0$}}%
      \put(5339,484){\makebox(0,0){\strut{}$20$}}%
      \put(6803,484){\makebox(0,0){\strut{}$40$}}%
      \put(1056,4599){\makebox(0,0){\strut{}$-0.75$}}%
      \put(1995,4599){\makebox(0,0){\strut{}$-0.5$}}%
      \put(2935,4599){\makebox(0,0){\strut{}$-0.25$}}%
      \put(3875,4599){\makebox(0,0){\strut{}$0$}}%
      \put(4814,4599){\makebox(0,0){\strut{}$0.25$}}%
      \put(5754,4599){\makebox(0,0){\strut{}$0.5$}}%
      \put(6693,4599){\makebox(0,0){\strut{}$0.75$}}%
    }%
    \gplbacktext
    \put(0,0){\includegraphics{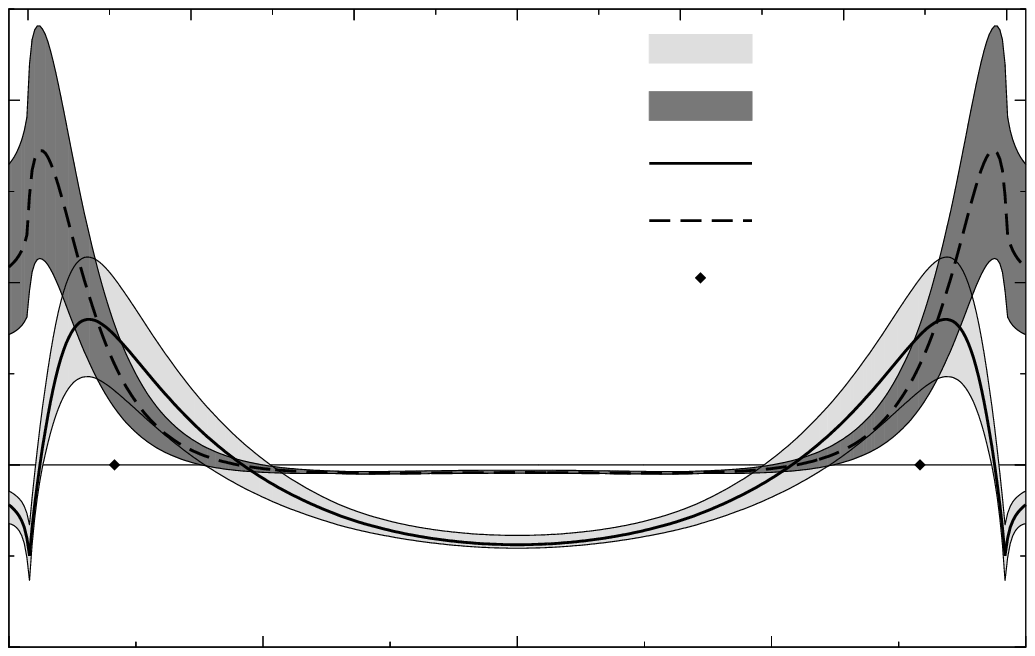}}%
    \gplfronttext
  \end{picture}%
\endgroup

%% file: plots/VES_adler_zero_DR4.tex
\begingroup
  \makeatletter
  \providecommand\color[2][]{%
    \GenericError{(gnuplot) \space\space\space\@spaces}{%
      Package color not loaded in conjunction with
      terminal option `colourtext'%
    }{See the gnuplot documentation for explanation.%
    }{Either use 'blacktext' in gnuplot or load the package
      color.sty in LaTeX.}%
    \renewcommand\color[2][]{}%
  }%
  \providecommand\includegraphics[2][]{%
    \GenericError{(gnuplot) \space\space\space\@spaces}{%
      Package graphicx or graphics not loaded%
    }{See the gnuplot documentation for explanation.%
    }{The gnuplot epslatex terminal needs graphicx.sty or graphics.sty.}%
    \renewcommand\includegraphics[2][]{}%
  }%
  \providecommand\rotatebox[2]{#2}%
  \@ifundefined{ifGPcolor}{%
    \newif\ifGPcolor
    \GPcolorfalse
  }{}%
  \@ifundefined{ifGPblacktext}{%
    \newif\ifGPblacktext
    \GPblacktexttrue
  }{}%
  \let\gplgaddtomacro\g@addto@macro
  \gdef\gplbacktext{}%
  \gdef\gplfronttext{}%
  \makeatother
  \ifGPblacktext
    \def\colorrgb#1{}%
    \def\colorgray#1{}%
  \else
    \ifGPcolor
      \def\colorrgb#1{\color[rgb]{#1}}%
      \def\colorgray#1{\color[gray]{#1}}%
      \expandafter\def\csname LTw\endcsname{\color{white}}%
      \expandafter\def\csname LTb\endcsname{\color{black}}%
      \expandafter\def\csname LTa\endcsname{\color{black}}%
      \expandafter\def\csname LT0\endcsname{\color[rgb]{1,0,0}}%
      \expandafter\def\csname LT1\endcsname{\color[rgb]{0,1,0}}%
      \expandafter\def\csname LT2\endcsname{\color[rgb]{0,0,1}}%
      \expandafter\def\csname LT3\endcsname{\color[rgb]{1,0,1}}%
      \expandafter\def\csname LT4\endcsname{\color[rgb]{0,1,1}}%
      \expandafter\def\csname LT5\endcsname{\color[rgb]{1,1,0}}%
      \expandafter\def\csname LT6\endcsname{\color[rgb]{0,0,0}}%
      \expandafter\def\csname LT7\endcsname{\color[rgb]{1,0.3,0}}%
      \expandafter\def\csname LT8\endcsname{\color[rgb]{0.5,0.5,0.5}}%
    \else
      \def\colorrgb#1{\color{black}}%
      \def\colorgray#1{\color[gray]{#1}}%
      \expandafter\def\csname LTw\endcsname{\color{white}}%
      \expandafter\def\csname LTb\endcsname{\color{black}}%
      \expandafter\def\csname LTa\endcsname{\color{black}}%
      \expandafter\def\csname LT0\endcsname{\color{black}}%
      \expandafter\def\csname LT1\endcsname{\color{black}}%
      \expandafter\def\csname LT2\endcsname{\color{black}}%
      \expandafter\def\csname LT3\endcsname{\color{black}}%
      \expandafter\def\csname LT4\endcsname{\color{black}}%
      \expandafter\def\csname LT5\endcsname{\color{black}}%
      \expandafter\def\csname LT6\endcsname{\color{black}}%
      \expandafter\def\csname LT7\endcsname{\color{black}}%
      \expandafter\def\csname LT8\endcsname{\color{black}}%
    \fi
  \fi
    \setlength{\unitlength}{0.0500bp}%
    \ifx\gptboxheight\undefined%
      \newlength{\gptboxheight}%
      \newlength{\gptboxwidth}%
      \newsavebox{\gptboxtext}%
    \fi%
    \setlength{\fboxrule}{0.5pt}%
    \setlength{\fboxsep}{1pt}%
\begin{picture}(7200.00,5040.00)%
    \gplgaddtomacro\gplbacktext{%
      \csname LTb\endcsname%
      \put(814,704){\makebox(0,0)[r]{\strut{}$-10$}}%
      \put(814,1754){\makebox(0,0)[r]{\strut{}$0$}}%
      \put(814,2804){\makebox(0,0)[r]{\strut{}$10$}}%
      \put(814,3854){\makebox(0,0)[r]{\strut{}$20$}}%
      \put(946,484){\makebox(0,0){\strut{}$-40$}}%
      \put(2410,484){\makebox(0,0){\strut{}$-20$}}%
      \put(3875,484){\makebox(0,0){\strut{}$0$}}%
      \put(5339,484){\makebox(0,0){\strut{}$20$}}%
      \put(6803,484){\makebox(0,0){\strut{}$40$}}%
      \put(1056,4599){\makebox(0,0){\strut{}$-0.75$}}%
      \put(1995,4599){\makebox(0,0){\strut{}$-0.5$}}%
      \put(2935,4599){\makebox(0,0){\strut{}$-0.25$}}%
      \put(3875,4599){\makebox(0,0){\strut{}$0$}}%
      \put(4814,4599){\makebox(0,0){\strut{}$0.25$}}%
      \put(5754,4599){\makebox(0,0){\strut{}$0.5$}}%
      \put(6693,4599){\makebox(0,0){\strut{}$0.75$}}%
    }%
    \gplgaddtomacro\gplfronttext{%
      \csname LTb\endcsname%
      \put(176,2541){\rotatebox{-270}{\makebox(0,0){\strut{}$\M(2M_{\pi}^{2},t,u)$}}}%
      \put(3874,154){\makebox(0,0){\strut{}$(t-u)$ in $M_{\pi}^{2}$}}%
      \put(3874,4928){\makebox(0,0){\strut{}$(t-u)$ in GeV$^{2}$}}%
      \csname LTb\endcsname%
      \put(4503,4151){\makebox(0,0)[r]{\strut{}$\Re\M$ uncertainty}}%
      \csname LTb\endcsname%
      \put(4503,3821){\makebox(0,0)[r]{\strut{}$\Im\M$ uncertainty}}%
      \csname LTb\endcsname%
      \put(4503,3491){\makebox(0,0)[r]{\strut{}$\Re\M$ central value}}%
      \csname LTb\endcsname%
      \put(4503,3161){\makebox(0,0)[r]{\strut{}$\Im\M$ central value}}%
      \csname LTb\endcsname%
      \put(4503,2831){\makebox(0,0)[r]{\strut{}SPT Adler zeros}}%
      \csname LTb\endcsname%
      \put(814,704){\makebox(0,0)[r]{\strut{}$-10$}}%
      \put(814,1754){\makebox(0,0)[r]{\strut{}$0$}}%
      \put(814,2804){\makebox(0,0)[r]{\strut{}$10$}}%
      \put(814,3854){\makebox(0,0)[r]{\strut{}$20$}}%
      \put(946,484){\makebox(0,0){\strut{}$-40$}}%
      \put(2410,484){\makebox(0,0){\strut{}$-20$}}%
      \put(3875,484){\makebox(0,0){\strut{}$0$}}%
      \put(5339,484){\makebox(0,0){\strut{}$20$}}%
      \put(6803,484){\makebox(0,0){\strut{}$40$}}%
      \put(1056,4599){\makebox(0,0){\strut{}$-0.75$}}%
      \put(1995,4599){\makebox(0,0){\strut{}$-0.5$}}%
      \put(2935,4599){\makebox(0,0){\strut{}$-0.25$}}%
      \put(3875,4599){\makebox(0,0){\strut{}$0$}}%
      \put(4814,4599){\makebox(0,0){\strut{}$0.25$}}%
      \put(5754,4599){\makebox(0,0){\strut{}$0.5$}}%
      \put(6693,4599){\makebox(0,0){\strut{}$0.75$}}%
    }%
    \gplbacktext
    \put(0,0){\includegraphics{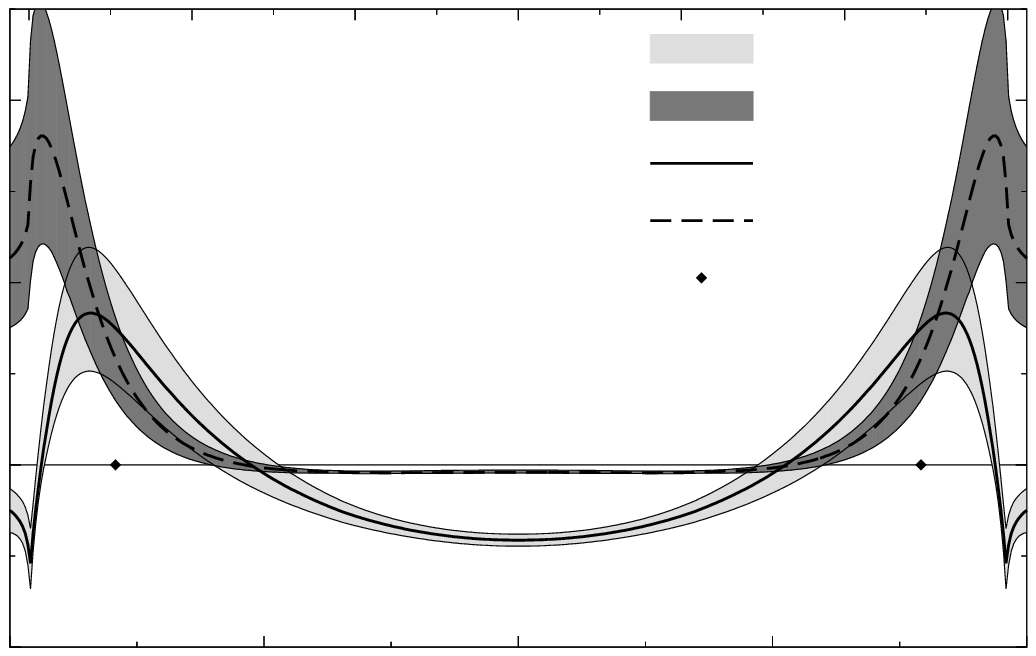}}%
    \gplfronttext
  \end{picture}%
\endgroup

%% file: plots/BES_Yproj_neutral_DR3.tex
\begingroup
  \makeatletter
  \providecommand\color[2][]{%
    \GenericError{(gnuplot) \space\space\space\@spaces}{%
      Package color not loaded in conjunction with
      terminal option `colourtext'%
    }{See the gnuplot documentation for explanation.%
    }{Either use 'blacktext' in gnuplot or load the package
      color.sty in LaTeX.}%
    \renewcommand\color[2][]{}%
  }%
  \providecommand\includegraphics[2][]{%
    \GenericError{(gnuplot) \space\space\space\@spaces}{%
      Package graphicx or graphics not loaded%
    }{See the gnuplot documentation for explanation.%
    }{The gnuplot epslatex terminal needs graphicx.sty or graphics.sty.}%
    \renewcommand\includegraphics[2][]{}%
  }%
  \providecommand\rotatebox[2]{#2}%
  \@ifundefined{ifGPcolor}{%
    \newif\ifGPcolor
    \GPcolorfalse
  }{}%
  \@ifundefined{ifGPblacktext}{%
    \newif\ifGPblacktext
    \GPblacktexttrue
  }{}%
  \let\gplgaddtomacro\g@addto@macro
  \gdef\gplbacktext{}%
  \gdef\gplfronttext{}%
  \makeatother
  \ifGPblacktext
    \def\colorrgb#1{}%
    \def\colorgray#1{}%
  \else
    \ifGPcolor
      \def\colorrgb#1{\color[rgb]{#1}}%
      \def\colorgray#1{\color[gray]{#1}}%
      \expandafter\def\csname LTw\endcsname{\color{white}}%
      \expandafter\def\csname LTb\endcsname{\color{black}}%
      \expandafter\def\csname LTa\endcsname{\color{black}}%
      \expandafter\def\csname LT0\endcsname{\color[rgb]{1,0,0}}%
      \expandafter\def\csname LT1\endcsname{\color[rgb]{0,1,0}}%
      \expandafter\def\csname LT2\endcsname{\color[rgb]{0,0,1}}%
      \expandafter\def\csname LT3\endcsname{\color[rgb]{1,0,1}}%
      \expandafter\def\csname LT4\endcsname{\color[rgb]{0,1,1}}%
      \expandafter\def\csname LT5\endcsname{\color[rgb]{1,1,0}}%
      \expandafter\def\csname LT6\endcsname{\color[rgb]{0,0,0}}%
      \expandafter\def\csname LT7\endcsname{\color[rgb]{1,0.3,0}}%
      \expandafter\def\csname LT8\endcsname{\color[rgb]{0.5,0.5,0.5}}%
    \else
      \def\colorrgb#1{\color{black}}%
      \def\colorgray#1{\color[gray]{#1}}%
      \expandafter\def\csname LTw\endcsname{\color{white}}%
      \expandafter\def\csname LTb\endcsname{\color{black}}%
      \expandafter\def\csname LTa\endcsname{\color{black}}%
      \expandafter\def\csname LT0\endcsname{\color{black}}%
      \expandafter\def\csname LT1\endcsname{\color{black}}%
      \expandafter\def\csname LT2\endcsname{\color{black}}%
      \expandafter\def\csname LT3\endcsname{\color{black}}%
      \expandafter\def\csname LT4\endcsname{\color{black}}%
      \expandafter\def\csname LT5\endcsname{\color{black}}%
      \expandafter\def\csname LT6\endcsname{\color{black}}%
      \expandafter\def\csname LT7\endcsname{\color{black}}%
      \expandafter\def\csname LT8\endcsname{\color{black}}%
    \fi
  \fi
    \setlength{\unitlength}{0.0500bp}%
    \ifx\gptboxheight\undefined%
      \newlength{\gptboxheight}%
      \newlength{\gptboxwidth}%
      \newsavebox{\gptboxtext}%
    \fi%
    \setlength{\fboxrule}{0.5pt}%
    \setlength{\fboxsep}{1pt}%
\begin{picture}(7200.00,5040.00)%
    \gplgaddtomacro\gplbacktext{%
      \csname LTb\endcsname%
      \put(814,1156){\makebox(0,0)[r]{\strut{}$0.8$}}%
      \put(814,2061){\makebox(0,0)[r]{\strut{}$0.9$}}%
      \put(814,2966){\makebox(0,0)[r]{\strut{}$1$}}%
      \put(814,3870){\makebox(0,0)[r]{\strut{}$1.1$}}%
      \put(814,4775){\makebox(0,0)[r]{\strut{}$1.2$}}%
      \put(1478,484){\makebox(0,0){\strut{}$-0.8$}}%
      \put(2599,484){\makebox(0,0){\strut{}$-0.4$}}%
      \put(3720,484){\makebox(0,0){\strut{}$0$}}%
      \put(4841,484){\makebox(0,0){\strut{}$0.4$}}%
      \put(5962,484){\makebox(0,0){\strut{}$0.8$}}%
    }%
    \gplgaddtomacro\gplfronttext{%
      \csname LTb\endcsname%
      \put(176,2739){\rotatebox{-270}{\makebox(0,0){\strut{}$\frac{\int \diff x\,|\bar\M(x,y)|^{2}}{\int\diff x\,\diff\bar\Phi(x,y)}$}}}%
      \put(3874,154){\makebox(0,0){\strut{}$y$}}%
      \csname LTb\endcsname%
      \put(4569,1592){\makebox(0,0)[r]{\strut{}fit uncertainty}}%
      \csname LTb\endcsname%
      \put(4569,1262){\makebox(0,0)[r]{\strut{}phase uncertainty}}%
      \csname LTb\endcsname%
      \put(4569,932){\makebox(0,0)[r]{\strut{}central value}}%
      \csname LTb\endcsname%
      \put(814,1156){\makebox(0,0)[r]{\strut{}$0.8$}}%
      \put(814,2061){\makebox(0,0)[r]{\strut{}$0.9$}}%
      \put(814,2966){\makebox(0,0)[r]{\strut{}$1$}}%
      \put(814,3870){\makebox(0,0)[r]{\strut{}$1.1$}}%
      \put(814,4775){\makebox(0,0)[r]{\strut{}$1.2$}}%
      \put(1478,484){\makebox(0,0){\strut{}$-0.8$}}%
      \put(2599,484){\makebox(0,0){\strut{}$-0.4$}}%
      \put(3720,484){\makebox(0,0){\strut{}$0$}}%
      \put(4841,484){\makebox(0,0){\strut{}$0.4$}}%
      \put(5962,484){\makebox(0,0){\strut{}$0.8$}}%
    }%
    \gplbacktext
    \put(0,0){\includegraphics{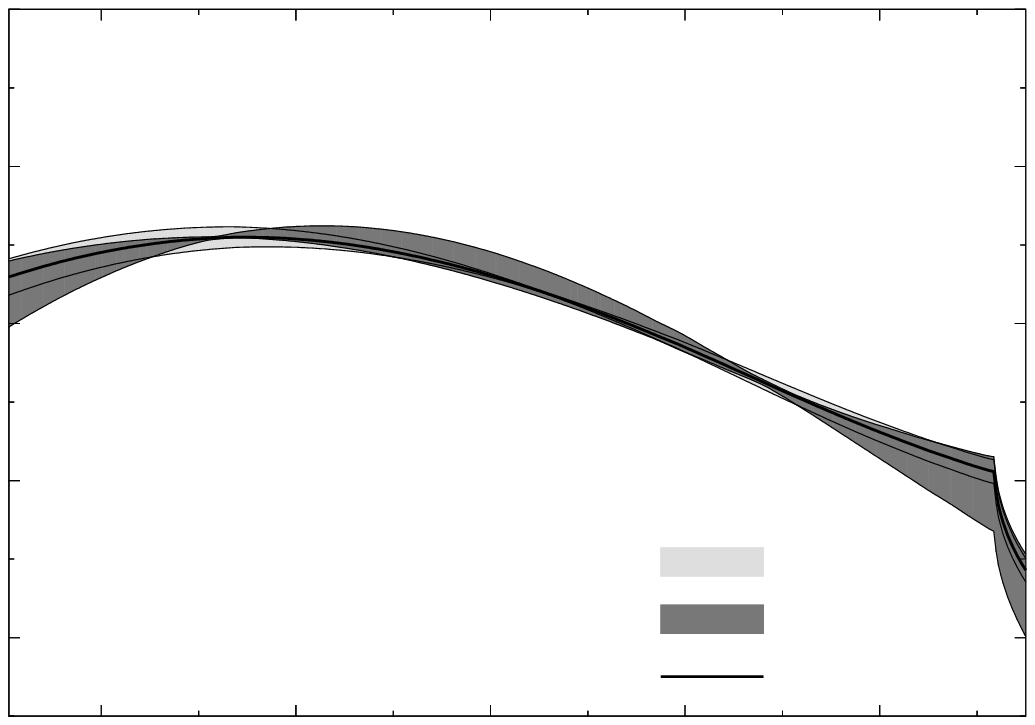}}%
    \gplfronttext
  \end{picture}%
\endgroup

%% file: plots/BES_Yproj_neutral_DR4.tex
\begingroup
  \makeatletter
  \providecommand\color[2][]{%
    \GenericError{(gnuplot) \space\space\space\@spaces}{%
      Package color not loaded in conjunction with
      terminal option `colourtext'%
    }{See the gnuplot documentation for explanation.%
    }{Either use 'blacktext' in gnuplot or load the package
      color.sty in LaTeX.}%
    \renewcommand\color[2][]{}%
  }%
  \providecommand\includegraphics[2][]{%
    \GenericError{(gnuplot) \space\space\space\@spaces}{%
      Package graphicx or graphics not loaded%
    }{See the gnuplot documentation for explanation.%
    }{The gnuplot epslatex terminal needs graphicx.sty or graphics.sty.}%
    \renewcommand\includegraphics[2][]{}%
  }%
  \providecommand\rotatebox[2]{#2}%
  \@ifundefined{ifGPcolor}{%
    \newif\ifGPcolor
    \GPcolorfalse
  }{}%
  \@ifundefined{ifGPblacktext}{%
    \newif\ifGPblacktext
    \GPblacktexttrue
  }{}%
  \let\gplgaddtomacro\g@addto@macro
  \gdef\gplbacktext{}%
  \gdef\gplfronttext{}%
  \makeatother
  \ifGPblacktext
    \def\colorrgb#1{}%
    \def\colorgray#1{}%
  \else
    \ifGPcolor
      \def\colorrgb#1{\color[rgb]{#1}}%
      \def\colorgray#1{\color[gray]{#1}}%
      \expandafter\def\csname LTw\endcsname{\color{white}}%
      \expandafter\def\csname LTb\endcsname{\color{black}}%
      \expandafter\def\csname LTa\endcsname{\color{black}}%
      \expandafter\def\csname LT0\endcsname{\color[rgb]{1,0,0}}%
      \expandafter\def\csname LT1\endcsname{\color[rgb]{0,1,0}}%
      \expandafter\def\csname LT2\endcsname{\color[rgb]{0,0,1}}%
      \expandafter\def\csname LT3\endcsname{\color[rgb]{1,0,1}}%
      \expandafter\def\csname LT4\endcsname{\color[rgb]{0,1,1}}%
      \expandafter\def\csname LT5\endcsname{\color[rgb]{1,1,0}}%
      \expandafter\def\csname LT6\endcsname{\color[rgb]{0,0,0}}%
      \expandafter\def\csname LT7\endcsname{\color[rgb]{1,0.3,0}}%
      \expandafter\def\csname LT8\endcsname{\color[rgb]{0.5,0.5,0.5}}%
    \else
      \def\colorrgb#1{\color{black}}%
      \def\colorgray#1{\color[gray]{#1}}%
      \expandafter\def\csname LTw\endcsname{\color{white}}%
      \expandafter\def\csname LTb\endcsname{\color{black}}%
      \expandafter\def\csname LTa\endcsname{\color{black}}%
      \expandafter\def\csname LT0\endcsname{\color{black}}%
      \expandafter\def\csname LT1\endcsname{\color{black}}%
      \expandafter\def\csname LT2\endcsname{\color{black}}%
      \expandafter\def\csname LT3\endcsname{\color{black}}%
      \expandafter\def\csname LT4\endcsname{\color{black}}%
      \expandafter\def\csname LT5\endcsname{\color{black}}%
      \expandafter\def\csname LT6\endcsname{\color{black}}%
      \expandafter\def\csname LT7\endcsname{\color{black}}%
      \expandafter\def\csname LT8\endcsname{\color{black}}%
    \fi
  \fi
    \setlength{\unitlength}{0.0500bp}%
    \ifx\gptboxheight\undefined%
      \newlength{\gptboxheight}%
      \newlength{\gptboxwidth}%
      \newsavebox{\gptboxtext}%
    \fi%
    \setlength{\fboxrule}{0.5pt}%
    \setlength{\fboxsep}{1pt}%
\begin{picture}(7200.00,5040.00)%
    \gplgaddtomacro\gplbacktext{%
      \csname LTb\endcsname%
      \put(814,1156){\makebox(0,0)[r]{\strut{}$0.8$}}%
      \put(814,2061){\makebox(0,0)[r]{\strut{}$0.9$}}%
      \put(814,2966){\makebox(0,0)[r]{\strut{}$1$}}%
      \put(814,3870){\makebox(0,0)[r]{\strut{}$1.1$}}%
      \put(814,4775){\makebox(0,0)[r]{\strut{}$1.2$}}%
      \put(1478,484){\makebox(0,0){\strut{}$-0.8$}}%
      \put(2599,484){\makebox(0,0){\strut{}$-0.4$}}%
      \put(3720,484){\makebox(0,0){\strut{}$0$}}%
      \put(4841,484){\makebox(0,0){\strut{}$0.4$}}%
      \put(5962,484){\makebox(0,0){\strut{}$0.8$}}%
    }%
    \gplgaddtomacro\gplfronttext{%
      \csname LTb\endcsname%
      \put(176,2739){\rotatebox{-270}{\makebox(0,0){\strut{}$\frac{\int \diff x\,|\bar\M(x,y)|^{2}}{\int\diff x\,\diff\bar\Phi(x,y)}$}}}%
      \put(3874,154){\makebox(0,0){\strut{}$y$}}%
      \csname LTb\endcsname%
      \put(4569,1592){\makebox(0,0)[r]{\strut{}fit uncertainty}}%
      \csname LTb\endcsname%
      \put(4569,1262){\makebox(0,0)[r]{\strut{}phase uncertainty}}%
      \csname LTb\endcsname%
      \put(4569,932){\makebox(0,0)[r]{\strut{}central value}}%
      \csname LTb\endcsname%
      \put(814,1156){\makebox(0,0)[r]{\strut{}$0.8$}}%
      \put(814,2061){\makebox(0,0)[r]{\strut{}$0.9$}}%
      \put(814,2966){\makebox(0,0)[r]{\strut{}$1$}}%
      \put(814,3870){\makebox(0,0)[r]{\strut{}$1.1$}}%
      \put(814,4775){\makebox(0,0)[r]{\strut{}$1.2$}}%
      \put(1478,484){\makebox(0,0){\strut{}$-0.8$}}%
      \put(2599,484){\makebox(0,0){\strut{}$-0.4$}}%
      \put(3720,484){\makebox(0,0){\strut{}$0$}}%
      \put(4841,484){\makebox(0,0){\strut{}$0.4$}}%
      \put(5962,484){\makebox(0,0){\strut{}$0.8$}}%
    }%
    \gplbacktext
    \put(0,0){\includegraphics{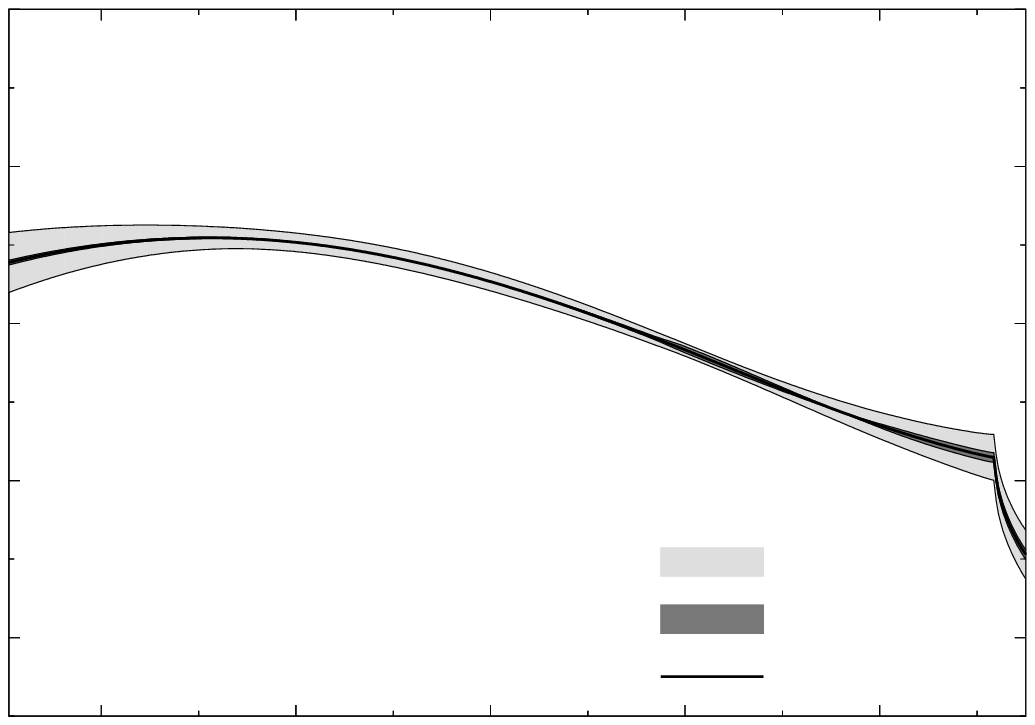}}%
    \gplfronttext
  \end{picture}%
\endgroup

%% file: plots/isobreak-phase.tex
\begingroup
  \makeatletter
  \providecommand\color[2][]{%
    \GenericError{(gnuplot) \space\space\space\@spaces}{%
      Package color not loaded in conjunction with
      terminal option `colourtext'%
    }{See the gnuplot documentation for explanation.%
    }{Either use 'blacktext' in gnuplot or load the package
      color.sty in LaTeX.}%
    \renewcommand\color[2][]{}%
  }%
  \providecommand\includegraphics[2][]{%
    \GenericError{(gnuplot) \space\space\space\@spaces}{%
      Package graphicx or graphics not loaded%
    }{See the gnuplot documentation for explanation.%
    }{The gnuplot epslatex terminal needs graphicx.sty or graphics.sty.}%
    \renewcommand\includegraphics[2][]{}%
  }%
  \providecommand\rotatebox[2]{#2}%
  \@ifundefined{ifGPcolor}{%
    \newif\ifGPcolor
    \GPcolorfalse
  }{}%
  \@ifundefined{ifGPblacktext}{%
    \newif\ifGPblacktext
    \GPblacktexttrue
  }{}%
  \let\gplgaddtomacro\g@addto@macro
  \gdef\gplbacktext{}%
  \gdef\gplfronttext{}%
  \makeatother
  \ifGPblacktext
    \def\colorrgb#1{}%
    \def\colorgray#1{}%
  \else
    \ifGPcolor
      \def\colorrgb#1{\color[rgb]{#1}}%
      \def\colorgray#1{\color[gray]{#1}}%
      \expandafter\def\csname LTw\endcsname{\color{white}}%
      \expandafter\def\csname LTb\endcsname{\color{black}}%
      \expandafter\def\csname LTa\endcsname{\color{black}}%
      \expandafter\def\csname LT0\endcsname{\color[rgb]{1,0,0}}%
      \expandafter\def\csname LT1\endcsname{\color[rgb]{0,1,0}}%
      \expandafter\def\csname LT2\endcsname{\color[rgb]{0,0,1}}%
      \expandafter\def\csname LT3\endcsname{\color[rgb]{1,0,1}}%
      \expandafter\def\csname LT4\endcsname{\color[rgb]{0,1,1}}%
      \expandafter\def\csname LT5\endcsname{\color[rgb]{1,1,0}}%
      \expandafter\def\csname LT6\endcsname{\color[rgb]{0,0,0}}%
      \expandafter\def\csname LT7\endcsname{\color[rgb]{1,0.3,0}}%
      \expandafter\def\csname LT8\endcsname{\color[rgb]{0.5,0.5,0.5}}%
    \else
      \def\colorrgb#1{\color{black}}%
      \def\colorgray#1{\color[gray]{#1}}%
      \expandafter\def\csname LTw\endcsname{\color{white}}%
      \expandafter\def\csname LTb\endcsname{\color{black}}%
      \expandafter\def\csname LTa\endcsname{\color{black}}%
      \expandafter\def\csname LT0\endcsname{\color{black}}%
      \expandafter\def\csname LT1\endcsname{\color{black}}%
      \expandafter\def\csname LT2\endcsname{\color{black}}%
      \expandafter\def\csname LT3\endcsname{\color{black}}%
      \expandafter\def\csname LT4\endcsname{\color{black}}%
      \expandafter\def\csname LT5\endcsname{\color{black}}%
      \expandafter\def\csname LT6\endcsname{\color{black}}%
      \expandafter\def\csname LT7\endcsname{\color{black}}%
      \expandafter\def\csname LT8\endcsname{\color{black}}%
    \fi
  \fi
    \setlength{\unitlength}{0.0500bp}%
    \ifx\gptboxheight\undefined%
      \newlength{\gptboxheight}%
      \newlength{\gptboxwidth}%
      \newsavebox{\gptboxtext}%
    \fi%
    \setlength{\fboxrule}{0.5pt}%
    \setlength{\fboxsep}{1pt}%
\begin{picture}(7200.00,5040.00)%
    \gplgaddtomacro\gplbacktext{%
      \csname LTb\endcsname%
      \put(946,704){\makebox(0,0)[r]{\strut{}$0$}}%
      \put(946,1439){\makebox(0,0)[r]{\strut{}$0.02$}}%
      \put(946,2174){\makebox(0,0)[r]{\strut{}$0.04$}}%
      \put(946,2909){\makebox(0,0)[r]{\strut{}$0.06$}}%
      \put(946,3644){\makebox(0,0)[r]{\strut{}$0.08$}}%
      \put(946,4379){\makebox(0,0)[r]{\strut{}$0.1$}}%
      \put(1078,484){\makebox(0,0){\strut{}$3.5$}}%
      \put(2509,484){\makebox(0,0){\strut{}$3.75$}}%
      \put(3941,484){\makebox(0,0){\strut{}$4$}}%
      \put(5372,484){\makebox(0,0){\strut{}$4.25$}}%
      \put(6803,484){\makebox(0,0){\strut{}$4.5$}}%
      \put(1613,4599){\makebox(0,0){\strut{}$0.07$}}%
      \put(3083,4599){\makebox(0,0){\strut{}$0.075$}}%
      \put(4552,4599){\makebox(0,0){\strut{}$0.08$}}%
      \put(6021,4599){\makebox(0,0){\strut{}$0.085$}}%
    }%
    \gplgaddtomacro\gplfronttext{%
      \csname LTb\endcsname%
      \put(176,2541){\rotatebox{-270}{\makebox(0,0){\strut{}$\text{arg}F_{0}(s),~\delta_{0}^{0}(s)~\text{in~rad}$}}}%
      \put(3940,154){\makebox(0,0){\strut{}$s$ in $M_{\pi}^{2}$}}%
      \put(3940,4928){\makebox(0,0){\strut{}$s$ in GeV$^{2}$}}%
      \csname LTb\endcsname%
      \put(2398,4151){\makebox(0,0)[r]{\strut{}$\text{arg}F_{0}(s)$}}%
      \csname LTb\endcsname%
      \put(2398,3821){\makebox(0,0)[r]{\strut{}$\delta_{0}^{0}(s)$}}%
      \csname LTb\endcsname%
      \put(946,704){\makebox(0,0)[r]{\strut{}$0$}}%
      \put(946,1439){\makebox(0,0)[r]{\strut{}$0.02$}}%
      \put(946,2174){\makebox(0,0)[r]{\strut{}$0.04$}}%
      \put(946,2909){\makebox(0,0)[r]{\strut{}$0.06$}}%
      \put(946,3644){\makebox(0,0)[r]{\strut{}$0.08$}}%
      \put(946,4379){\makebox(0,0)[r]{\strut{}$0.1$}}%
      \put(1078,484){\makebox(0,0){\strut{}$3.5$}}%
      \put(2509,484){\makebox(0,0){\strut{}$3.75$}}%
      \put(3941,484){\makebox(0,0){\strut{}$4$}}%
      \put(5372,484){\makebox(0,0){\strut{}$4.25$}}%
      \put(6803,484){\makebox(0,0){\strut{}$4.5$}}%
      \put(1613,4599){\makebox(0,0){\strut{}$0.07$}}%
      \put(3083,4599){\makebox(0,0){\strut{}$0.075$}}%
      \put(4552,4599){\makebox(0,0){\strut{}$0.08$}}%
      \put(6021,4599){\makebox(0,0){\strut{}$0.085$}}%
    }%
    \gplbacktext
    \put(0,0){\includegraphics{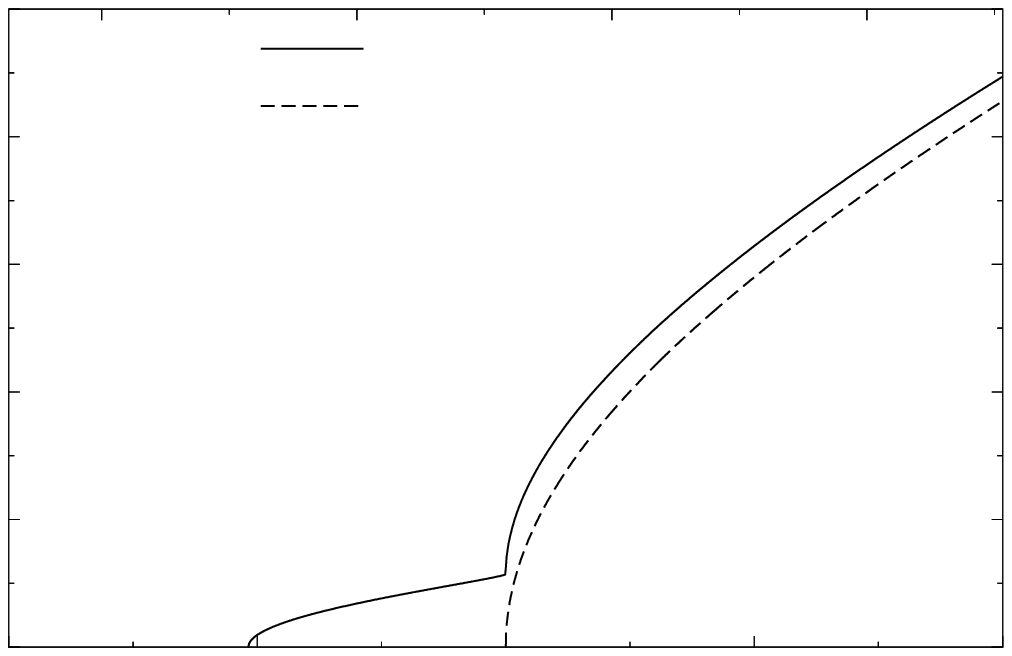}}%
    \gplfronttext
  \end{picture}%
\endgroup

%% file: plots/isobreak-phase-difference.tex
\begingroup
  \makeatletter
  \providecommand\color[2][]{%
    \GenericError{(gnuplot) \space\space\space\@spaces}{%
      Package color not loaded in conjunction with
      terminal option `colourtext'%
    }{See the gnuplot documentation for explanation.%
    }{Either use 'blacktext' in gnuplot or load the package
      color.sty in LaTeX.}%
    \renewcommand\color[2][]{}%
  }%
  \providecommand\includegraphics[2][]{%
    \GenericError{(gnuplot) \space\space\space\@spaces}{%
      Package graphicx or graphics not loaded%
    }{See the gnuplot documentation for explanation.%
    }{The gnuplot epslatex terminal needs graphicx.sty or graphics.sty.}%
    \renewcommand\includegraphics[2][]{}%
  }%
  \providecommand\rotatebox[2]{#2}%
  \@ifundefined{ifGPcolor}{%
    \newif\ifGPcolor
    \GPcolorfalse
  }{}%
  \@ifundefined{ifGPblacktext}{%
    \newif\ifGPblacktext
    \GPblacktexttrue
  }{}%
  \let\gplgaddtomacro\g@addto@macro
  \gdef\gplbacktext{}%
  \gdef\gplfronttext{}%
  \makeatother
  \ifGPblacktext
    \def\colorrgb#1{}%
    \def\colorgray#1{}%
  \else
    \ifGPcolor
      \def\colorrgb#1{\color[rgb]{#1}}%
      \def\colorgray#1{\color[gray]{#1}}%
      \expandafter\def\csname LTw\endcsname{\color{white}}%
      \expandafter\def\csname LTb\endcsname{\color{black}}%
      \expandafter\def\csname LTa\endcsname{\color{black}}%
      \expandafter\def\csname LT0\endcsname{\color[rgb]{1,0,0}}%
      \expandafter\def\csname LT1\endcsname{\color[rgb]{0,1,0}}%
      \expandafter\def\csname LT2\endcsname{\color[rgb]{0,0,1}}%
      \expandafter\def\csname LT3\endcsname{\color[rgb]{1,0,1}}%
      \expandafter\def\csname LT4\endcsname{\color[rgb]{0,1,1}}%
      \expandafter\def\csname LT5\endcsname{\color[rgb]{1,1,0}}%
      \expandafter\def\csname LT6\endcsname{\color[rgb]{0,0,0}}%
      \expandafter\def\csname LT7\endcsname{\color[rgb]{1,0.3,0}}%
      \expandafter\def\csname LT8\endcsname{\color[rgb]{0.5,0.5,0.5}}%
    \else
      \def\colorrgb#1{\color{black}}%
      \def\colorgray#1{\color[gray]{#1}}%
      \expandafter\def\csname LTw\endcsname{\color{white}}%
      \expandafter\def\csname LTb\endcsname{\color{black}}%
      \expandafter\def\csname LTa\endcsname{\color{black}}%
      \expandafter\def\csname LT0\endcsname{\color{black}}%
      \expandafter\def\csname LT1\endcsname{\color{black}}%
      \expandafter\def\csname LT2\endcsname{\color{black}}%
      \expandafter\def\csname LT3\endcsname{\color{black}}%
      \expandafter\def\csname LT4\endcsname{\color{black}}%
      \expandafter\def\csname LT5\endcsname{\color{black}}%
      \expandafter\def\csname LT6\endcsname{\color{black}}%
      \expandafter\def\csname LT7\endcsname{\color{black}}%
      \expandafter\def\csname LT8\endcsname{\color{black}}%
    \fi
  \fi
    \setlength{\unitlength}{0.0500bp}%
    \ifx\gptboxheight\undefined%
      \newlength{\gptboxheight}%
      \newlength{\gptboxwidth}%
      \newsavebox{\gptboxtext}%
    \fi%
    \setlength{\fboxrule}{0.5pt}%
    \setlength{\fboxsep}{1pt}%
\begin{picture}(7200.00,5040.00)%
    \gplgaddtomacro\gplbacktext{%
      \csname LTb\endcsname%
      \put(1210,704){\makebox(0,0)[r]{\strut{}$-0.002$}}%
      \put(1210,1229){\makebox(0,0)[r]{\strut{}$0$}}%
      \put(1210,1754){\makebox(0,0)[r]{\strut{}$0.002$}}%
      \put(1210,2279){\makebox(0,0)[r]{\strut{}$0.004$}}%
      \put(1210,2804){\makebox(0,0)[r]{\strut{}$0.006$}}%
      \put(1210,3329){\makebox(0,0)[r]{\strut{}$0.008$}}%
      \put(1210,3854){\makebox(0,0)[r]{\strut{}$0.01$}}%
      \put(1210,4379){\makebox(0,0)[r]{\strut{}$0.012$}}%
      \put(1590,484){\makebox(0,0){\strut{}$4$}}%
      \put(2583,484){\makebox(0,0){\strut{}$8$}}%
      \put(3576,484){\makebox(0,0){\strut{}$12$}}%
      \put(4569,484){\makebox(0,0){\strut{}$16$}}%
      \put(5562,484){\makebox(0,0){\strut{}$20$}}%
      \put(6555,484){\makebox(0,0){\strut{}$24$}}%
      \put(1872,4599){\makebox(0,0){\strut{}$0.1$}}%
      \put(3146,4599){\makebox(0,0){\strut{}$0.2$}}%
      \put(4420,4599){\makebox(0,0){\strut{}$0.3$}}%
      \put(5694,4599){\makebox(0,0){\strut{}$0.4$}}%
    }%
    \gplgaddtomacro\gplfronttext{%
      \csname LTb\endcsname%
      \put(176,2541){\rotatebox{-270}{\makebox(0,0){\strut{}$\text{arg}F_{0}(s)-\delta_{0}^{0}(s)~\text{in~rad}$}}}%
      \put(4072,154){\makebox(0,0){\strut{}$s$ in $M_{\pi}^{2}$}}%
      \put(4072,4928){\makebox(0,0){\strut{}$s$ in GeV$^{2}$}}%
      \csname LTb\endcsname%
      \put(5816,4151){\makebox(0,0)[r]{\strut{}$\text{arg}F_{0}(s)-\delta_{0}^{0}(s)$}}%
      \csname LTb\endcsname%
      \put(1210,704){\makebox(0,0)[r]{\strut{}$-0.002$}}%
      \put(1210,1229){\makebox(0,0)[r]{\strut{}$0$}}%
      \put(1210,1754){\makebox(0,0)[r]{\strut{}$0.002$}}%
      \put(1210,2279){\makebox(0,0)[r]{\strut{}$0.004$}}%
      \put(1210,2804){\makebox(0,0)[r]{\strut{}$0.006$}}%
      \put(1210,3329){\makebox(0,0)[r]{\strut{}$0.008$}}%
      \put(1210,3854){\makebox(0,0)[r]{\strut{}$0.01$}}%
      \put(1210,4379){\makebox(0,0)[r]{\strut{}$0.012$}}%
      \put(1590,484){\makebox(0,0){\strut{}$4$}}%
      \put(2583,484){\makebox(0,0){\strut{}$8$}}%
      \put(3576,484){\makebox(0,0){\strut{}$12$}}%
      \put(4569,484){\makebox(0,0){\strut{}$16$}}%
      \put(5562,484){\makebox(0,0){\strut{}$20$}}%
      \put(6555,484){\makebox(0,0){\strut{}$24$}}%
      \put(1872,4599){\makebox(0,0){\strut{}$0.1$}}%
      \put(3146,4599){\makebox(0,0){\strut{}$0.2$}}%
      \put(4420,4599){\makebox(0,0){\strut{}$0.3$}}%
      \put(5694,4599){\makebox(0,0){\strut{}$0.4$}}%
    }%
    \gplbacktext
    \put(0,0){\includegraphics{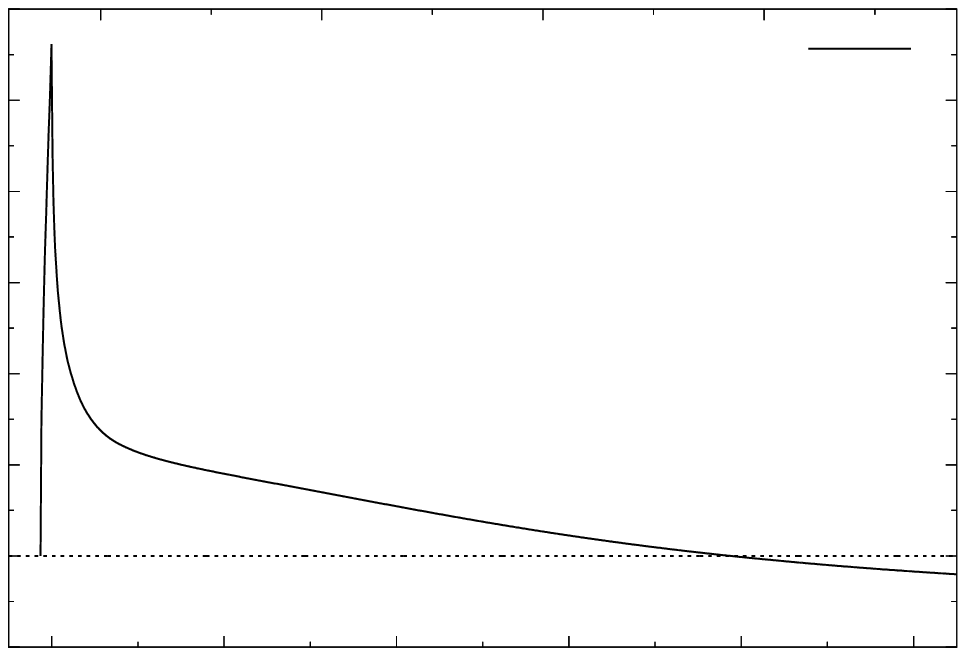}}%
    \gplfronttext
  \end{picture}%
\endgroup